\documentclass[american,english,prd,url,aps,preprintnumbers,superscriptaddress,nofootinbib]{revtex4-1}
\usepackage[T1]{fontenc}
\usepackage[utf8]{inputenc}
\usepackage{color}
\usepackage{amsmath}
\usepackage{amssymb}
\usepackage{graphicx}
\usepackage{subfig}
\usepackage[unicode=true,pdfusetitle,
 bookmarks=true,bookmarksnumbered=false,bookmarksopen=false,
 breaklinks=false,pdfborder={0 0 1},colorlinks=false]
 {hyperref}
\usepackage{breakurl}

\makeatletter

\newcommand{\lyxmathsym}[1]{\ifmmode\begingroup\def\b@ld{bold}
  \text{\ifx\math@version\b@ld\bfseries\fi#1}\endgroup\else#1\fi}

\providecommand{\tabularnewline}{\\}

\@ifundefined{textcolor}{}
{%
 \definecolor{BLACK}{gray}{0}
 \definecolor{WHITE}{gray}{1}
 \definecolor{RED}{rgb}{1,0,0}
 \definecolor{GREEN}{rgb}{0,1,0}
 \definecolor{BLUE}{rgb}{0,0,1}
 \definecolor{CYAN}{cmyk}{1,0,0,0}
 \definecolor{MAGENTA}{cmyk}{0,1,0,0}
 \definecolor{YELLOW}{cmyk}{0,0,1,0}
}

\usepackage{mathrsfs}\usepackage{amsfonts}\usepackage{array}\usepackage{epsfig}

\@ifundefined{definecolor}
 {\usepackage{color}}{}
\usepackage{multirow}

\makeatother

\begin{document}

\preprint{SMU-HEP-19-02}

\title{
Charting the coming synergy between lattice QCD and high-energy phenomenology
}

\author{T. J. Hobbs}
\email{tjhobbs@smu.edu}

\affiliation{Department of Physics, Southern Methodist University,\\
 Dallas, TX 75275-0175, U.S.A. }

\affiliation{Jefferson Lab, EIC Center, Newport News, VA 23606, U.S.A.}

\author{Bo-Ting Wang}
\email{botingw@mail.smu.edu}

\author{Pavel~M. Nadolsky}
\email{nadolsky@physics.smu.edu}

\author{Fredrick I. Olness}
\email{olness@smu.edu}

\affiliation{Department of Physics, Southern Methodist University,\\
 Dallas, TX 75275-0175, U.S.A. }

\begin{abstract}
	Building upon the \texttt{PDFSense} framework developed in Ref.~\cite{Wang:2018heo},
	we perform a comprehensive analysis of the sensitivity of present
	and future high-energy data to a number of quantities commonly evaluated
	in lattice gauge theory, with a particular focus on the integrated
	Mellin moments of nucleon parton distribution functions (PDFs), such
	as $\langle x \rangle_{u^+ - d^+}$ and $\langle x \rangle_{g}$, as
	well as $x$-dependent quark quasi-distributions --- in particular,
	that of the isovector combination. Our results
	demonstrate the potential for lattice calculations and phenomenological
	quark distributions informed by high-energy experimental data
	to cooperatively improve the picture of the nucleon's collinear
	structure. This will increasingly be the case as computational resources
	for lattice calculations further expand, and QCD global analyses continue to 
	grow in sophistication. Our sensitivity analysis suggests that a future
	lepton-hadron collider would be especially instrumental in providing phenomenological
	constraints to lattice observables.
\end{abstract}

\date{\today}

\maketitle
\tableofcontents{}

\newpage

\section{Introduction}
\label{sec:intro}
Owing to steady theoretical progress and the growing availability of computational resources, the
ability of perturbative QCD (pQCD) to predict parton-level processes at high energies
has continued to improve in recent years, with accuracies now reaching next-to-next-to-leading
order (N$^2$LO) in many circumstances. Inevitably, however, predictions
for experiments involving hadronic collisions require precise knowledge
of the structure of initial-state hadrons
at comparatively small energy scales similar to the nucleon mass, $\Lambda\! \sim\! M$,
at which $\alpha_s(\Lambda)\! \sim\! 1$ is too large to permit a converging diagrammatic expansion
of the relevant amplitudes. This general consequence of the negative $\beta$-function of QCD
is realized in the theory of spin-averaged deeply-inelastic lepton-nucleon scattering, for example, in the
factorization of physical cross sections into perturbatively calculable
short-distance matrix elements and inherently nonperturbative long-distance parton distribution functions (PDFs),
$q(x,\mu)$, of the quark-to-hadron light-front momentum fraction $x = k^+/p^+$ and factorization scale $\mu$.

Given the nonperturbative nature of the collinear PDFs, the prevailing recourse has traditionally been either to fit them in comprehensive
analyses of global data using flexible parametric forms \cite{Accardi:2016qay,Harland-Lang:2014zoa,Ball:2017nwa,Dulat:2015mca,Hou:2016nqm}, or to calculate them
in the context of models or effective theories
\cite{Speth:1996pz,Kumano:1997cy,Bourrely:2001du,Holt:2010vj,Avakian:2010br,Bednar:2018htv,Hobbs:2014lea,Melnitchouk:1999mv,Hobbs:2013bia,Hobbs:2017fom}
that aim to capture specific aspects of QCD --- {\it e.g.}, its pattern of dynamical chiral symmetry breaking. Determination of precise effective parametrizations of the collinear PDFs in the $\overline{MS}$ factorization scheme from the global QCD analysis of experimental measurements has grown into a multifaceted research field -- for reviews of its status see, {\it e.g.}, Refs.~\cite{Gao:2017yyd, Kovarik:2019xvh}.
Parallel to these efforts, the past couple of decades have seen a complementary effort
founded in the use of lattice gauge theory techniques to either indirectly compute the $x$ dependence
of the PDFs themselves, or, at minimum, determine the integrated moments of the parton distributions
in Mellin space. (For a comprehensive review, we refer the reader to the recent white paper, Ref.~\cite{Lin:2017snn}.)

By definition, the PDFs are intrinsically nonlocal correlation functions constructed between parton fields
with lightlike spacetime separation along the $\xi^-$ direction at fixed light-front time $\xi^+ = 0$: {\it viz.},
$\sim\!\! \langle p| \bar{q}(\xi^-)\, \hat{\mathcal{O}}\, q(0) |p \rangle$. Dynamically simulating such matrix elements
on a hypercubic lattice, however, is numerically problematic, given the fact
\footnote{In defining light-front variables, we assume a convention in which the components of
an arbitrary $4$-vector $a^\mu$ transform as $a^\pm = a^0 \pm a^3$.}
that $x^2 = x^+ x^- - x^2_\perp = 0$ can only trivially hold at the origin in a Euclidean spacetime, for which
$x^2_\mathrm{E} = x^2_1 + \dots + x^2_4$.
In contrast, the integrated {\it Mellin moments} of the quark distributions have a direct interpretation
in terms of the matrix elements of local operators and can be accessed on a Euclidean lattice
via an operator product expansion (OPE). Moments computed in this fashion are informative
in the sense that they encapsulate aspects of the nonperturbative dynamics responsible for
a hadron's low energy structure --- for instance, the magnitude of the nucleon's collinear
magnitude carried by its total $u$-quark content,
\begin{equation}
\langle x \rangle_{u^+} = \int_0^1 dx\, x [ u + \overline{u} ] \big( x, \mu_F \big)\ .
\label{eq:umom}
\end{equation}
Lattice calculations generally evaluate moments like Eq.~(\ref{eq:umom}) using a scheme
and renormalization scale $\mu^\mathrm{lat}$ chosen to match the
$\overline{\mathrm{MS}}$ scheme usually employed by phenomenologists. Most often in the literature, this scale is taken to be
$\mu^\mathrm{lat} = 2$ GeV, and in this analysis we shall for consistency compute moments at a matching factorization scale,
$\mu_F = \mu^\mathrm{lat} = 2$ GeV, unless otherwise indicated.
Various attempts have been made to determine the $x$ dependence of the PDFs by computing a sufficient
number of moments in Mellin space that the transform into PDF space can be determined (typically with the help of some
parametrization). In practice, however, the mixing among operators of successively higher spin
and the resulting signal-to-noise issues become less controlled as additional covariant derivatives are inserted to
obtain PDF moments of higher order. In effect, only a small number of moments can be accessed on the lattice --- presently,
up to the quark distributions' third moment, $\langle x^3 \rangle_{q^+}$ (although there are recent suggestions that perhaps several more may become
available in the near future). It should be noted that the uncertainties of the lattice moments typically
grow with increasing order.
Still, the ostensible ability of lattice gauge theory to access even several moments of the PDFs has
long presented the possibility of determining (or at least constraining) the parton distributions directly from a first-principles
QCD calculation. Indeed, with a sufficiently restrictive parametric form for the
quark distribution of a given flavor, the latter can be fully determined given enough moments \cite{Detmold:2003rq}; for example,
if the PDF of the $u$-quark distribution is taken to have a very simple $x$ dependence
given by
$u(x, Q_0) = \alpha\, x^\beta\, (1-x)^\gamma$,
knowledge of 3 distinct moments would in principle be adequate to parametrically determine
(up to some uncertainty) the above-noted distribution. At the same time, however, both the
diversity of the experimental data inputs and sophistication of modern QCD analyses are such that
much more flexible parametric forms are required, and lattice calculations remain
far below the requisite level of precision across the many flavors and moment orders needed
to be competitive in a complete determination of the PDFs according to such a
procedure.

More recently, a promising method which may allow the calculation of the PDFs' $x$ dependence on
the QCD lattice in terms of parton {\it quasi-distribution} functions (qPDFs) has been introduced by
Ji \cite{Ji:2013dva}, as well as the more recent pseudo-PDF concept first developed in
Ref.~\cite{Radyushkin:2017cyf} of Radyushkin.
Extracting information from quasi-distributions requires an accompanying large momentum effective theory
(LaMET) for performing the necessary ultraviolet matchings that are realized as convolutional relations
of the form 
\begin{equation}
	\widetilde{q}(x, P_z, \widetilde{\mu})\ = \int\, dy\, Z\left( {x \over y}, {\Lambda \over P_z}, {\mu \over P_z} \right) q(y, \mu)\,
	+\, \mathcal{O} \left( {\Lambda^2 \over P^2_z}\, , {M^2 \over P^2_z} \right)\ ,
\label{eq:match}
\end{equation}
which relate the quasi-distribution $\widetilde{q}$ to the traditional phenomenological PDF $q$ with the usual support
over $x \in [0,1]$.\footnote{Traditional PDFs are correlations at a fixed light-front time $\xi^+ = 0$
and functions of the parton-to-hadron momentum fraction $x = k^+/P^+$. The qPDFs, on the other
hand, are analogous fractions of longitudinal momentum $x = k_z/P_z$, which can be matched
in the context of LaMET via Eq.~(\ref{eq:match}). In the subsequent plots mapping the sensitivity of
CTEQ-TEA data to qPDFs, we compute the qPDF at a specific $x = k_z/P_z$ by means of
Eq.~(\ref{eq:match}), with the sensitivity computed on the basis of the variation
of the underlying PDF appearing inside the convolution. For this reason, sensitivity
plots are always shown as maps giving the pull of data at a given, matched $x_i \in [0,1]$
and $\mu_i$ to the qPDF at a specific $x = k_z/P_z \in (-\infty, \infty)$.
 }
This matching depends critically upon the pQCD-calculable ultraviolet matching function, $Z$.
In practice, the quasi-distribution $\widetilde{q}(x, P_z, \widetilde{\mu})$ [{\it i.e.}, the
left-hand side of Eq.~(\ref{eq:match})] may be evaluated on the lattice for a specific
choice of the longitudinal hadron momentum $P_z$, and the usual PDF extracted by
numerical inversion of Eq.~(\ref{eq:match}). This method therefore has the potential to yield information
on the $x$ dependence of the PDFs themselves, up to knowledge of dynamical and mass-dependent corrections,
the perturbative order of the matching kernel $Z$, and technical details of the actual lattice calculation
--- for instance, artifacts arising from the finite lattice spacing or signal-to-noise problems.
In addition, it should be pointed out that limitations to this procedure remain, especially given the
fact that the lattice calculations and LaMET procedure are in a relatively early stage
of theoretical development --- much as there are limitations to the lattice
computed PDF moments as well.

For the reasons noted above, as computational resources continue to grow, it will be necessary to reconcile
the output of lattice-based methods (especially, concerning the PDF moments and quasi-distributions) with work
in the context of QCD global analyses. This will necessarily go both directions: benchmarking the lattice calculations with knowledge of the PDFs
from phenomenological analyses, and constraining QCD analyses with the output of the lattice.
Laying the groundwork for this synergy will require a comprehensive understanding
of the relation between phenomenological constraints placed on the PDFs determined
in fits (and, by extension, the $x$-weighted moments computed therefrom) and information
obtained from the lattice.

\begin{figure}
\includegraphics[scale=0.7]{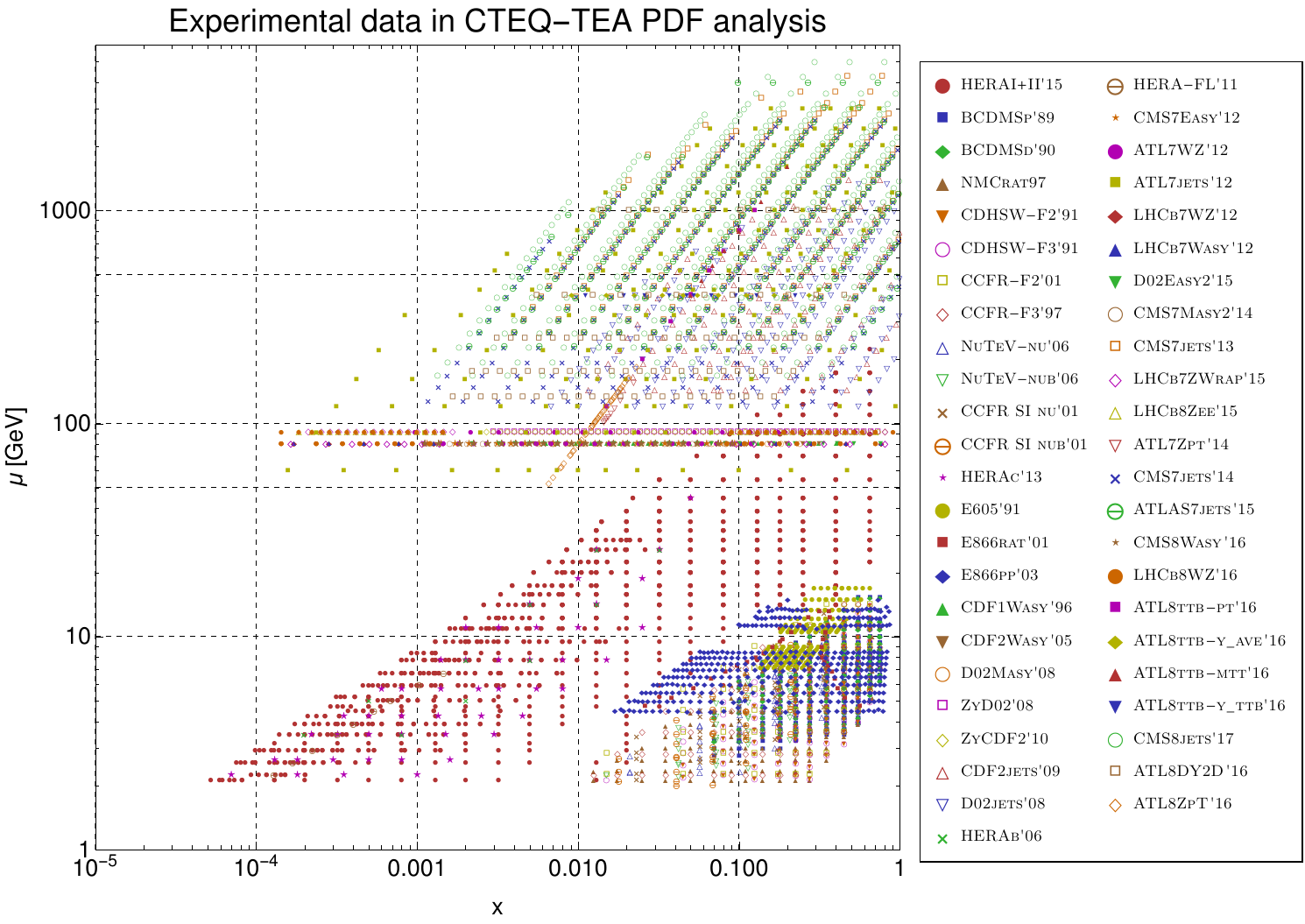}
\caption{A graphical representation of the space of $(x,\mu)$ points probed
by the full data set treated in Ref.~\cite{Wang:2018heo} as well as the present
analysis, designated ``CTEQ-TEA.'' It corresponds to an expansion of the CT14HERA2 NNLO data
fitted in the most recent CT14 framework, including
measurements from Run II of HERA. We stress that the $x$ and $\mu$
which define the axes in this and subsequent plots represent the momentum
fraction $x$ and scale $\mu$ matched from empirical data as described in Appendix
A of Ref.~\cite{Wang:2018heo}. These $(x,\mu)$ values are not to be confused
with the $x$ of the qPDF in Eq.~(\ref{eq:match}) or $\mu_F$ used to evaluate
PDF moments. We direct the reader to the footnote near Eq.~(\ref{eq:match})
for further clarification of this point.
}
\label{fig:xQ_CT14HERA2_LHC}
\end{figure}

In this analysis we systematically canvass these issues, using the recently developed \texttt{PDFSense}
framework of Ref.~\cite{Wang:2018heo} to present a comprehensive summary of the differential impact modern
data have upon knowledge of the PDF Mellin moments evaluated from phenomenological fits --- as well as which
data give the leading contributions to present understanding of one of the typical quasi-distributions, that of the isovector
combination, $u\!-\!d$. The remainder of the paper is therefore as
follows.

After a brief review of the formalism and a
description of the \texttt{PDFSense} methodology, in Sec.~\ref{sec:moments}
we review the constraints HEP data place on the lowest moments for several light quark $q^\pm$ distributions
and the gluon. We also discuss the degree of the agreement among the
main experiments that constrain the relevant PDF flavors at
$x>0.1$. For this purpose, we introduce a new statistical indicator,
$L_2$ sensitivity, which is especially convenient for exploring the
aggregate constraints of the individual experiments on the PDFs, and
the pulls they exert on the PDFs in various momentum fraction intervals.

In Sec.~\ref{sec:quasi}, we illustrate the constraints that the data place on the $u\!-\!d$ quasi-distributions
by computing them from the CT14HERA2 NNLO PDFs \cite{Hou:2016nqm}
at several choices of the momentum fraction, $\pm x$, and hadronic boost momentum, $P_z$, using Eq.~(\ref{eq:match}).
Sec.~\ref{sec:EIC} demonstrates the sizable potential impact future measurements at a high-luminosity
lepton-hadron collider will have on these quantities by analyzing pseudodata generated from a combination
of Monte Carlo-generated information with cross section predictions based upon CT14HERA2 NNLO. 
Sec.~\ref{sec:implem} contains a number of conclusions drawn from our analysis of
the PDF moments and qPDFs regarding expected consequences of implementing lattice information
into future global fits.
In Sec.~\ref{sec:conc} we provide a number of closing
observations, focusing on points that will allow this work to be leveraged in the future. Lastly,
Appendix~\ref{sec:append} collects several tables --- counterpart to
Figs.~\ref{fig:rank_plot} and~\ref{fig:rank_plot_qPDF} --- summarizing the aggregated impact on lattice
QCD observables of the HEP experiments considered in this work.

\section{The Sensitivity of HEP data to PDF Mellin Moments}
\label{sec:moments}
\subsection{Theory of PDF Mellin moments}
\label{sec:thy-mell}
The $x$-weighted moments of the PDFs have long been of interest to practitioners of QCD
analyses on the logic that they may provide the necessary input to
either reconstruct or at least constrain the PDFs determined in
global fits.
The accessibility of these moments to lattice gauge techniques is
facilitated by the OPE \cite{Zimmermann:1972tv,Wilson:1969zs,Christ:1972ms,Gross:1973ju,Gross:1974cs,Georgi:1951sr},
which allows an expansion of the PDF moments in terms of matrix elements of
well-defined, local twist-$2$ operators which can be evaluated in a
discretized Euclidean spacetime. Subsequently, the Mellin moments themselves
may be derived via algebraic relations from the matrix elements of these twist-$2$
operators. In principle, it is possible to reconstruct
a given PDF's $x$ dependence via an inverse Mellin transform {\it if}
its moments $\langle x^n \rangle_q$ are known to all orders, as noted in
Sec.~\ref{sec:intro}.

The crucial relation that connects $x$-dependent parton distributions $q(x)$ to
an $n$-dependent tower of integrals in Mellin space is the inverse Mellin transform, which
enables one in principle to reconstruct integrated Mellin moments of the PDFs.
These PDF Mellin moments have the general form
\begin{equation}
\langle x{}^{n}\rangle_q=\int_{0}^{1}dx\ x^{n}\, [ q(x)+(-1)^{n+1}\,\overline{q}(x) ]\, .
\label{eq:moment}
\end{equation}
Using Eq.~(\ref{eq:moment}), it is possible to define a collection of PDF moments $\langle x^n \rangle_{q^\pm}$
which are actually calculable on the QCD lattice, such as $\langle x \rangle_u^+$ of Eq.~(\ref{eq:umom}). These are
\begin{subequations}
\begin{equation}
	\langle x^n\rangle_{q^+} = \langle x^n\rangle_{q}\ \ \mathrm{for}\ \ n = 2\ell-1,
\end{equation}
\begin{equation}
	\langle x^n\rangle_{q^-} = \langle x^n\rangle_{q}\ \ \mathrm{for}\ \ n = 2\ell,
\end{equation}
\label{eq:momlat}
\end{subequations}
where $\ell \in \mathbb{Z}^+$ such that the lattice may provide, for instance,
$\langle x \rangle_{u^+}$, $\langle x^2 \rangle_{u^-}$, $\langle x^3 \rangle_{u^+}$,
{\it etc.}
Moreover, the PDFs themselves can be unfolded from a complete set of Mellin moments
via the inverse Mellin transform,
\begin{equation}
q(x)+(-1)^{n+1}\,\overline{q}(x)=\frac{1}{2\pi i}\int_{c-i\infty}^{c+i\infty}dn\,x^{-n-1}\, \langle x{}^{n}\rangle_{q}\ .
\end{equation}
In practice, however, lattice QCD techniques currently enable the calculation of the few lowest PDF moments.
Once the accuracy of these computations improves, theoretical constraints in the form of lattice-calculated
PDF moments (or $x$-dependent information unfolded from parton quasi-distributions,
discussed in Sec.~\ref{sec:quasi}) might eventually be implemented as $\chi^2$ penalties
in global QCD analyses --- essentially,
taking lattice data as theoretical priors to constrain the likelihood
function of a global fit. For example, exploratory studies based on Bayesian
profiling have suggested that lattice calculations even with somewhat
sizable uncertainties can still provide powerful constraints to
PDFs in regions that are relatively unconstrained by experimental data
(see Ref.~\cite{Lin:2017snn} and references therein).
In the remainder of this subsection, we review the essential theory
for accessing the integrated moments of the PDFs, with a
special focus on unpolarized distributions, given their
importance to high-energy phenomenology.
(Although it is worth noting that, given the comparative paucity of
spin-dependent data, it is reasonable to expect that lattice calculations for
the moments of helicity distributions may more quickly become competitive
against fitted spin-dependent PDFs; we defer such considerations, however, to future
work.)

The OPE expands hadronic matrix elements of non-local
products of field operators in terms of local operator matrix elements weighted by Wilson
coefficients that obey renormalization group evolution \cite{Wilson:1969zs,Georgi:1976ve}. It is then possible
to calculate these coefficients in the context of QCD perturbatively ({\it i.e.}, they embody
the relevant short-distance dynamics), while the local operator matrix
elements are nonperturbative (depending on the details of the long-distance physics).
In the case relevant for the present study, the matrix elements of twist-$2$ operators may be expanded by
the OPE as
\begin{equation}
\frac{1}{2}\sum_{s}\langle p,s|\mathcal{O}_{\{\mu_{1},\cdots,\mu_{n+1}\}}^{q}|p,s\rangle={}2v_{q}^{n+1}\,[p_{\mu_{1}}\cdots p_{\mu_{n+1}}-{\rm traces}]\ ,
\label{eq:OPE_unp}
\end{equation}
\begin{equation}
\langle p,s|\mathcal{O}_{\{\sigma\mu_{1},\cdots,\mu_{n+1}\}}^{5\,q}|p,s\rangle={}\frac{1}{n+2}a_{q}^{n+1}\,[s_{\sigma}p_{\mu_{1}}\cdots p_{\mu_{n+1}}-{\rm traces}]\ ,
\label{eq:OPE_pol}
\end{equation}
where $p$ and $s$ represent the nucleon $4$-momentum and spin, respectively, $q$ indicates the flavor
of the relevant quark field, and $\{\}$ stands for
index symmetrization.
Higher-twist terms enter as power suppressed corrections in $1/Q^2$ \cite{Blumlein:2008kz},
so here we only consider the contribution of Mellin moments from twist-$2$ operators. 
For the quark fields, the twist-$2$ operators occurring in the OPE expressions above are constructed
from the usual bilinears as
\begin{equation}
	\mathcal{O}_{\{\mu_{1},\cdots,\mu_{n+1}\}}^{q} = \ensuremath{{\displaystyle \left({i \over 2}\right)^{n}\bar{q}(x)\gamma_{\{\mu_{1}}\overleftrightarrow{D}_{\mu_{2}}...\overleftrightarrow{D}_{\mu_{n+1}\}}q(x)}}
\label{eq:ops}
\end{equation}
and
\begin{equation}
	\mathcal{O}_{\{\sigma\mu_{1},\cdots,\mu_{n+1}\}}^{5\,q} = \ensuremath{{\displaystyle \left({i \over 2}\right)^{n+1}\bar{q}(x)\gamma_{\{\sigma}\gamma_{5}\overleftrightarrow{D}_{\mu_{1}}...\overleftrightarrow{D}_{\mu_{n+1}\}}q(x)}},
\label{eq:pol2}
\end{equation}
where 
\begin{equation}
\overleftrightarrow{\cal D}_{\mu}\,=\,\frac{1}{2}\left(\overrightarrow{\cal D}_{\mu}-\overleftarrow{\cal D}_{\mu}\right),\qquad\overrightarrow{\cal D}_{\mu}\,=\,\overrightarrow{\partial}_{\mu}-\,ig\,t_{a}A_{\mu}^{a}(z),\qquad\overleftarrow{\cal D}_{\mu}\,=\,\overleftarrow{\partial}_{\mu}+\,ig\,t_{a}A_{\mu}^{a}(z)
\end{equation}
are the gauge covariant derivatives, $A_{\mu}^{a}(z)$ denotes gluon fields, and $t_a$
represents the $8$ standard generators of the $\mathrm{SU}(3)$ color group. We note that
$\{ \cdots \}$ in the expressions above denotes symmetrization over the enclosed indices.
In Eqs.~(\ref{eq:OPE_unp}) and~(\ref{eq:OPE_pol}), $v_{q}^{n+1}$ and $a_{q}^{n}$ are identifiable
with the $n^\mathit{th}$-order moments of the twist-$2$ PDFs of unpolarized and longitudinally polarized
nucleons, respectively \cite{Gockeler:1995wg,Lin:2017snn}:
\begin{subequations}
\begin{equation}
	v_{q}^{n+1}\ =\ \int_0^1 dx\, x^n\, q(x,\, \mu)\ ,
\end{equation}
\begin{equation}
	a_{q}^{n}\   =\ \int_0^1 dx\, x^n\, \Delta q(x,\, \mu)\ .
\label{eq:mom-intsb}
\end{equation}
\label{eq:mom-ints}
\end{subequations}
Lastly, we note that analogous matrix elements are responsible for moments of the gluon distribution,
with the lowest lattice-accessible moment $\langle x \rangle_g$ for the total gluon momentum fraction
given by the insertion of a twist-$2$ operator constructed from the gluon field strength as
$\mathcal{O}^g_{\mu_1 \mu_2} = -\mathrm{Tr} (\mathcal{G}_{\mu_1 \alpha} \mathcal{G}_{\mu_2 \alpha})$.

Insofar as the LHS expressions of Eqs.~(\ref{eq:OPE_unp}) and~(\ref{eq:OPE_pol}) can be formulated
in terms of lattice gauge theory and evaluated on a discretized Euclidean spacetime, the PDF moments in Eq.~(\ref{eq:mom-ints})
are themselves directly accessible on the QCD lattice through
the direct evaluation of nucleonic matrix elements of twist-$2$ operators.
For reasons that we sketch below, however, the extraction of higher moments is complicated
by operator-mixing effects, and modern lattice calculations can reliably extract
only the lowest few moments in practice \cite{Detmold:2001dv,Detmold:2003rq,Detmold:2003tm}.
Various systematic errors in generic lattice calculations are reviewed and assessed in
Ref.~\cite{1607.00299}, with the dominant systematic errors in evaluations of Mellin moments
arising from power-divergent operator mixing and renormalization effects.
Power-divergent mixing is associated with an $O(4)$ symmetry breaking inherent
to the Euclidean spacetime discretization of lattice calculations: the lattice regulator breaks Lorentz symmetry,
causing radiative divergences in operators of different mass dimensions
mix together \cite{Gockeler:1996mu}. The renormalization of non-local
operators on a discretized lattice induces another error: the renormalized
fields that are nonperturbatively determined on the lattice are power-divergent.
In addition to these, a number of other systematic effects generally enter lattice
QCD calculations, including corrections from the unphysically large quark (or pion) masses
often used as well as the associated chiral extrapolations to the physical pion mass. Moreover,
details involved in the extraction of lattice signals as a function of
lattice time contribute to the landscape of systematic uncertainties. The
effort to control these and other lattice artifacts partially depends upon
the ability of lattice practitioners to benchmark their calculations against
phenomenological knowledge. Exploring the capacity of high-energy data
to tighten these benchmarks is a primary motivation for the
present study.
\begin{table}
\begin{tabular*}{0.42\textwidth}{c||c||c}
\hline
PDF moment                             &   {\bf Lattice}                            &    {\bf CT14HERA2 NNLO}       \tabularnewline
\hline                                                                                                      
$\langle x \rangle_{u^+-d^+}$          &   $0.184(32) \star$                        &    $0.159(6)$                              \tabularnewline
$\langle x^2 \rangle_{u^--d^-}$        &   $0.107(98) \star$                        &    $0.055(2)$                              \tabularnewline
$\langle x^3 \rangle_{u^+-d^+}$        &     {\it N/A}                              &    $0.022(1)$                              \tabularnewline
\hline                                                                                                                           
$\langle x \rangle_{g}$                &   $0.267(35)$ \cite{Alexandrou:2017oeh}    &    $0.415(8)$                              \tabularnewline
\hline                                                                                                                           
$\langle x \rangle_{u^+}$              &   $0.453(75)$ \cite{Alexandrou:2017oeh}    &    $0.351(5)$                              \tabularnewline
$\langle x \rangle_{d^+}$              &   $0.259(74)$ \cite{Alexandrou:2017oeh}    &    $0.193(6)$                              \tabularnewline
$\langle x \rangle_{s^+}$              &   $0.092(41)$ \cite{Alexandrou:2017oeh}    &    $0.031(8)$                              \tabularnewline
\hline                                                                                                                           
$\langle x^2 \rangle_{u^-}$            &   $0.117(18)$ \cite{Deka:2008xr}           &    $0.085(1)$                              \tabularnewline
$\langle x^2 \rangle_{d^-}$            &   $0.052(9)$ \cite{Deka:2008xr}            &    $0.030(1)$                              \tabularnewline
$\langle x^2 \rangle_{s^-}$            &      {\it N/A}                             &      ---                                   \tabularnewline
\hline                                                                                                                           
$\langle 1 \rangle_{\bar{d}-\bar{u}}$  &     ---                                    &    $-0.367(410)$                           \tabularnewline
$\kappa^s$                             &   $0.795(95)$ \cite{Liang:2019xdx}         &    $0.459(132)$                            \tabularnewline
\end{tabular*}\caption{
	A table comparing the most recent PDF moment results obtained from lattice QCD calculations (central column)
	to the analogous results determined from fitted PDFs, here, CT14HERA2 NNLO (rightmost column).
	For the former, many reported results can be found in the recent community white paper in Ref.~\cite{Lin:2017snn}, and we show here those lattice results that
	were designated as having ``benchmark'' status, where possible. Those lattice entries corresponding to
	single calculations are given with the associated reference, whereas those which result from an average of several
	lattice extractions are indicated with ``$\star$.'' In particular, the result for $\langle x \rangle_{u^+-d^+}$
	follows from averaging the calculations in Refs.~\cite{Bali:2014gha,Green:2012ud,Alexandrou:2017oeh}, while
	the corresponding result for $\langle x^2 \rangle_{u^--d^-}$ is an average over the result in
	Ref.~\cite{Gockeler:2004wp} and two separate calculations reported in Ref.~\cite{Dolgov:2002zm}.
}
\label{tab:moments}
\end{table}
As for the present status of lattice QCD calculations for some of the lowest
PDF moments, we present in Table~\ref{tab:moments} a numerical comparison of
several recent lattice extractions (rightmost column) with the results of computing
the same moments from CT14HERA2 NNLO \cite{Hou:2016nqm} (middle column).
Notably, despite significant improvements in recent years, the uncertainties of modern
lattice QCD calculations remain considerably larger than the analogous
errors for the moments computed from fitted PDFs, which we quote at $90\%$ C.L.~in
Table~\ref{tab:moments}. There is also an apparent tendency for the lattice-computed moments
to overestimate the phenomenologically determined ones, a behavior that is likely especially
driven by the effects of excited-state contamination. Despite this, however, we typically
find reasonable alignment with lattice moments, partly owing to their larger uncertainties.
As continued improvements tame the lattice effects described above, these uncertainties
are expected to contract, leading to growing utility to global fits of unpolarized
distributions.
It is worth mentioning that, due to a comparatively smaller set of experimental data, the degree to which
lattice calculations must improve to impact QCD analyses for spin-polarized
quantities via Eqs.~(\ref{eq:OPE_pol}), (\ref{eq:pol2}), and (\ref{eq:mom-intsb}) is less pronounced.

\subsection{Analysis procedure}
\label{sec:analysis}
\subsubsection{Correlations $C_f$ and sensitivities $S_f$}
To explore the sensitivity of high-energy data to the PDF Mellin moments and
qPDFs accessible in lattice QCD, in this work we extend the analysis of Ref.~\cite{Wang:2018heo}
that applied the recently developed \texttt{PDFSense} framework to weigh the impact of an extended amalgam of
HEP experimental data (the `CTEQ-TEA'
data, plotted in Fig.~\ref{fig:xQ_CT14HERA2_LHC}) as a part of the preparation for the release of the CT18 PDF
global analysis \cite{Hou:2019efy}.
In this case, special emphasis was placed on the impact
these data might have on the unpolarized collinear PDFs themselves and on observables derived directly
therefrom, including the 14 TeV inclusive Higgs production cross section,
$\sigma_H$. We note that Fig.~\ref{fig:xQ_CT14HERA2_LHC} is reproduced
from Ref.~\cite{Wang:2018heo} and provided here for ease of reference.
Given the fact that a number of lattice QCD observables
are calculable from phenomenological PDFs along the lines
described in Secs.~\ref{sec:intro}, \ref{sec:thy-mell}, and~\ref{sec:quasi}
below, we repeat our analysis to illustrate the constraints a typical
experimental data set can impose on our phenomenological knowledge of
such lattice observables.

We refer the reader to Secs.~II and III of Ref.~\cite{Wang:2018heo} for
a systematic presentation of the details of the \texttt{PDFSense} framework.
Still, it is worthwhile to summarize the particulars of a sensitivity
analysis dedicated to the PDF moments $\langle x^n\rangle_{q^\pm}$.
Whereas in Ref.~\cite{Wang:2018heo} we primarily concentrated on the sensitivities
of data to the local values of the collinear PDFs $q(x_i,\mu_i)$ at
the typical $x_i$ and $\mu_i$ of the high-energy data points
(see Appendix A of the same reference), here we are chiefly concerned with
the sensitivity to Mellin moments for which the $x$ dependence has
been integrated away, and in general at a scale $\mu_F = \mu^\mathrm{lat} = 2$ GeV at which
moments are commonly computed in lattice QCD. Whether by
a Hessian or Monte Carlo error procedure, a PDF global
analysis typically produces a central PDF set and a finite ensemble replicas of the error PDFs,
$q^{j\in{\{2N\}}}(x,\mu_F)$. Given this ensemble, it is then
possible to evaluate a respective error set for values of integrated PDF moments. In the case of the CT
fitting approach, from the underlying Hessian error sets --- of which there
are $2N$ for an $N$-parameter global fit [leading to $1 (56)$ central (error)
PDFs in the $28$-dimensional CT14HERA2 NNLO fit] --- we directly compute error replicas for the
moments by integrating over the CT fitted distributions. Namely,
\begin{equation}
q^{j\in{\{2N\}}}(x,\mu^\mathrm{lat})\ \longrightarrow\
	\langle x^n \rangle^{j\in{\{2N\}}}_{q^\pm,\,\mu^\mathrm{lat}} = \int_0^1 dx\, x^n\, \Big( q(x,\mu^\mathrm{lat}) \pm \overline{q} (x,\mu^\mathrm{lat}) \Big)_{j\in{\{2N\}}}\ .
\label{eq:rep_mom}
\end{equation}
In practice, we evaluate the integrals of Eq.~(\ref{eq:rep_mom}) numerically from grids in which the
bounds of integration are chosen to ensure stable convergence. For instance, we generally evaluate
\begin{equation}
	\int_{x_a}^{x_b} dx\, x^n\, \Big( q(x,\mu^\mathrm{lat}) \pm \overline{q} (x,\mu^\mathrm{lat}) \Big)_{j\in{\{2N\}}}\ ,
\end{equation}
where typically we choose $x_a \sim 10^{-7}$ and $x_b \sim 0.999$. The stability of our results
against variations about these choices has been explicitly verified. As can be seen in
Fig.~\ref{fig:xQ_CT14HERA2_LHC}, the lowest reach of data in the CTEQ-TEA set is
$x\! \sim\! 5 \cdot 10^{-5}$, such that the very low-$x$ contributions to
the moments analyzed here are to be regarded as parametrization dependent. Generally, however, these
contributions are heavily suppressed in moments, $\langle x^n \rangle$, and the overall
parametrization dependence is minimal.

With replica sets for lattice observables like the PDF Mellin moments as in Eq.~(\ref{eq:rep_mom}),
we may deploy the statistical framework of Ref.~\cite{Wang:2018heo}, computing the Pearson
correlation between the residual $r_i(\vec{a})$ of the $i^\mathit{th}$ measurement of
our CTEQ-TEA set (again, evaluated over the $1\ [56]$ central [error] sets of CT14HERA2 NNLO)
and the corresponding ensemble for $\langle x^n
\rangle^{j\in{\{2N\}}}_{q^\pm,\,\mu^\mathrm{lat}}$.
The residual is defined by 
\begin{align}
r_{i}(\vec{a}) & =\frac{1}{s_{i}}\,\big(T_{i}(\vec{a})-D_{i,sh}(\vec{a})\big)\ \label{eq:residual}
\end{align}
as the difference
between the theoretical prediction $T_i(\vec a)$, dependent on the PDF
parameters $\vec a$, and central data value
$D^\mathit{sh}_i(\vec a)$ that is shifted to accommodate systematic error
correlations. This difference is weighted by the total uncorrelated
uncertainty, $s_i$ \cite{Pumplin:2002vw}.

In practice, the correlation is computed using
\begin{align}
	C_f(x_i,\mu_i) =\  &\mathrm{\mbox{Corr}}[f, r_i(x_i,\mu_i)]\ , \nonumber \\
	&\mathrm{\mbox{Corr}}[X,Y]  =\frac{1}{4\Delta X\Delta Y}\sum_{l=1}^{N}(X_{l}^{+}-X_{l}^{-})(Y_{l}^{+}-Y_{l}^{-})\ ,
\label{eq:corr-def}
\end{align}
in which $f$ is a generic function of the PDFs ({\it e.g.}, a PDF of given flavor at the matched
$(x_i,\mu_i)$ of the $i^\mathit{th}$ data point as in Ref.~\cite{Wang:2018heo}, or a PDF moment
computed from the PDFs), and the $N$ $l^\pm$ pairs of Eq.~(\ref{eq:corr-def}) may be identified with the $2N$ Hessian
error sets just described; the uncertainty quantities in the denominator of Eq.~(\ref{eq:corr-def}) are evaluated
from the Hessian error sets as
\begin{equation}
\Delta X = \frac{1}{2}\sqrt{\sum_{l=1}^{N}\left(X_{l}^{+}-X_{l}^{-}\right)^{2}}\ .
\label{eq:hess_unc}
\end{equation}
Equations (\ref{eq:corr-def}) and (\ref{eq:hess_unc}) are 
motivated by the
observation that, given the approximately Gaussian dependence of the probability
distribution on the PDF parameters $\vec a$ in the close vicinity of the global
minimum of $\chi^2$, small displacements of a PDF-dependent quantity
$X$ from its value for the best-fit combination of PDF parameters can
be computed using the $N$-dimensional {\it vector} of the
gradient $\vec\nabla X$ with respect to $\vec a$
\cite{Pumplin:2001ct,Pumplin:2002vw}. Therefore, we can write, using
the angle $\varphi$ between $\vec \nabla X$ and $\vec \nabla Y$: 
\begin{eqnarray}
 \Delta X & = & |\vec \nabla X|,\nonumber \\
\mathrm{\mbox{Corr}}[X,Y] & =& \cos{\varphi}=\frac{\vec \nabla X \cdot
  \vec \nabla Y}{\Delta X \Delta Y}. 
\label{eq:DXCorr}
\end{eqnarray}

One of the principal results of Ref.~\cite{Wang:2018heo} was the demonstration that the Pearson correlation
of Eq.~(\ref{eq:corr-def}) cannot fully capture the phenomenological weight of individual measurements, given the fact that it
does not explicitly depend upon the {\it magnitudes} of the PDF or experimental uncertainties.
For this reason, we introduced a generalization of the correlation we call the \textit{sensitivity},
$S_{f}$, of the $i^\mathit{th}$ point in experiment $E$ to PDF flavor (or PDF-derived quantity) $f$:
\begin{equation}
S_{f}\ =\ \frac{\Delta r_{i}}{\langle r_{0}\rangle_{E}}\,C_{f}\ .
\label{eq:sens}
\end{equation}
 $\Delta r_{i}$ is calculated
using Eq.~(\ref{eq:hess_unc}), and $\langle r_{0}\rangle_{E}$ represents the point-averaged residuals of each of the
points of experiment $E$ computed with the central PDF set.

With the family of PDF error sets of CT14HERA2 NNLO and the statistical
formalism and metric embodied by Eq.~(\ref{eq:sens}), the sensitivities of data points
shown in Fig.~\ref{fig:xQ_CT14HERA2_LHC} may be assessed and mapped in a plane of
typical partonic momentum fraction $x_i$ and factorization scale $\mu_i$
at which a PDF, $q(x_i,\mu_i)$, might be evaluated.
As developed in detail in Appendix A of Ref.~\cite{Wang:2018heo}, the matching of experimental
data to specific values of $x_i$ and $\mu_i$ is based upon a leading-order identification of $x_i$
and $\mu_i$ with scales dictated by external kinematics. For instance, in DIS, we match
the partonic fraction $x_i$ with Bjorken-$x_B$: $x_i\! \approx\! x_B|_i$. Similarly, the QCD factorization
scale associated with a specific DIS measurement is chosen to be
the corresponding virtuality of the exchanged $\gamma$, $\mu_i\! \approx\! Q|_i$. Analogous
relations for data generated by the other processes appearing in Fig.~\ref{fig:xQ_CT14HERA2_LHC}, such as
high-mass Drell-Yan scattering or jet production, appear in Appendix A of Ref.~\cite{Wang:2018heo}.
By mapping sensitivities in this fashion, we may isolate processes and individual data sets with strong
phenomenological pull on lattice observables, while also identifying the most constraining regions of $(x,\mu)$.
The size of the absolute sensitivities
$|S_{\langle x^n\rangle_{q^\pm}}|$ for highlighted points ($|S_{\langle x^n\rangle_{q^\pm}}|>0.25$)
are identified by the ``rainbow stripe'' color palette: hot colors
(red, orange) correspond to strong sensitivities, and cool colors (yellow,
green) correspond to weak sensitivities. Unhighlighted points ---
{\it i.e.}, those with relatively minimal expected impact, ($|S_{\langle x^n\rangle_{q^\pm}}|<0.25$),
are represented with gray colors.

% ~ ~ ~ ~ ~ ~ ~ ~ ~ ~ ~ ~ ~ ~ ~ ~ ~ ~ ~ ~ ~ ~ ~ ~ ~ ~ ~ ~ ~ ~ ~ ~ ~ ~ ~ ~ ~ ~ ~ ~ ~ ~ ~ ~ ~ ~ ~ ~ ~ ~ ~
%
\subsubsection{The $L_2$ sensitivity \label{sec:L2}}
\begin{figure}[b]
\hspace*{-0.75cm}
\includegraphics[scale=0.9]{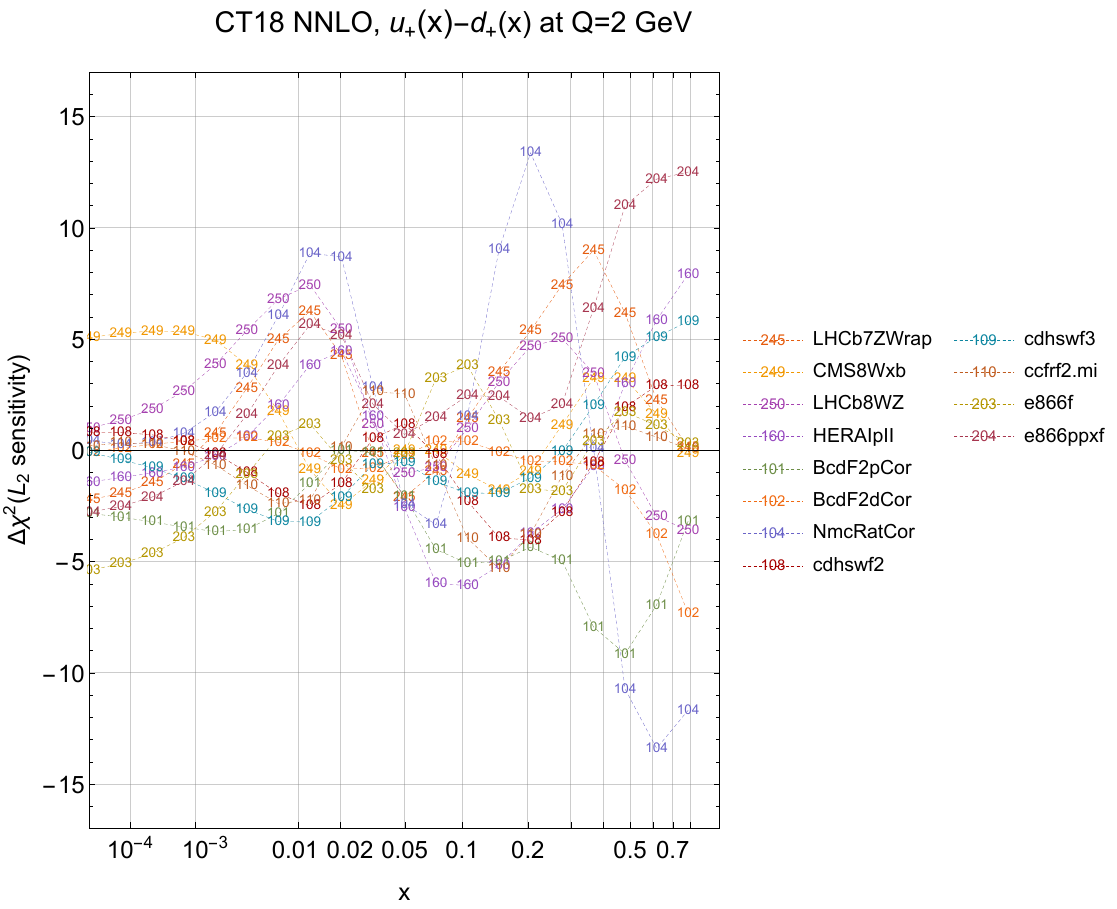}
	\caption{
		An illustration of the $L_2$ sensitivities of the CTEQ-TEA with the strongest
		pulls on the isovector PDF combination, $u^+(x)\! -\! d^+(x)$, at the
		scale $\mu = 2$ GeV as computed using Eq.~(\ref{eq:L2}). Here, we evaluate the $L_2$ sensitivity using a subset of the CTEQ-TEA data prepared for the CT18 NNLO PDF fit \cite{Hou:2019efy}, based on the implementation of LHC Run-1
		data explored in Ref.~\cite{Wang:2018heo}. The curves for
		the fitted experiments reveal a number of competing pulls on the
		isovector $q^+$ distribution: for instance, at $x > 0.1$, 
between
		the E866 $pp$ (204), combined HERA (160), and CDHSW
                (109) data sets, on the
		one hand, and the NMC $F_2^d/F_2^p$ ratio measurements (104) and BCDMS $F^d_2$ data
		(102), on the other.
	}
\label{fig:L2} 
\end{figure}

The quantity $|S_f|$ defined in Eq.~(\ref{eq:sens}) is a convenient
figure of merit that we will use to quantify
the sensitivity of experimental data point residuals to the PDF
dependence of select Mellin moments and qPDFs. It
represents a normalized variation of the data residual $r_i$ when the PDF parameters
are displaced along the direction of maximal variation of $f(x,Q)$, i.e.,
along $\vec\nabla f$, by an amount corresponding to the PDF error
$\Delta f$. 

To further elucidate the meaning of $S_f$, let us briefly discuss a related
quantity, defined by 
\begin{align}
	S_{f,L2}(E)\, &\equiv\, \vec{\nabla} \chi^2_E \cdot {\vec{\nabla} f \over |\vec{\nabla} f|} \nonumber \\
	&= \Delta \chi^2_E(\widehat a_f) = \Delta \chi^2_E \, \cos{\varphi(f,\chi^2_E)}.
\label{eq:L2}
\end{align}
In this equation,
\begin{align}
\chi_{E}^{2}(\vec{a}) & =\sum_{i=1}^{N_{\mathit{pt}}}\,r_{i}^{2}(\vec{a})+\sum_{\alpha=1}^{N_{\lambda}}\overline{\lambda}_{\alpha}^{2}(\vec{a})\ \label{eq:chi2}
\end{align}
is the log-likelihood function for experiment $E$, equal to the
quadratic sum ($L_2$-norm sum) of $N_{\mathit{pt}}$ shifted data point residuals
$r_i$ and $N_{\lambda}$ best-fit nuisance parameters $\overline{\lambda}_{\alpha}$
representing the sources of correlated systematic errors.

In accord with the discussion above, $S_{f,L2}(E)$ is equal to $\Delta
\chi^2_E(\widehat a_f)$, the
variation of $\chi^2_E$ when the PDF
parameters are displaced by a unit vector 
$\widehat a_f\equiv \vec\nabla f/|\vec\nabla f|$ along the gradient of
$f$. As before, the  
displacement $\widehat a_f$ from the best fit changes $f(x,Q)$ by its PDF error $\Delta
f$. But the respective change $\Delta \chi^2_E(\widehat a_f)$ is not
equal to the PDF error $\Delta \chi^2_E$ on $\chi^2_E$.
By applying Eqs.~(\ref{eq:DXCorr}),  one arrives
at a different expression shown as the last equality
on the second line of Eq.~(\ref{eq:L2}), i.e., a product of the PDF
uncertainty $\Delta \chi^2_E$ estimated by Eq.~(\ref{eq:hess_unc}),
and the correlation cosine between $f(x,Q)$ and $\chi^2_E$ given by
Eq.~(\ref{eq:corr-def}).
Thus, $S_{f,L_2}(E)$ measures the $\chi^2$ change for experiment $E$ when the
PDF parameters are displaced from their best-fit value so that $f(x,Q)$ at the
chosen $x$ and $Q$ is increased by its Hessian PDF error, which we take to
correspond to the 68\% probability level (p.l.)~in this presentation.
By computing $S_{f,L2}(E)$
for every fitted experiment $E$, we can easily compare
the pulls of the individual experiments on $f(x,Q)$, solving essentially the same
task as a Lagrange multiplier scan on $f(x,Q)$ \cite{Stump:2001gu}, but
extracting most of the same information much faster, within the limits
of the Hessian approximation.
This approach can explore consistency of individual experiments along much the
same lines as the Hessian data set diagonalization technique \cite{Pumplin:2009sc},
with the advantage that it performs this analysis along multiple directions in
the Hessian PDF parameter space at once.
We may call $S_{f,L2}(E)$ as {\it the $L_2$
  (norm) sensitivity}, to distinguish it from the $L_1$ (norm)
  sensitivity $S_f$ that is constructed from the residuals
themselves, rather than from their squares. 

Figure~\ref{fig:L2} illustrates the application of the $L_2$
sensitivity to the CTEQ-TEA experimental data sets that were selected for
the upcoming CT18 global fit. We compute the $L_2$ sensitivity at the lattice
scale $\mu = 2$ GeV for the isovector combination $u^+(x,\mu)\! -\!
d^+(x,\mu)$. Due to the dependence of $S_{f,L_2}$ on
$\cos{\varphi(f,\chi^2_E)}$, it concisely represents the competing
pulls of the CTEQ-TEA data sets.
The respective $\Delta \chi^2_E(\widehat a_f)$ can be either correlated ($S_{f,L_2} > 0$) or 
anti-correlated ($S_{f,L_2} < 0$) with $u^+(x)\! -\! d^+(x)$. Large
experimental errors or weak correlations correspond to $S_{f,L_2}(E)
\approx 0$ and imply vanishing pulls for a particular experiment on
the PDF at a given $x$. 

In Fig.~\ref{fig:L2}, we included only the curves for 12 experiments
with the highest $L_2$ sensitivities, selected out of 40 eligible
experiments by the requirement that the shown experiments must have
$|S_{f,L_2}(E)|>4$ in some interval of $x$. Recall that the textbook
parameter-fitting criterion associates $\Delta \chi^2_E=1$ (or 4) 
with the uncertainty at the 68\% (or 95\%) probability level. The
$L_2$ sensitivity plots like the one shown in Fig.~\ref{fig:L2} reveal
far larger latent tensions between the experiments in some $x$
regions, 
evident as deviations in opposite directions with magnitudes $|S_{f,L_2}(E)|\!>\! 5$.
That is, the $\chi^2$ value changes by more than five units when the PDFs are
varied by the 68\% p.l.~error along the respective direction.
In the region $x>0.1$ that
contributes the most to the Mellin moments,
Fig.~\ref{fig:L2} suggests that the sharpest positive
pulls on the $u^+(x)\! -\! d^+(x)$ distribution come from
the E866 $pp$ (204), combined HERA (160), and CDHSW (109) data, whose $\chi^2_E$
are positively correlated with the isovector PDF; and the opposite
pulls by the NMC ratio measurements (104) and BCDMS $F^d_2$ data
(102), for which the $L_2$ sensitivity is negative.
Similar tensions exist at smaller values of $x$ as well.

The upshot of the figure is that a variety
of DIS and Drell-Yan experiments on nucleon and nuclear fixed targets
continue to play the prominent role in constraining the isovector
combination, surpassing in this regard
the available LHC measurements that still are to grow in their
importance. Similar observations apply to the Mellin moments and qPDFs that we will consider. The
sensitivity studies, as well as related methods such as the Lagrange
multiplier scan, reveal tensions among the miscellaneous experiments
in the CTEQ-TEA data set that constrain the PDFs at $x > 0.1$. 
The $x$-dependent $L_2$ sensitivity shown in Fig.~\ref{fig:L2}
should be consulted in parallel with the $S_f$ sensitivity
calculations shown for the moments and quasi-PDFs
in the sections below.
%
% ~ ~ ~ ~ ~ ~ ~ ~ ~ ~ ~ ~ ~ ~ ~ ~ ~ ~ ~ ~ ~ ~ ~ ~ ~ ~ ~ ~ ~ ~ ~ ~ ~ ~ ~ ~ ~ ~ ~ ~ ~ ~ ~ ~ ~ ~ ~ ~ ~ ~ ~

\subsection{Structure of the presentation}
In the next subsection, we will compute and investigate
the $S_f$ sensitivity of the CTEQ-TEA data set to
the lowest moments of the unpolarized light quark and gluon distributions, $|S_{\langle x^n\rangle_{q^\pm,g}}|$,
in the $(x, \mu)$ plane with the \texttt{PDFSense} package \cite{Wang:2018heo}.
We use the CT14HERA2 NNLO PDF set
\cite{Hou:2016nqm} in the theoretical predictions and residuals of
experimental data. Our data sets measurements in the
CT14HERA2 NNLO fit and the latest LHC jet, $t\bar{t}$, $W/Z$ production data sets.

The sensitivities from the existing experiments discussed in this
section can be further confronted by the projected sensitivities to the
same PDF-dependent quantities in Sec.~\ref{sec:EIC} for a set of pseudodata for
inclusive neutral-current (NC) and charge-current (CC)-mediated $e^\pm
p$ deeply inelastic scattering (DIS) at a future EIC-like collider.
The scientific program at an Electron-Ion Collider (EIC) or a similar
machine is anticipated to significantly complement lattice QCD in
learning about the 3-dimensional structure of hadrons.
The procedure of the preliminary impact study in Sec.~\ref{sec:EIC}
is broadly similar to that for the analysis of the real CTEQ-TEA data, but
is based on the DIS pseudodata obtained by generating Gaussian fluctuations
about the CT14HERA2 NNLO theoretical prediction for the reduced cross sections
according to an assumed precision.

The flavor combinations primarily discussed in this paper
are $u^\pm\!-\!d^\pm, u^\pm, d^\pm, s^+$ and $g$
in which $q^+$ here refers to the C-even combination of (anti-)quark
distributions, $q+\bar{q}$; correspondingly, we also consider C-odd quantities,
$q^- = q-\bar{q}$, as defined in Eqs.~(\ref{eq:moment})-(\ref{eq:momlat}). The PDF moment(s) for the light flavor combinations $u^\pm\!-\!d^\pm$, $u^\pm$,
$d^\pm$, and $s^+$ are computed on the lattice by the $\mathcal{O}_{\{\mu_{1},\cdots,\mu_{n}\}}^{F}$
operators noted above in Eq.~(\ref{eq:ops}), whereas for the gluon distribution $g$,
the operator noted immediately after Eq.~(\ref{eq:mom-ints}) is required.
The present status of the lattice QCD calculations of these parton moments is
widely varied, with some moments (especially for the isovector combination
$u\!-\!d$) evaluated by multiple groups with various systematic treatments;
on the other hand, lattice information on the second moments of the individual light quark flavors
$\langle x^2 \rangle_{q^-}$, for instance,
is comparatively sparse. At the same time, the corpus of lattice computations is
growing, and the availability of calculations for
the moments considered here (and beyond) will increase.

We note that many numerical results for PDF moments computed both on
the lattice and from different QCD global analyses are detailed in
Appendices B and C of the recent white paper in Ref.~\cite{Lin:2017snn}.

\subsection{Numerical results}
\begin{figure}[b]
\hspace*{-0.75cm}
\includegraphics[scale=0.59]{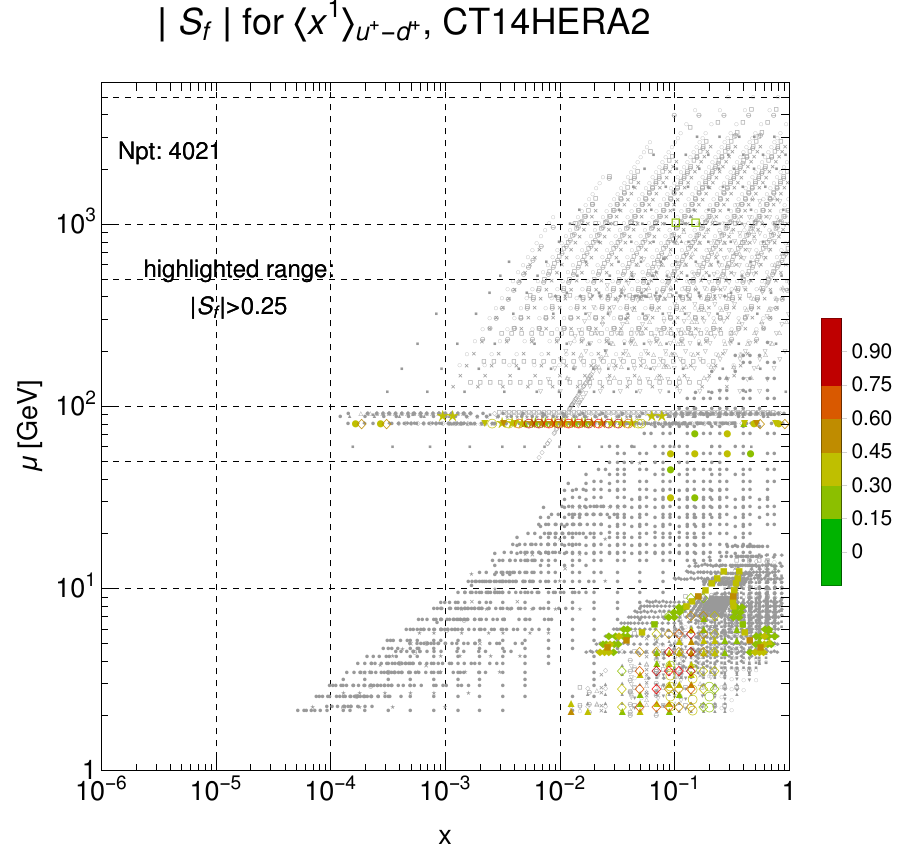} \ \ \                      
\includegraphics[scale=0.59]{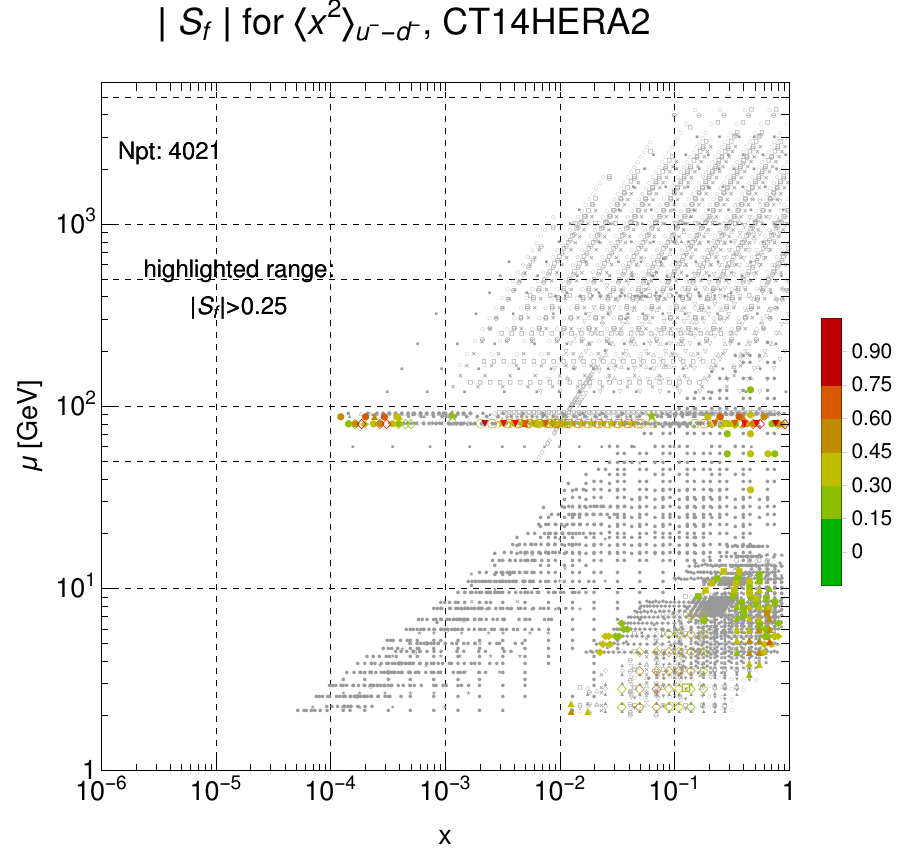} 
\caption{Sensitivity of the CTEQ-TEA data sets to $\langle x \rangle_{u^+-d^+}$
and $\langle x^{2}\rangle_{u^--d^-}$. The factorization scales
of Mellin moments and PDFs are $\mu_F = \mu^\mathrm{lat} = 2$ GeV.
As described in the text below, we colorize only those data points with significant
individual sensitivity $|S_f| > 0.25$, so as to highlight the pulls
of the most constraining measurements. Points falling below this
highlighting threshold (i.e., with $|S_f| < 0.25$) are indicated
in grayscale.
}
\label{fig:mom_iso1} 
\end{figure}

We now present the calculated sensitivity maps for the CTEQ-TEA data to each of the lowest
moments of the light quark ($u\!-\!d, u, d, s$) distributions and the gluon.
We also examine the aggregated impact of the experiments in the
CTEQ-TEA set on each of these quantities and consider the implications
for unraveling the nucleon's flavor structure and benchmarking
lattice QCD output of the same objects.

\subsubsection{Moments of nucleon quark distributions}
\label{sec:q-moms}
%

%%%%%%%%%%%%%%%%%%%%%%%%%%%%%%%%%%%%%%%%%%%%%%%%%%%%%%%%%%%%%%%%%%%%%%%%%%%%%%%%%%

\paragraph{{\bf Moments of Isovector Flavor Distributions.}}
\label{para_iso}
Historically, computation of isovector PDF combinations in $\mathrm{SU}(2)$
isospin space has represented an especially fertile proving ground for
lattice gauge methods --- particularly given that gluon and singlet
quark densities mix evenly with $u^+$ and $d^+$ distributions under
DGLAP evolution, such that $u^+\! -\! d^+$ has a nonsinglet scale
dependence. A consequence specific to lattice QCD is the
fact that contributions from disconnected insertions vanish in the
difference (assuming parton-level charge symmetry), and a much less
computationally costly calculation based purely on connected insertions
is generally adequate.

\begin{figure}[p]
\hspace*{-0.75cm}
\includegraphics[scale=0.59]{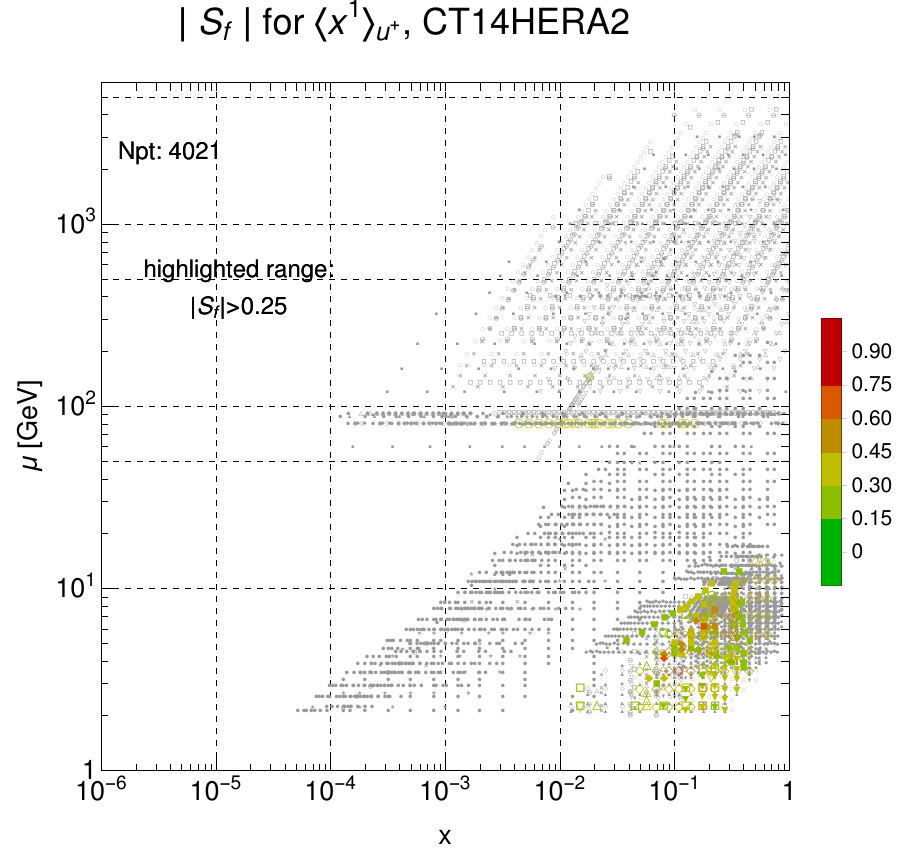}
\ \ \
\includegraphics[scale=0.59]{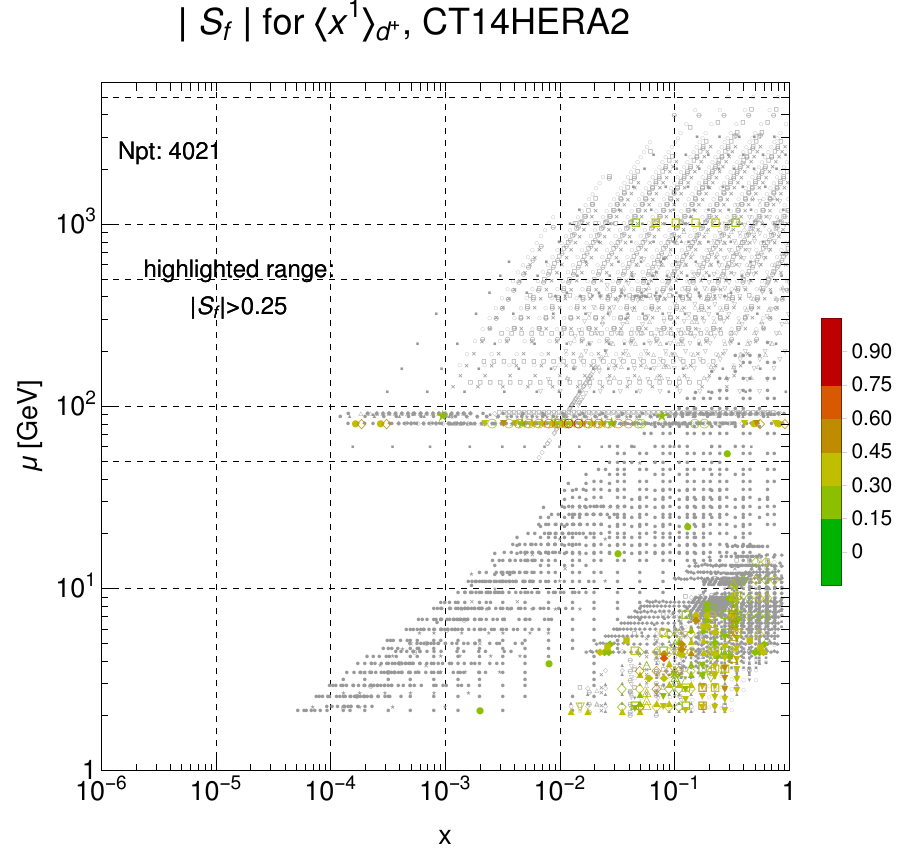} 
	\caption{Sensitivity of the CTEQ-TEA data sets to $\langle x\rangle_{u^+}$ (left panel)
	and $\langle x\rangle_{d^+}$ (right panel). As in Fig.~\ref{fig:mom_iso1}, sensitivities are with respect to moments
	evaluated at $\mu_F = \mu^\mathrm{lat} = 2$ GeV. }
\label{fig:Mellin_ud+moments} 
\end{figure}
\begin{figure}[p]
\hspace*{-0.75cm}
\includegraphics[scale=0.59]{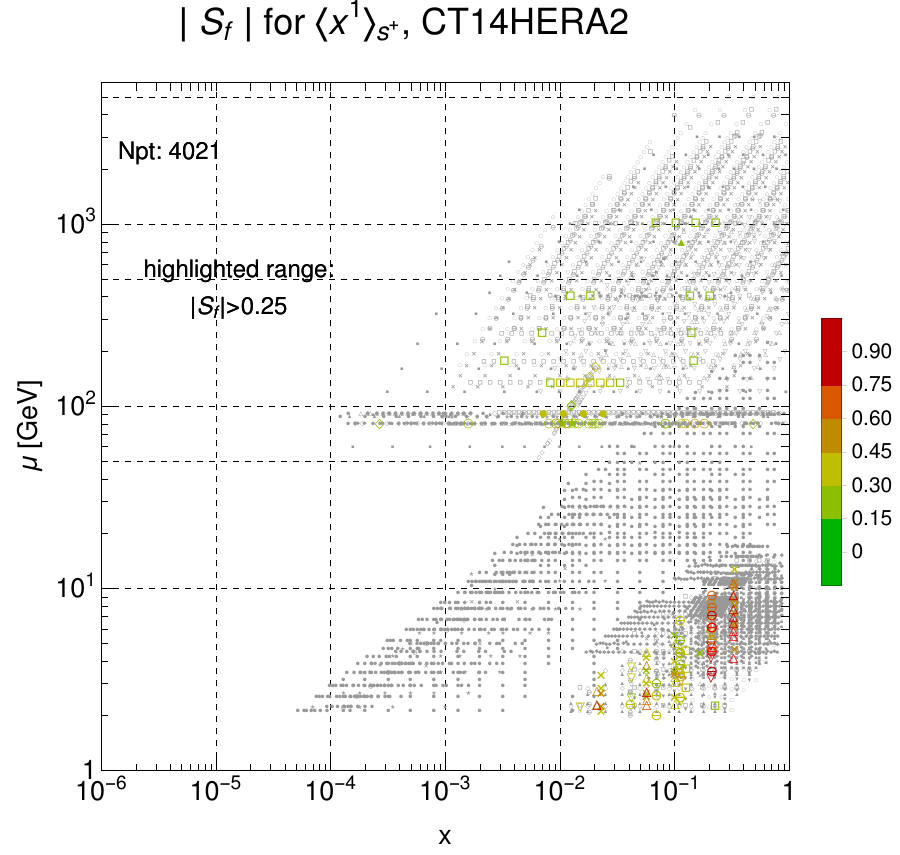} 
\ \ \                        
\includegraphics[scale=0.59]{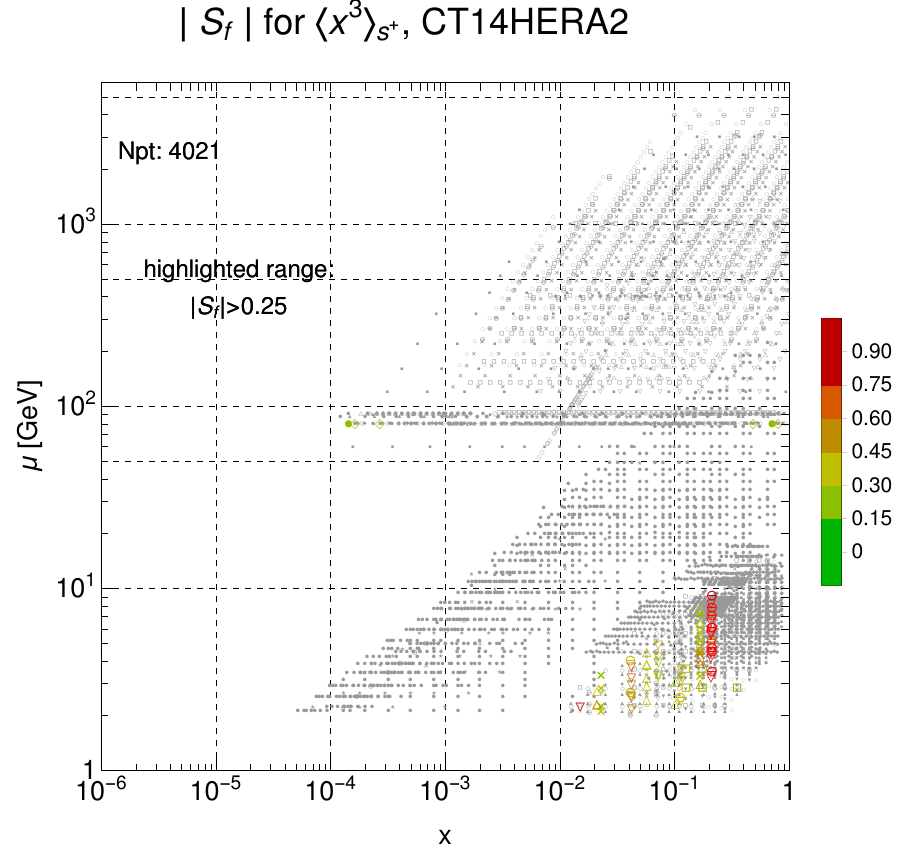} 
\caption{CTEQ-TEA sensitivity to the 1$^\mathit{st}$- and 3$^\mathit{rd}$-order Mellin moments of the $s^+$ distribution.
	As elsewhere, the factorization scales of Mellin moments and PDFs are $\mu_F = 2$ GeV.
	}
\label{fig:Mellin_s} 
\end{figure}
\clearpage

For this main reason, the isovector combinations
have been a focus of lattice calculations for both the PDF Mellin
moments --- and, more recently, the quasi-distributions in both
the nucleon \cite{Aoki:2010xg,Lin:2014zya,Chen:2017mzz,Gupta:2018qil,Liu:2018uuj,Alexandrou:2018pbm}
and pion \cite{Chen:2018fwa,Sufian:2019bol,Oehm:2018jvm,Hobbs:2017xtq}.

We plot the sensitivity map of the CTEQ-TEA data to two of the lower moments
of the nucleon isovector distribution in Fig.~\ref{fig:mom_iso1} --- namely, 
the sensitivities to the total isovector momentum $\langle x\rangle_{u^+-d^+}$ (left panel) and
the second-order asymmetry moment $\langle x^2 \rangle_{u^--d^-}$ (right panel). These plots have the same
basic configuration as developed in Ref.~\cite{Wang:2018heo}, with
emphasis placed on specific measurements of individual experimental
data sets with particularly strong pulls in the global analysis. The
predicted pull of these measurements as quantified by the sensitivity $|S_f|$
is represented by the color scheme shown in the offset to the right for
each panel in Fig.~\ref{fig:mom_iso1}. As in
Ref.~\cite{Wang:2018heo}, we draw attention to the most impactful data
and physical processes by imposing a highlighting cut, $|S_f| > 0.25$,
and selecting a coloration scheme which suitably reveals the dependence
of data sensitivities on the kinematical matching parameters.
On the basis of our sensitivity analysis, we are in a position to make a
number of observations regarding the empirical information that drives
the current knowledge of the lower $\langle x^n \rangle_{u-d}$ Mellin
moments.

The highlighted points emphasized in the panels of Fig.~\ref{fig:mom_iso1}
often rather closely correspond to the CTEQ-TEA experiments which enjoy the highest per-datum
sensitivities,
\begin{equation}
	\langle | S_f | \rangle =  {1 \over N_\mathit{pt}}\, \sum_{i \in N_\mathit{pt}} |S^i_f|\ .
\end{equation}
In decreasing order, these are CMS7Masy2'14 (0.557), E866rat'01 (0.365), CMS7Easy'12 (0.333),
CCFR-F3'97 (0.307), and NMCrat'97 (0.212),
where the quantity in parentheses is the computed average of each experiment {\it per measured point}.

For the purpose of enumerating this information, we include only those experimental measurements with
point-averaged sensitivities exceeding lower bound $\langle |S_f| \rangle > 0.2$.
On the other hand, by the total sensitivity metric $\sum_{i \in N_\mathit{pt}} |S^i_f|$, we
identify a somewhat different collection of
experiments with leading impact on the first isovector moment; {\it viz.}
HERAI+II'15 (37.8); CCFR-F3'97 (26.4); NMCrat'97 (26.1); E866pp'03 (20.7); BCDMSp'89 (19.3).
We point out that the point-averaged and total sensitivities for all CTEQ-TEA experiments
are summarized in Appendix~\ref{sec:append} using ranking tables, as well as in Figs.~\ref{fig:rank_plot}
and~\ref{fig:rank_plot_qPDF}, which are further described below. Additional numerical details
can be found on the companion website, Ref.~\cite{Website}.
In this context, there is a pronounced influence of the combined HERAI+II experiment due to the extremely
large number of measurements ($N_\mathit{pt} = 1120$) taken --- and despite the fact that only a minimal number
of these exhibit per-point sensitivities that exceed the highlighting cut $|S_f| > 0.25$ imposed on the impact
maps in this analysis.

Continuing, the right panel of Fig.~\ref{fig:mom_iso1} also shows the corresponding
distribution of CTEQ-TEA sensitivities in $(x,\,\mu)$ space for the second isovector
moment $\langle x^2 \rangle_{u^--d^-}$, for which the constraints arising from individual
experiments fitted by CT are somewhat different.
In this instance, we find the distribution of point-averaged sensitivities within the CTEQ-TEA
data set to be driven primarily by electroweak boson production measurements:
CMS7Masy2'14 (0.492), D02Easy2'15 (0.416), CMS7Easy'12 (0.282), LHCb7Wasy'12 (0.250), CCFR-F3'97 (0.224), and
D02Masy'08 (0.211).
In contrast to the total isovector momentum considered above, we therefore again find a leading role
for the 7 TeV CMS lepton asymmetry measurements of $A_\mu(\eta)$ [CMS7Masy2'14] and $A_e(\eta)$ [CMS7Easy'12],
although the size of the sets ($N_\mathit{pt} = 11$) is such that their aggregated pull on $\langle x^2 \rangle_{u^--d^-}$
is dominated by larger fixed-target data sets identified by an analysis of the summed sensitivities, as we point out below.
In addition to the CMS measurements, a number of other electroweak boson sets evidently have
stronger pull on the $\langle x^2 \rangle$ isovector moment, including the corresponding D$\emptyset$
measurement of $A_e(\eta)$ (D02Easy2'15), which probes higher $x$, as well as LHCb.
The evaluation according to the aggregated sensitivities reveals a different hierarchy. In
this case, fixed-target measurements of DIS cross sections and structure functions --- as well
as a couple Drell-Yan sets --- are dominant, namely,
HERAI+II'15 (36.5), BCDMSp'89 (33.1), E866pp'03 (22.2), CCFR-F3'97 (19.3), and NMCrat'97 (18.4),
with a rapid falloff in the aggregated sensitivity below $\sum |S_f|$ beyond these experiments.
It should be noted, however, that were the boson production data sets with especially strong
per-datum sensitivities indicated above combined into a single collection, the resulting aggregated
impact of this collection would approach $\sum |S_f| \sim 34$, placing this combination of $150$ data points
just beyond the BCDMS $F^p_2$ data ($N_\mathit{pt} = 337$) and only behind the HERAI+II'15
set ($N_\mathit{pt} = 1120$) in total sensitivity.

%
%%%%%%%%%%%%%%%%%%%%%%%%%%%%%%%%%%%%%%%%%%%%%%%%%%%%%%%%%%%%%%%%%%%%%%%%%%%%%%%%%%

\vspace*{0.1cm}

\paragraph{{\bf Moments of $q^+$ distributions.}}
\label{para_q+}
As we pointed out in the discussion of the $u-d$ moments at the start of Sec.~\ref{para_iso} above,
the fact that the disconnected insertions contribute equally to
$u$ and $d$-type distributions implies their vanishing for isovector ($\tau_3$) charges.
Unlike these combinations, the moments of flavor-separated
distributions like $u^+(x,\mu^\mathrm{lat})$ and $d^+(x,\mu^\mathrm{lat})$ receive
contributions from both connected and disconnected insertions. The disconnected
insertions arise from Wick contractions of quark fields not explicitly present in interpolation
operators used to construct the $2$-point function associated with the nucleon propagator; disconnected insertions are
therefore essentially equally present in both $u$-type and $d$-type flavor-separated moments. Unfortunately, evaluating
disconnected insertions on the lattice is computationally expensive and, historically, has proved challenging.
In the case of the higher moments, they are generally quite small ---
{\it e.g.}, Ref.~\cite{Deka:2008xr} found $\langle x^2 \rangle_{u^-,d^-}$ to be consistent with zero, and, along these lines,
the disconnected contributions in these instances will themselves be fairly small. In fact, even for the
larger first moments $\langle x \rangle_q^+$, the differences between calculations with and without
disconnected insertions are within uncertainties, suggesting that these contributions
may not be so large for the $u$ and $d$-type distribution moments. Nucleon strangeness, on the other hand,
necessarily originates exclusively with disconnected insertions, since
the proton possesses no valence strange content, and, consequently, no strange
quark fields are explicitly present in the nucleon interpolation operators from which
two-point correlation functions are evaluated.
Precise lattice data involving each of these flavors and for multiple Mellin moment orders would be instrumental
in disentangling the interplay of connected vs.~disconnected insertions and helping to resolve the underlying
dynamics. This observation also motivates a comprehensive assessment of the same
moments as computed from phenomenological PDFs as well as a reckoning of the
various pulls from experimental data that act upon them.

\vspace{0.2cm}

% - - - - -  u^+ - - - - - - -
%
\paragraph{\underline{$u^+$-quark moments.}}
For $\langle x\rangle_{u+}$, we consider the CTEQ-TEA sensitivity contained in the
map of the LHS panel of Fig.~\ref{fig:Mellin_ud+moments}; as is the case fairly generically for
the the leading moments of the light quark distributions, the most concentrated locus of high-sensitivity
data are found in the fixed-target sector in the lower right quadrant of the
$(x,\mu)$ plot --- particularly, for $x \gtrsim 0.01$ and $\mu \lesssim 10$ GeV. Upon inspection,
these points arise from measurements at BCDMS (on the proton --- BCDMSp'89 --- as well as the
deuteron, BCDMSd'90) and the E866 data. Empirical information with especially larger
per-datum sensitivities can again be identified by listing
the leading experiments in descending order of their point-averaged sensitivities. These are
CCFR-F3'97 (0.337), E866rat'01 (0.277), D02Masy'08 (0.250), CMS7Masy2'14 (0.248), and NuTeV-nu'06 (0.221).
While for the total sensitivities we find
HERAI+II'15 (40.8), BCDMSp'89 (39.5), CCFR-F3'97 (29.0), BCDMSd'90 (24.8), CDHSW-F2'91 (16.5), CDHSW-F3'91 (15.1), E866pp'03 (10.3).

% - - - - -  d^+ - - - - - - -
%
\paragraph{\underline{$d^+$-quark moments.}}
As an illustration of the flavor dependence of the PDF moments, we compare in the
right panel of Fig.~\ref{fig:Mellin_ud+moments} with the corresponding sensitivities
for $\langle x\rangle_{d+}$, shown in the right panel. Here we find again
a strong role again for charged-current processes from lepton charge asymmetry data and $\nu A$ DIS, if the leading per-datum sensitivities are considered:
CMS7Masy2'14 (0.419), NuTeV-nu'06 (0.238), CMS7Easy'12 (0.228), CCFR-F3'97 (0.227), CDHSW-F2'91 (0.225).
On the basis of the total sensitivities of these experiments, however, we again find a hierarchy dominated by
the combined HERA data set, for which the charge-current (CC) $e^\pm p$ channels show somewhat enhanced sensitivity to moments
of $d(x)$ relative to $u(x)$ according to both the $\langle |S_f| \rangle$ and $\sum |S_f|$ metrics illustrated in
Fig.~\ref{fig:rank_plot}; this is particularly true of the CC $e^+p$ HERAI+II information, for which the LO reduced cross section
$\sigma_r (x, Q^2)$ is closely driven by the behavior of $d$-type quark distributions, especially at larger $x$. Beyond
the HERA measurements, the descending list of the experiments with high total sensitivities has a trailing collection of fixed-target measurements, namely,
HERAI+II'15 (54.2), BCDMSd'90 (26.5), NMCrat'97 (22.6), CCFR-F3'97 (19.5), CDHSW-F2'91 (19.1), BCDMSp'89 (18.5), E866pp'03 (14.8).
In this instance, the second most influential measurement is the deuteron target structure functions extractions 
from BCDMS (BCDMSd'90) --- a fact consistent with the traditional importance ascribed to deuteron measurements for
performing nucleon flavor separations.

\vspace{0.2cm}

% - - - - -  s^+ - - - - - - -
%
\paragraph{\underline{$s^+$-quark moments.}}

The sensitivities to the moments of the $s^+$ distribution are presented in Fig.~\ref{fig:Mellin_s}. For $\langle x \rangle_{s^+}$, the measurements with leading point-averaged sensitivities are found to be
NuTeV-nu'06 (0.429), CCFR SI nub'01 (0.344), CCFR SI nu'01 (0.313), NuTeV-nub'06 (0.302), D02Masy'08 (0.274);
while those with the highest predicted total impact based on aggregated sensitivity
are
HERAI+II'15 (31.4), NuTeV-nu'06 (16.3), CCFR SI nub'01 (13.1), CCFR SI nu'01 (12.5), NuTeV-nub'06 (10.0).
Across both the aggregated and average per-point sensitivities, the decisive role of neutrino scattering data is evident, despite the still
leading role of the combined HERA measurements --- especially noting the fact that the summed sensitivity of
the $4$ leading $\nu$ experiments mentioned above is $\sum |S^\nu_f| = 51.9$, exceeding the HERA accumulated
impact by $\sim\!65\%$.

In the CT14HERA2 NNLO PDF set, strangeness was parametrized symmetrically ({\it i.e.},
under the assumption $s(x) = \bar{s}(x)$; as a result, the moments of
the $s^-$-type distributions, including $\langle x^2 \rangle_{s^-}$, are
identically zero. For that reason, we instead consider here the next highest moment of
the strangeness distribution, {\it i.e.}, the third moment $\langle x^3 \rangle_{s^+}$; for which we find the point-averaged
sensitivities of the leading experiments (again, cutting at $\langle |S_f| \rangle > 0.2$) to be
NuTeV-nub'06 (0.568), CCFR SI nub'01 (0.387), and NuTeV-nu'06 (0.269),
clearly suggesting the very important role of the NuTeV $\bar{\nu}$ dimuon production measurements (NuTeV-nub'06), which
show especially enhanced sensitivity to the higher $\langle x^3 \rangle_{s^+}$ moment than was seen for the total strange
momentum $\langle x \rangle_{s^+}$.
For the total sensitivities, the constraints imposed by the CTEQ-TEA data set come primarily
from several experiments
HERAI+II'15 (20.3), NuTeV-nub'06 (18.7), CCFR SI nub'01 (14.7), and NuTeV-nu'06 (10.2).
Thus, for both Mellin moments of the $s^+$ distribution, the fixed-target $\nu$ DIS experiments enjoy a clear advantage in their sensitivity compared to the rest of the CT14HERA2 experimental data sets. 

\begin{figure}[t]
\hspace*{-0.75cm}
\includegraphics[scale=0.59]{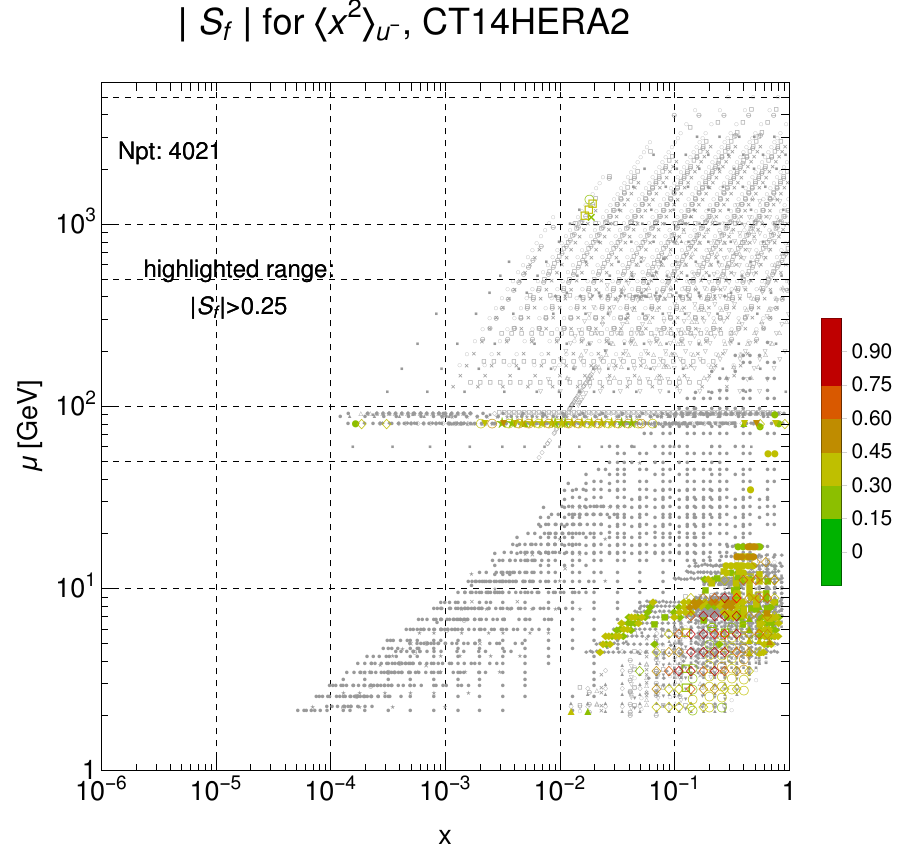}
\ \ \                        
\includegraphics[scale=0.59]{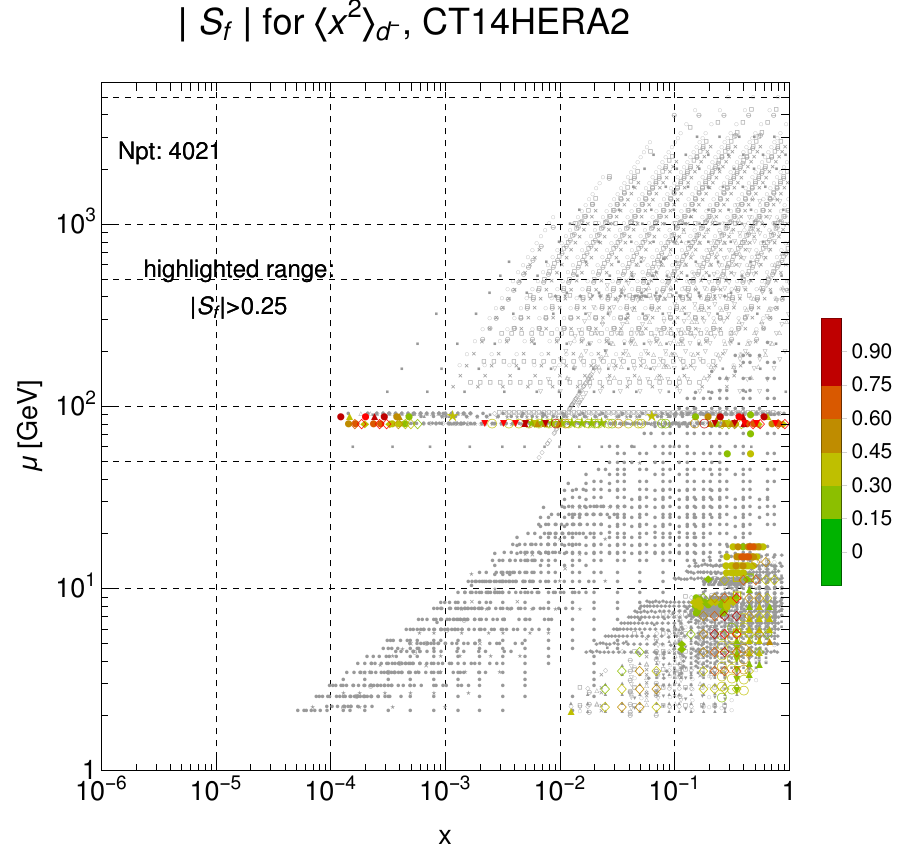} 
\caption{Sensitivity of the CTEQ-TEA data to the $v^3_{u,d}$ moments
	$\langle x^{2}\rangle_{u^-}$ (left) and $\langle x^{2}\rangle_{d^-}$ (right),
	computed for $\mu_F = 2$ GeV.
}
\label{fig:Mellin_ud-moments} 
\end{figure}

%
%
%%%%%%%%%%%%%%%%%%%%%%%%%%%%%%%%%%%%%%%%%%%%%%%%%%%%%%%%%%%%%%%%%%%%%%%%%%%%%%%%%%
% - - - - - - - - - - - - - - - - - - - - - - - - - - - - - - - - - - - - - - - - - - - - - - - -
%
\paragraph{{\bf Moments of $q^-$ distributions.}}
\label{para_q-}

At present, lattice determinations for the next highest $\langle x^2 \rangle_q$ moments of the light-quark distributions
have not matured to the level of extant calculations of the first moments $\langle x \rangle_q$, particularly in the sense that these have
been computed thus far only in Ref.~\cite{Deka:2008xr} in the quenched approximation ({\it i.e.}, excluding dynamical
quark loops). Nonetheless, such determinations are likely forthcoming, and can yield vital information regarding asymmetric
$x$ dependence in the light quark distributions.

We plot the sensitivity maps to the $\langle x^2 \rangle_{q^-}$ moments of the $u$ and $d$ quark distributions
in the left and right panels of Fig.~\ref{fig:Mellin_ud-moments}, respectively. As elsewhere, these
panels examine the sensitivity of the CTEQ-TEA set to moments evaluated at the typical lattice scale
$\mu = \mu^\mathrm{lat} = 2$ GeV.

\vspace{0.2cm}

\paragraph{\underline{$u^-$-quark moment.}}
For the second moment of the $u^-$ distribution, the leading point-averaged sensitivities
are due to fixed-target DIS experiments and the $7$ TeV CMS lepton charge asymmetries,
led by
CCFR-F3'97 (0.503); beyond this, experiments with $\langle | S_f | \rangle > 0.2$ are
CMS7Masy2'14 (0.413), CDHSW-F3'91 (0.248), and CMS7Easy'12 (0.244).
In this context, the fact that information on the parity-odd structure function $F^p_3$
--- especially as provided by CCFR-F3'97 --- shows such sizable influence over $\langle x^2 \rangle_{u^-}$
is consistent with the leading-order $\sim\! q\!-\!\bar{q}$ behavior of $F^p_3$ in the quark-parton model. As
such, thorough knowledge of the $x$ dependence of $xF_3$ facilitates an unraveling of the $C$-odd
distributions of the $q^-$ type, and constrains their higher moments.
As was the case, however, for the $\langle x \rangle_{q^+}$ moments, consideration of the
aggregated sensitivities reveals a larger spread of experiments with strongest pulls belonging
again to
the combined HERA data set HERAI+II'15 (43.9), the $\nu$DIS measurements of $xF_3$ from CCFR [CCFR-F3'97 (43.2)]
identified by the point-averaged ranking above, BCDMSp'89 (39.2), and E866pp'03 (32.7).
Having somewhat diminished but still significant pulls are several of the other fixed-target experiments
involving both neutrino and $\mu$ DIS as well as the Drell-Yan process; namely, these are
CDHSW-F3'91 (23.8), E605'91 (18.4), BCDMSd'90 (13.6), and NMCrat'97 (13.4).

\vspace{0.2cm}

\paragraph{\underline{$d^-$-quark moment.}}
As observed above for the lower $\langle x \rangle_{q^+}$ moments imaged in Fig.~\ref{fig:Mellin_ud+moments},
there are evident differences between the sensitivity maps for $d$- vs.~$u$-quark moments,
and this holds again for the explicit comparison of $\langle x^2 \rangle_{d^-,u^-}$
illustrated in Fig.~\ref{fig:Mellin_ud-moments}. In fact, these systematic differences are
especially marked for the $\langle x^2 \rangle$ moments, as inspection of Fig.~\ref{fig:Mellin_ud-moments}
attests. Especially notable in the right panel of Fig.~\ref{fig:Mellin_ud-moments} is the very
strong sensitivity $|S_f| \gtrsim 0.75$ for a select subset of the gauge production data, particularly
for $x \gtrsim 10^{-4}$ and separately for $x \gtrsim 0.1$. These especially strong constraints to
$\langle x^2 \rangle_{d^-}$ originate from an amalgam of electroweak data sets, among which we find
the $8$ TeV forward $W^\pm, Z$ production cross section data of LHCb (LHCb8WZ'16), the analogous information
at $7$ TeV (LHCb7ZWrap'15), as well as the forward-backward $e^+e^-$ asymmetry in $W^\pm, Z$ production
at Runs 1 and 2 of CDF, CDF1Wasy'96 and CDF2Wasy'05.
Compared with $\langle x^2 \rangle_{u^-}$, on the other hand, for the second moment of $d^-(x)$ we find a
substantially more restricted outlay of individual high-impact measurements in the fixed-target region,
with significantly fewer data belonging to very high $x \gtrsim 0.4$ or $x \lesssim 0.2$ identified. Of these,
the E605, NMCrat, and CCFR-F3 points enjoy special prominence.
Many of these trends revealed by the sensitivity map in the right panel of Fig.~\ref{fig:Mellin_ud-moments}
are further confirmed by quantitative ranking of the CTEQ-TEA experiments, especially based on the
per-point sensitivities.
For the  second moment of the $d^-$ distribution, the point-averaged sensitivity ranked experiments
are
D02Easy2'15 (0.519), CCFR-F3'97 (0.381), LHCb7Wasy'12 (0.362), CMS7Masy2'14 (0.328), D02Masy'08 (0.293),
CDF1Wasy'96 (0.252), LHCb8WZ'16 (0.217), LHCb7ZWrap'15 (0.214), and E605'91 (0.207).

For the aggregated sensitivities, here also we find the knowledge of the $d^-$ second moment to be driven foremost
by $xF_3$ data from CCFR and the combined HERA data, CCFR-F3'97 (32.8) and HERAI+II'15 (32.3), respectively.
We note, however, that the total sensitivity of these leading experiments to the $d^-$ distribution is reduced
roughly $\sim 30\%$ relatively to what was found for the corresponding $u$-quark sensitivities. 
Beyond these leading measurements, an assortment of $\mu$ and $\nu$DIS and Drell-Yan experiments have the
tightest pulls. Again in descending order, these are E605'91 (24.7), CDHSW-F3'91 (18.5), BCDMSd'90 (15.2),
NMCrat'97 (14.8), BCDMSp'89 (14.4), E866pp'03 (13.8), and CDHSW-F2'91 (11.7).

\begin{figure}[b]
\includegraphics[scale=0.7]{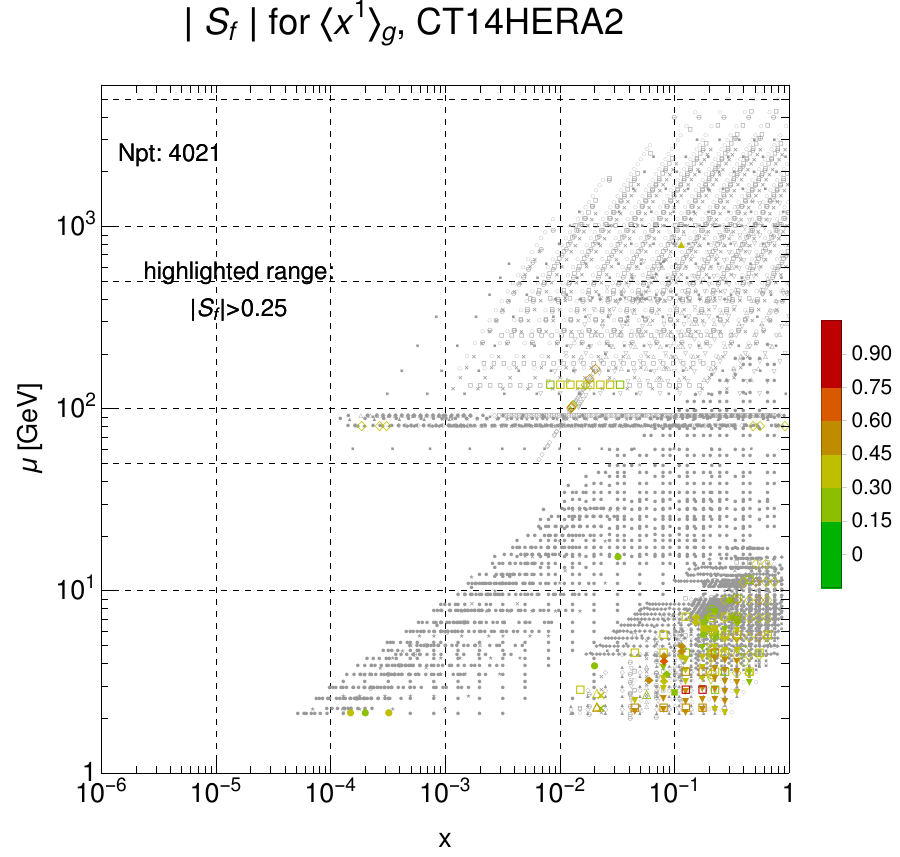}
\caption{Sensitivity of the CTEQ-TEA data sets to $\langle x\rangle_{g}$.
The factorization scale at which the moment is evaluated is $\mu_F = 2$ GeV.
Here we have only a single panel for $\langle x \rangle_g$, given that
lattice computations thus far only exist for $\langle \mathcal{G}_{\mu\nu}\mathcal{G}^{\mu\nu} \rangle$.
}
\label{fig:Mellin_gS} 
\end{figure}

%
% - - - - - - - - - - - - - - - - - - - - - - - - - - - - - - - - - - - - - - - - - - - - - -
%%%%%%%%%%%%%%%%%%%%%%%%%%%%%%%%%%%%%%%%%%%%%%%%%%%%%%%%%%%%%%%%%%%%%%%%%%%%%%%%%%
%

\subsubsection{The gluon momentum fraction}
\label{sec:gluon}

We can extend this program to the gluonic sector, considering the
total nucleon momentum carried by gluons as characterized by the first moment
of the gluon distribution, $\langle x \rangle_g$; for the time being, this is
the only moment of the gluon PDF which has been evaluated by multiple lattice
groups, and we therefore concentrate on it primarily. Fig.~\ref{fig:Mellin_gS}
illustrates the sensitivity to $\langle x\rangle_g$ of the CTEQ-TEA data
considered in the plots of the preceding section. 

From our analysis of the sensitivities, we arrive to perhaps not an entirely expected conclusion that a combination of DIS experiments holds the strongest cumulative sensitivity to $\langle x\rangle_g$, acquired through QCD radiative contributions at NLO and higher orders. Indeed, while neither the neutral-current DIS nor charged-current DIS probe the $g(x,Q)$ at the Born level, the degree of the Bjorken scaling violation in DIS cross sections at $x > 0.1$ is known to be driven by the magnitude of the gluon PDF. Consequently the extensive DIS data at high $x$ provide the dominant constraints on $g(x,Q)$ in the $x$ region giving the largest contribution to $\langle x\rangle_g$.

In contrast, the hadron-hadron collider measurements like production of inclusive jets or $t\bar t$ pairs, while probing the gluon PDFs already at the lowest order in $\alpha_s$, do not compete yet with DIS in their sensitivity to $\langle x\rangle_g$. Thus, unlike what was generally observed for the quark distribution moments reported above, only two experiments 
within the CTEQ-TEA set lie above the $\langle |S_f| \rangle > 0.2$ ranking cut for $\langle x \rangle_g$. Based on
their point-averaged sensitivities, these are both measurements of $F^p_2$
(albeit extracted from nuclear data), specifically,
CDHSW-F2'91 (0.312) and CCFR-F2'01 (0.237). Immediately beyond these most valuable `per-point' measurements of
$F_2$, several other experiments fall immediately below the cut with slightly weaker averaged sensitivities,
including the $\nu$DIS measurement of $F_3(x,Q^2)$ recorded by CCFR-F3'97 (0.188), the $7$ TeV ATLAS high-$p_T$ $Z$
production data of ATL7ZpT'14 (0.184), and the $8$ TeV $t\bar{t}$ measurements from ATLAS, ATL8ttb-mtt'16 (0.172).
\begin{figure}
\includegraphics[scale=0.7]{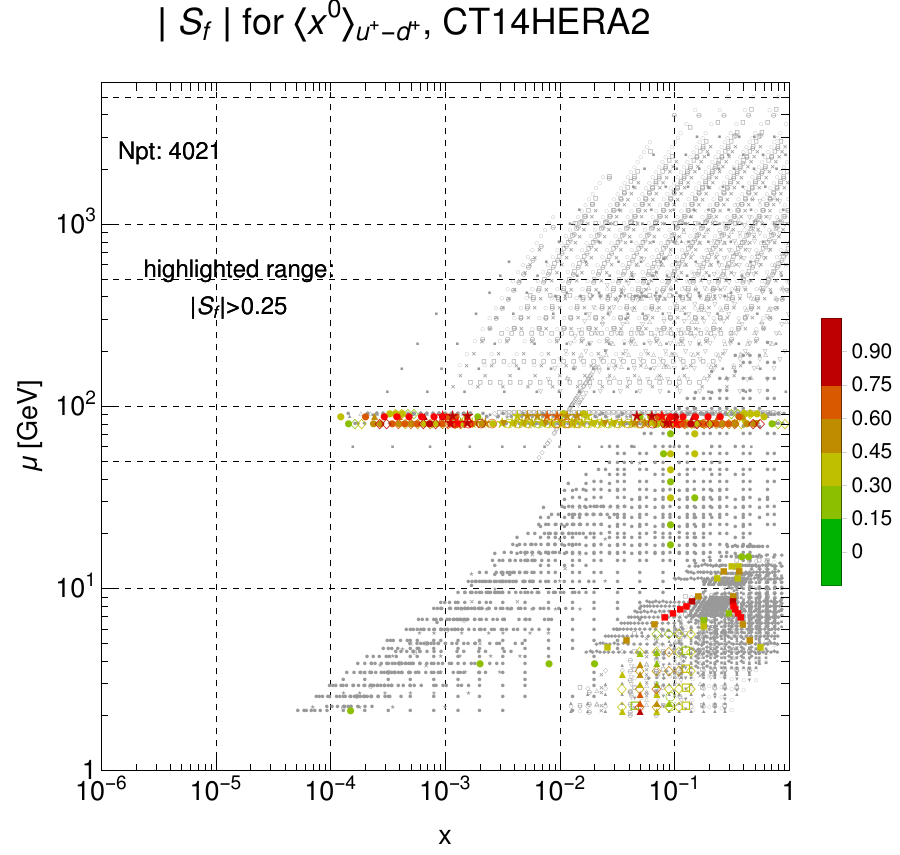}
\caption{Sensitivity of the CTEQ-TEA data to the first moment $\langle 1 \rangle_{(u^{+}-d^{+})}$. The factorization
scale taken for the Mellin moment is $\mu = 2$ GeV. We stress that, while this combination is not directly calculable
by the usual lattice methods, its appearance in the Gottfried Sum Rule motivates its study, as well as a focus upon
higher moments.
	}
\label{fig:Mellin_F2} 
\end{figure}

Once again, in terms of the the aggregated sensitivities,
we observe a distinctly important role for the 
combined HERA data set --- HERAI+II'15 (49.2) --- a result consistent with the significant precision and
very wide coverage over $x$ and $Q^2$ of these cross section data. This wide coverage in $Q$ acts as a crucial
lever arm to constrain the QCD evolution in the CT (or indeed any) parametrization, and thereby restricts
the phenomenological behavior of the singlet and gluon distributions. After the reduced cross section
measurements of HERA, a cascading series of nucleon or deuteron structure function $F^{p,d}_2$ measurements 
obtained on either hydrogen or nuclear targets contain the greatest share of information on the integrated
gluon distribution. In descending order, these are the $\nu-\mathrm{Fe}$ DIS data of CDHSW-F2'91 (26.5),
followed by $\mu$ scattering data from BCDMS, first on the deuteron, BCDMSd'90 (25.8), as well as on a
hydrogen target, BCDMSp'89 (24.9). Lastly, neutrino data from CCFR on $F_2$ [CCFR-F2'01 (16.3)] and $xF_3$
[CCFR-F3'97 (16.2)] have comparable pull between these two structure function measurement channels, and
important influence in the wider fit. It is intriguing to notice that, while the aggregated pull of HERAI+II'15
(49.2) strongly dominates the spread of CTEQ-TEA experiments considered in isolation, were the leading $\nu\!-\!\mathrm{Fe}$
experiments above regarded as a single experiment and their accumulated sensitivities simply combined
directly, the result ($59 \gtrsim 49.2$) surpasses the very large combined HERA data set, which is
based on $N_\mathit{pt} = 1120$ cross section measurements. A similar observation holds for the BCDMS
data. We therefore again stress the observation made above in the context of the aggregated CTEQ-TEA
sensitivities to, {\it e.g.}, $\langle x^2 \rangle_{u^--d^-}$: while the great extent of the combined HERA data set's
kinematical coverage frequently awards it a leading role in terms of its aggregated effect, agglomerations of
much smaller, targeted data sets can have a comparable or greater combined effect, in principle.

Although they do not appear among the core of most decisive experiments detailed above, some of the
newer LHC data sets canvassed in Ref.~\cite{Wang:2018heo} are nonetheless among the top $\sim\! 10$
most sensitive experiments to $\langle x \rangle_g$ --- particularly the inclusive jet data found in
Ref.~\cite{Wang:2018heo} to provide important constraints to the gluon distribution overall. Specifically,
these are the $8$ and $7$ TeV CMS inclusive jet production data, CMS8jets'17 (7.1) and CMS7jets'14 (6.1),
respectively. The future potential of the LHC jet data to constrain the gluon at large $x$ crucially depends on improvements in understanding of significant systematic errors in these measurements.
%
%
% - - - - - - - - - - - - - - - - - - - - - - - - - - - - - - - - - - - - - - - - - - - - - -
%%%%%%%%%%%%%%%%%%%%%%%%%%%%%%%%%%%%%%%%%%%%%%%%%%%%%%%%%%%%%%%%%%%%%%%%%%%%%%%%%%
%
%
\subsubsection{Flavor asymmetries of the nucleon sea}
\label{sec:SFs}

As a final consideration in this section, we examine the sensitivities to linear combinations of the Mellin moments that quantify breaking of flavor $SU(3)$ symmetry of the nucleon's light-quark sea.
The flavor structure of the proton's quark
sea has for decades attracted sustained focus, especially regarding the dynamical
origin of the observed charge-flavor asymmetry embodied by the breaking of
the $\mathrm{SU}(3)_\mathrm{flavor}$ PDF relation $\bar{u}(x)\! =\! \bar{d}(x)\! =\! s(x)\! =\! \bar{s}(x)$ often
assumed in the earliest phenomenological QCD global fits. Much formal interest in this
topic attends to the fact that the $x$-dependent breaking of the
$\mathrm{SU}(2)$ symmetry relation $\bar{d}(x,\mu)\! -\! \bar{u}(x,\mu) = 0$ 
at low scales is principally understood as a feature of nonperturbative QCD \cite{Thomas:1983fh,Signal:1991ug}; for
instance, patterns of dynamical chiral symmetry breaking in QCD favor hadronic
dissociations of the proton having the form $p \to \pi^+ n$ at
low energies, which are thought to produce generic excesses of $\bar{d}$ over $\bar{u}$ in
contributing to the nucleon's flavor structure \cite{Schreiber:1991qx,Alberg:2012wr,Salamu:2014pka}. It should be noted, however,
that accounting for the detailed $x$ dependence of
$\bar{d}(x)\! -\! \bar{u}(x)$ (or, equivalently, of deviations of the
flavor ratio from $\bar{d}/\bar{u}\! =\! 1$) in the context of meson-cloud models
informed by this physical picture has been challenging.

Historically, much of the empirical information on parton-level flavor symmetry
violation in the nucleon sea has been garnered through examinations of the unpolarized
DIS structure functions. Formally, the structure
functions can be described using
well-established factorization theorems in terms of which they may be separated
via convolution integrals over the long-distance PDFs and the perturbative coefficient
functions.

In this context, a crucial observable first measured systematically by NMC \cite{Amaudruz:1991at,Arneodo:1994sh} is the Gottfried Sum
Rule \cite{Gottfried:1967kk}, which is sensitive to nonperturbative dynamics leading to the $\mathrm{SU}(2)$ flavor
asymmetries in the light quark sea mentioned above. The canonical expression
of the sum rule can be obtained by applying the leading-order QPM to the isovector
structure function difference:
\begin{align}
	\int_0^1\frac{dx}{x}(F_{2}^{p}-F_{2}^{n})|_\mathrm{QPM} = \frac{1}{3}\int_{0}^{1} dx\, (u^{+}-d^{+}) &\equiv \frac{1}{3} \langle 1 \rangle_{u^{+}-d^{+}}\ \nonumber \\
								 &=\frac{1}{3} - \frac{2}{3}\int_{0}^{1} dx\, (\bar{d}-\bar{u})\ ,
\label{eq:Gott}
\end{align}
where we have used isospin and the identities $q^+ = q^- + 2\bar{q}$ and $\int dx (u^-\!-\!d^-) = 1$ to rearrange the first line into the standard statement of the sum
rule on the second.
Most importantly, we highlight the fact that the zeroth moment of the isovector PDF, $\langle 1 \rangle_{u^{+}-d^{+}}$ [the RHS of the first line of Eq.~(\ref{eq:Gott})],
is directly related to the behavior of $\bar{d}\!-\!\bar{u}$, deviating from unity when $\langle 1 \rangle_{\bar{d}-\bar{u}} \neq 0$.
While this latter sea quark PDF moment appearing on the far RHS of Eq.~(\ref{eq:Gott}) is not directly accessible
on the lattice as a zeroth unpolarized moment, we are nonetheless able to compute the sensitivity of the CTEQ-TEA set to
$\langle 1 \rangle_{u^{+}-d^{+}}$ and the related violation of the symmetric sea $\bar{u}=\bar{d}$ scenario
formulated in terms of Mellin moments; this connection crucially motivates lattice measurement of the higher
isovector moments $\langle x^{1,3} \rangle_{u^+-d^+}$ treated in Secs.~\ref{sec:q-moms} and \ref{sec:implem},
which would constrain the behavior of the phenomenological isovector distribution and inform its zeroth moment
and analyses of the Gottfried Sum Rule. Moreover, the $x\! <\! 0$ region of the isovector quasi-distribution presented
in Sec.~\ref{sec:quasi} is immediately related to $\bar{d}(x)\!-\!\bar{u}(x)$, again implying a complementary avenue
for lattice sensitivity to the light quark sea.

In Fig.~\ref{fig:Mellin_F2}, we map the calculated sensitivities of the CTEQ-TEA high-energy data set
to the zeroth isovector moment $\langle 1 \rangle_{u^{+}-d^{+}}$. While the general pattern of sensitivities
in Fig.~\ref{fig:Mellin_F2} is consistent with what we observed for the higher isovector
moments illustrated in Fig.~\ref{fig:mom_iso1}, the
sensitivity is especially substantial here for $W^\pm$ and $Z$ production and E866 cross section ratio (E866rat'01). To a lesser extent, we observe notable pulls
from an assembly of fixed-target measurements, including several
DIS experiments: the NMC structure function ratio information (NMCrat'97), the CCFR measurements
of $xF^p_3$ (CCFR-F3'97), and the BCDMS $F_2^p$ data (BCDMSp'89).

\begin{figure}
\hspace*{-0.75cm}
\includegraphics[scale=0.87]{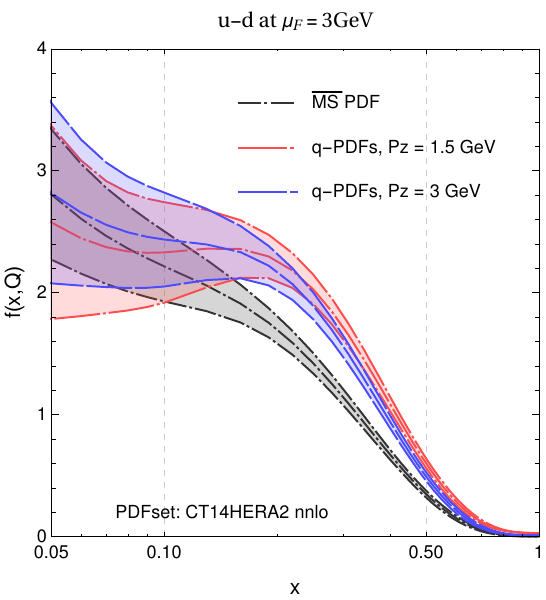} \ \
\includegraphics[scale=0.87]{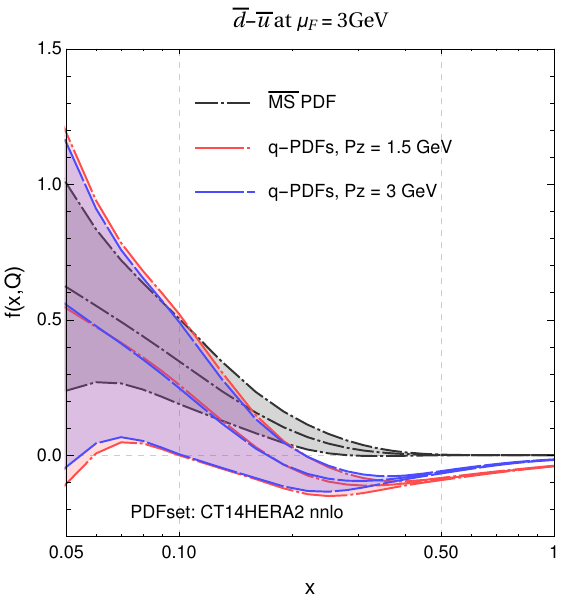}
	\caption{
		The parton quasi-distribution function $[\widetilde{u}-\widetilde{d}](x,P_z,\widetilde{\mu})$
		for $x > 0$ (left panel) and $x < 0$ (right panel), with the latter computed
		from the $\overline{\mathrm{MS}}$ PDF $\bar{u}-\bar{d}$ as given by CT14HERA2 NNLO.
	}
\label{fig:qPDF} 
\end{figure}

These visible features of the \texttt{PDFSense} sensitivity map are largely borne out by the point-averaged CTEQ-TEA
sensitivities to the zeroth isovector moment; like the moments of the higher isovector moments and $d^\pm$ distributions
explored above, the list of leading experiments ranked by this metric is again led by the $7$ TeV $\mu$ asymmetry data
recorded by CMS [CMS7Masy2'14 (0.645)], followed closely by the deuteron-proton cross section ratios measured by E866 [E866rat'01 (0.600)];
for the latter, this strong pull is notably consistent with E866's aim of probing the $x$ dependence of $\bar{d}(x) / \bar{u}(x)$ ---
a topic which continues to motivate modern experiments like SeaQuest. Following these, the per-datum sensitivities of the CTEQ-TEA data are dominated by
an amalgam of electroweak experiments represented by the rows of gauge boson data shown in Fig.~\ref{fig:Mellin_F2}. Again in descending
order, these include LHCb7Wasy'12 (0.546), LHCb8WZ'16 (0.432), CMS7Easy'12 (0.381), CMS8Wasy'16 (0.370), LHCb7ZWrap'15 (0.351),
D02Easy2'15 (0.323), D02Masy'08 (0.252), and ATL7WZ'12 (0.219).

Ordered according to their aggregated impact, on the other hand, only $9$ experiments
exceed $\sum |S_f| > 10$. These now include the usual DIS information from HERA and fixed-target
data from NMC, CCFR, and BCDMS, as well as the E866 $pp$ Drell-Yan cross section data --- again
owing to the aggregated pull of these enlarged data sets. In order of total sensitivity, these
most decisive experiments are
HERAI+II'15 (51.0), BCDMSp'89 (21.2), LHCb8WZ'16 (18.1), CCFR-F3'97 (15.7), NMCrat'97 (15.2), E866pp'03 (14.8), CMS8Wasy'16 (12.2),
LHCb7ZWrap'15 (11.6), and BCDMSd'90 (10.2).
Of these, there is again a pronounced effect from DIS experiments led by the
combined HERA data which contribute by merit of their marginal per-datum
sensitivity $\sim\! 0.05-0.1$ and the magnitude $N_\mathit{pt}$ of the data sets  
to which they belong, much as we observed for many of the other light quark moments above.

\begin{figure}
	\begin{tabular}{cc}
		\subfloat[Sensitivity to $\lbrack\widetilde{u}-\widetilde{d}\rbrack(x=-0.5,P_z=3.0\,\mathrm{GeV})$.]{\includegraphics[width=3.5in]{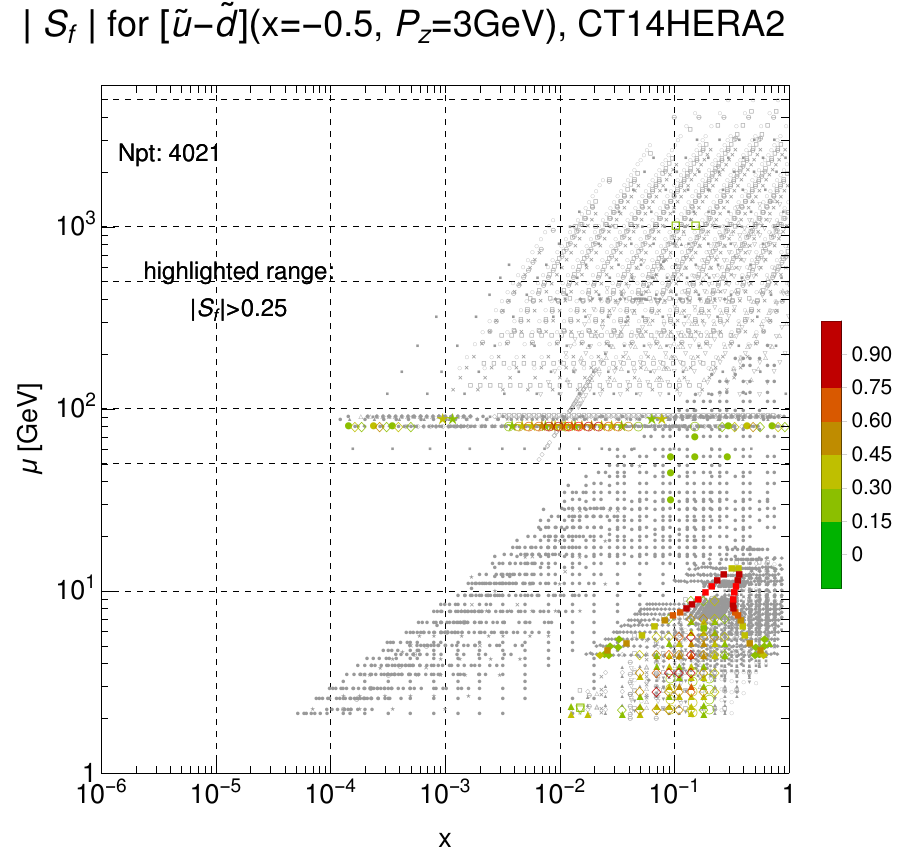}} &
		\subfloat[Sensitivity to $\lbrack\widetilde{u}-\widetilde{d}\rbrack(x=0.85,P_z=3.0\,\mathrm{GeV})$.]{\includegraphics[width=3.5in]{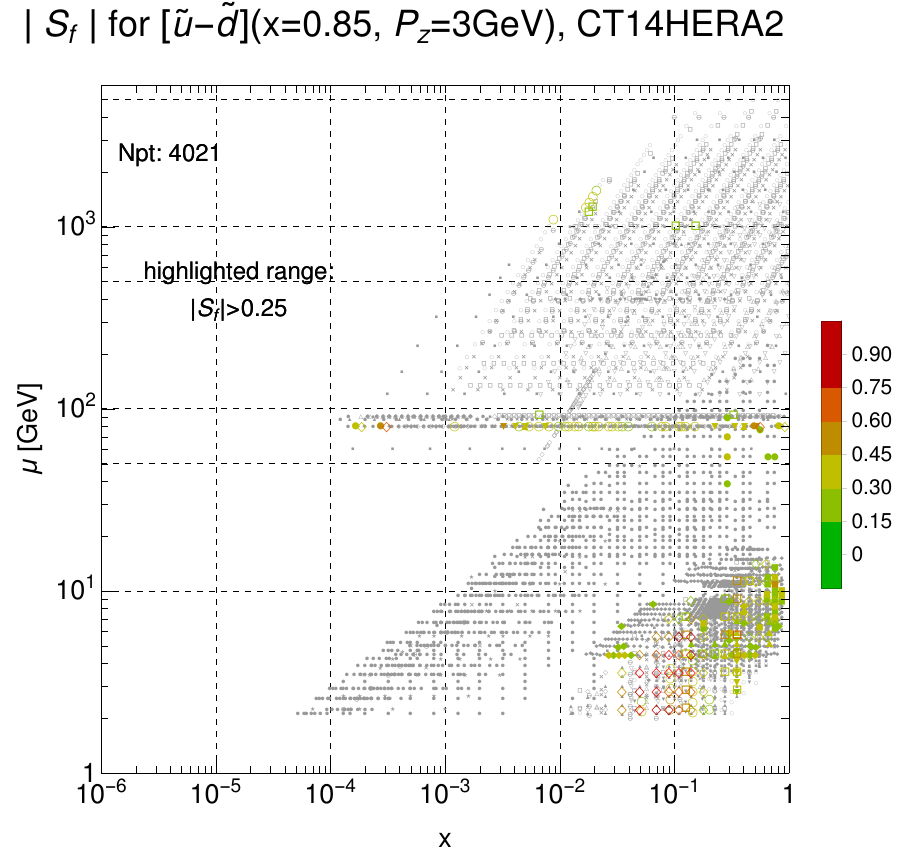}}\\
		\subfloat[Sensitivity to $\lbrack\widetilde{u}-\widetilde{d}\rbrack(x=-0.5,P_z=1.5\,\mathrm{GeV})$.]{\includegraphics[width=3.5in]{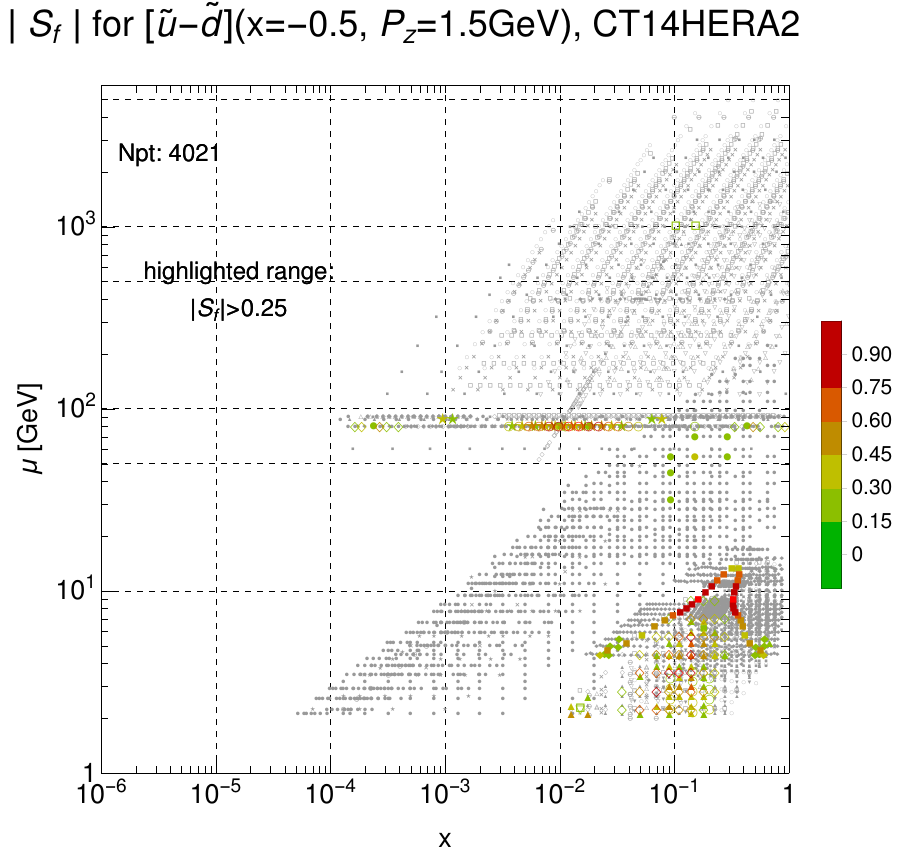}} &
		\subfloat[Sensitivity to $\lbrack\widetilde{u}-\widetilde{d}\rbrack(x=0.85,P_z=1.5\,\mathrm{GeV})$.]{\includegraphics[width=3.5in]{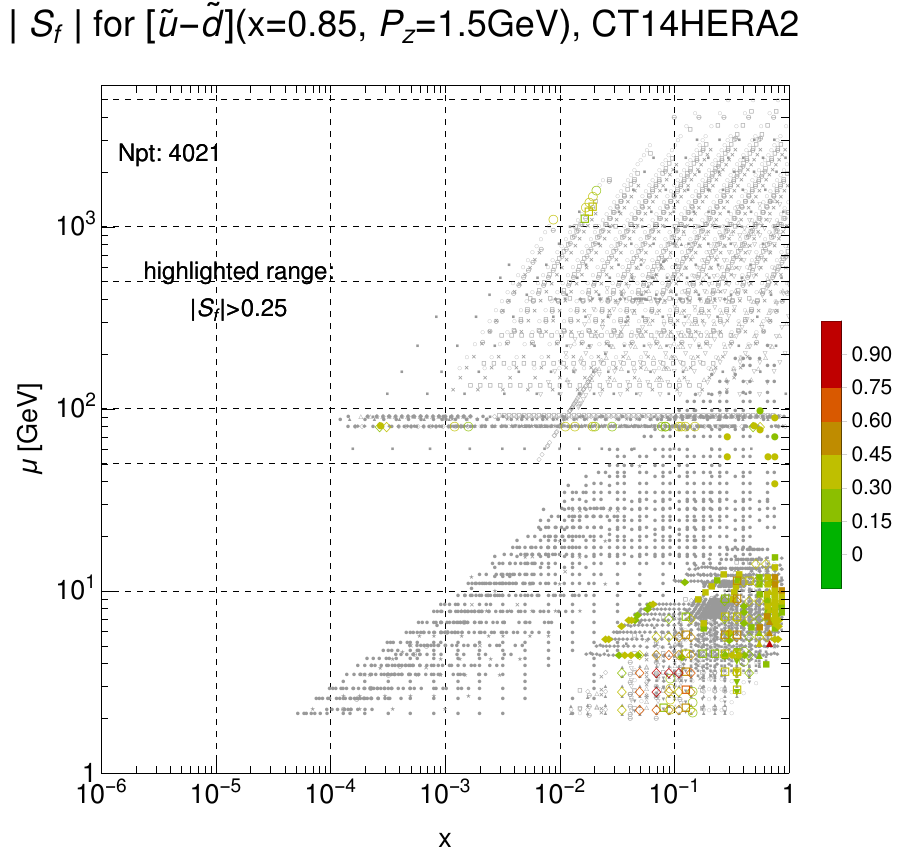}}
	\end{tabular}
\caption{
	The isovector quark quasi-distribution at large values of $\pm x = k_z / P_z$,
	{\it i.e.}, $x = -0.5$ (left panels) and $x = 0.85$ (right panels) for a relatively
	fast moving proton, boosted to $P_z = 3$ GeV (top panels), and comparatively
	slow protons boosted to $1/2$ this momentum, $P_z = 1.5$ GeV (lower panels).
}
\label{fig:quasi}
\end{figure}
%

%
% - - - - - - - - - - - - - - - - - - - - - - - - - - - - - - - - - - - - - - - - - - - - - -

%
%
\section{Sensitivities to quark quasi-distributions}
\label{sec:quasi}
In addition to the PDF Mellin moments we analyzed in Sec.~\ref{sec:moments}, it has recently been proposed \cite{Ji:2013dva}
that lattice QCD may evaluate parton ``quasi-distributions'' (qPDFs) over the quark-hadron longitudinal
momentum fraction $x = k_z / P_z$ by evaluating matrix elements of the form
\begin{equation}
	\widetilde{q}(x,P_z,\widetilde{\mu}) = \int_{-\infty}^{\infty}\, \frac{dz}{4\pi} e^{ix P_z z}
	\langle P | \overline{\psi}(z) \gamma^z U(z,0) \psi(0) | P \rangle\ ,
\label{eq:quasi-def}
\end{equation}
where $U(z,0)$ is a gauge link along the longitudinal $z$ direction, and the argument $\widetilde{\mu}$
represents the scales in the RI/MOM scheme \cite{Martinelli:1994ty}; in practice, this involves the introduction of the
parameters $p^R_z$ and $\mu_R$, which, for the purpose of this analysis, we fix to the values given
in Ref.~\cite{Liu:2018uuj}, $p^R_z = 2.2$ GeV and $\mu_R = 3.7$ GeV. Given its status as a matrix
element of correlation functions along a spacelike longitudinal direction (unlike the ordinary $\overline{\mathrm{MS}}$ PDFs),
the quasi-distribution of Eq.~(\ref{eq:quasi-def}) can be computed on the lattice\footnote{
  While the expression appearing in Eq.~(\ref{eq:quasi-def}) is standard in the quasi-PDF
literature, we clarify that in practice it can be advantageous to compute matrix elements with the replacement $\gamma^z \to \gamma^t$ \cite{Liu:2018uuj}.
While quasi-distributions computed with $\gamma^t$ have similar limiting
behavior for $P_z \to \infty$ as those determined using $\gamma^z$, lattice calculations carried out with $\gamma^t$
enjoy greater stability against operator mixing \cite{Constantinou:2017sej}, and the numerical results shown in this
section therefore assume this procedure.} and ultimately
matched to the traditional phenomenological PDFs via an inversion of the expression given in
Eq.~(\ref{eq:match}). On the other hand, rather than inverting Eq.~(\ref{eq:match}) to obtain
the $\overline{\mathrm{MS}}$ PDF from the lattice qPDF output, it is also possible to use Eq.~(\ref{eq:match})
to compute the $P_z$-dependent qPDF from a phenomenological $\overline{\mathrm{MS}}$ PDF. Before lattice output
matures to a sufficient level to help specify the $x$ dependence of PDFs through the combination
of qPDF calculations and LaMET, it will be crucial to benchmark lattice calculations against
knowledge derived from the fitted PDFs, and further improve perturbative matching and power-correction
formalism used to unfold PDFs from qPDF calculations on the lattice. Practically, these improvements
are informed by direct comparisons of lattice-calculated qPDFs and those matched from phenomenological
$\overline{\mathrm{MS}}$-PDFs computed according to Eq.~(\ref{eq:match}). As such, if the PDF
uncertainty of matched qPDFs determined from Eq.~(\ref{eq:match}) could be further
reduced by additional constraints from the appropriate experimental data, the ability to test
and refine the LaMET formalism would be substantially enhanced. This logic extends especially
to an understanding of the $P_z$ dependence of the matched qPDFs, which can be particularly
sensitive to power-corrections and the perturbative order of the matching formalism.

Thus, to illustrate the current knowledge of qPDFs {\it predicted}
using Eq.~(\ref{eq:match}), Fig.~\ref{fig:qPDF} displays the
qPDFs derived from the CT14HERA2 NNLO PDFs for $P_z=1.5$ and $3$ GeV,
together with the bands of current uncertainties on these qPDFs
estimated at the 90\% probability level. The uncertainties are
computed according to the Hessian master formula (\ref{eq:hess_unc})
from CT14HERA2 NNLO error sets. The error bands for the collinear
CT14HERA2 NNLO PDFs are also shown, labeled as
``$\overline{\mathrm{MS}}$ PDFs''.  
In fact, a Hessian error set for the qPDFs at a
given $P_z$, $\tilde \mu$, and $\mu$
can be obtained by applying Eq.~(\ref{eq:match}) to each error set of the CT14HERA2 NNLO ensemble. This algorithm is entirely 
analogous to the calculation of the error ensemble
for the Mellin moments using Eq.~(\ref{eq:rep_mom}).
The qPDF Hessian set $\widetilde{q}_{j\in{\{2N\}}}(x, P_z, \widetilde{\mu})$
may then be used to compute the sensitivities of the CTEQ-TEA set
to the quark quasi-distributions along the lines described
in Sec.~\ref{sec:analysis}.

\paragraph{\bf Sensitivity maps.} For this purpose, we again deploy \texttt{PDFSense}, this time to study in a proof-of-principle demonstration showing
the constraints from the present data on the $P_z$-dependent quasi-distributions
computed according to Eq.~(\ref{eq:match}) from the underlying phenomenological PDFs, given the current knowledge
of the perturbative matching coefficient $Z$ in Eq.~(\ref{eq:quasi-def}), computed using the one-loop
formalism of Ref.~\cite{Liu:2018uuj}. In the present Section, we assume an $\overline{\mathrm{MS}}$ factorization
scale of $\mu_F = 3$ GeV, as in Ref.~\cite{Liu:2018uuj}.

We wish to highlight both the dependence upon $x$ and $P_z$ of the quasi-distribution of the CTEQ-TEA
sensitivities, and we therefore plot in this section four panels in Fig.~\ref{fig:quasi} showing the
behavior of the quasi-distribution $[\widetilde{u}-\widetilde{d}](x,P_z,\widetilde{\mu})$ at two
representative values at relatively large $|x|$: $x = -0.5,\, 0.85$ for $P_z = 1.5$ and $3$ GeV. For
the quasi-distributions evaluated for $x\! <\! 0$, we note the implementation of the canonical relation
$\overline{q}(x) = -q(-x)$, such that the negative $x$ region of the quasi-distribution is related to the $x$ dependence of the
phenomenological anti-quark PDFs (on the logic that backward-moving quarks with longitudinal momenta
$k_z = x P_z < 0$ are identifiable with forward-moving anti-quarks).
\begin{figure}
	\begin{tabular}{cc}
		\subfloat[Sensitivity to $\lbrack\widetilde{u}-\widetilde{d}\rbrack(x=-0.05,P_z=1.5\,\mathrm{GeV})$.]{\includegraphics[width=3.5in]{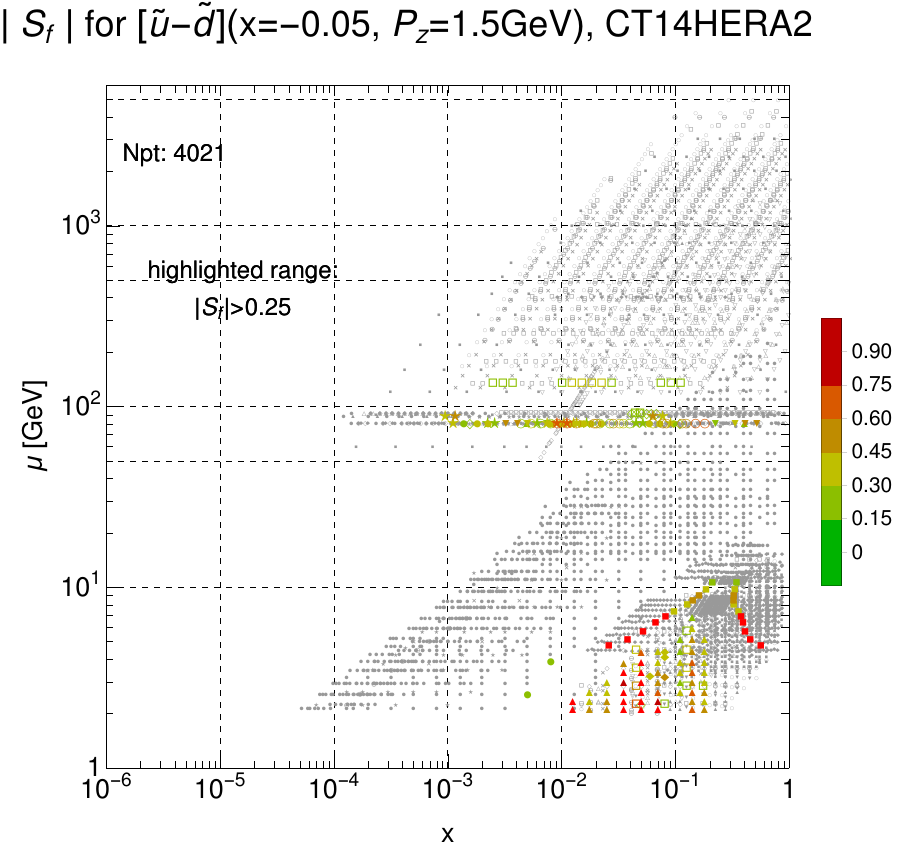}} &
		\subfloat[Sensitivity to $\lbrack\widetilde{u}-\widetilde{d}\rbrack(x=0.05,P_z=1.5\,\mathrm{GeV})$.]{\includegraphics[width=3.5in]{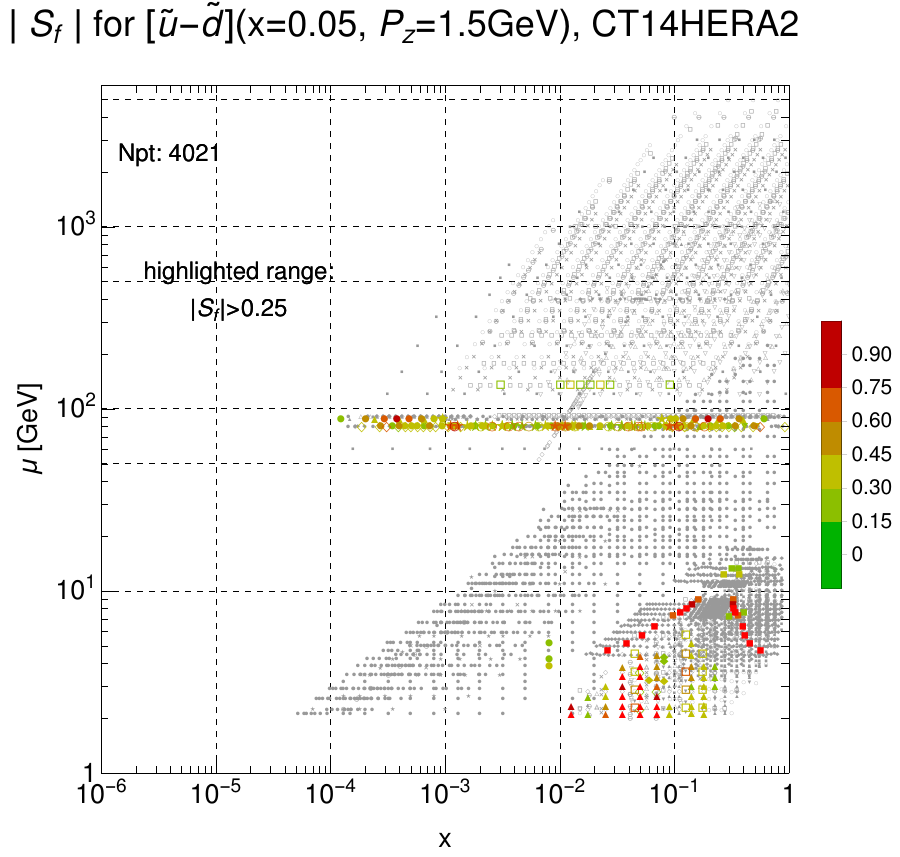}}
	\end{tabular}
\caption{
	Like Fig.~\ref{fig:quasi} for the isovector quark quasi-distribution, but now for
	comparatively small values of $|x|$, in this case, $x = -0.05$ (left panel) and $x = 0.05$ (right panel).
	Here, we plot $|S_f|$ maps only for the smaller boost scale, $P_z = 1.5$ GeV, as we find the $P_z$ dependence
	of the sensitivities at these smaller values of $|x|$ to be very mild.
}
\label{fig:quasi_smallx}
\end{figure}

The essential point that emerges from Fig.~\ref{fig:quasi} is the fact that a common cluster of experiments,
mostly of higher $x$ fixed-target and $W^\pm$ production and asymmetry measurements, represent the primary
constraint to the $\overline{u}-\overline{d}$ quasi-distribution in a fashion that is largely
independent of the boosted hadron's momentum $P_z$. Some intriguing $P_z$ dependence does begin
to emerge, however, for the CTEQ-TEA sensitivities to the highest $x$ region of the isovector quasi-distribution,
evident in Fig.~\ref{fig:quasi} by comparing the $x=0.85$ maps obtained for $P_z = 3.0$ and $1.5$ GeV in
the upper-right (b) and lower-right(d) panels.
In particular, the $P_z$ dependence appearing in the $|S_f|(x,\mu)$ distributions of Fig.~\ref{fig:quasi} is signaled
by the enhancement in the sensitivity to $[\widetilde{u}-\widetilde{d}](x=0.85)$ of the highest $x \gtrsim 0.5$ $\mu p$
DIS points of BCDMSp'89 and NMCrat'97 found for the $P_z = 1.5$ GeV [Panel (d)]  compared to the analogous calculation,
for the sensitivities to the $P_z = 3$ GeV quasi-distribution [Panel (b)].
This relative increase the sensitivity of the high-$x$ DIS information is offset by an accompanying relative reduction in
the general sensitivity of the $W^\pm, Z$ production data, which for $P_z = 3$ GeV exhibited significant pulls on
$[\widetilde{u}-\widetilde{d}](x=0.85)$, especially for the $7$ TeV $A_\mu(\eta)$ asymmetry data taken by CMS,
CMS7Masy2'14.
The implication of these observations is the fact that a careful exploration of the
nucleon structure function at high $x$ may provide crucial information for constraining the
$P_z$ dependence of the quasi-distributions required for a robust application of LaMET.
The qPDFs do not have the usual $x \in [0,1]$ support of the light-front PDFs, and are finite at
$x > 1$. For $P_z = 1.5,\,3$ GeV, the isovector qPDFs considered in this analysis are already rapidly vanishing
in the $x > 1$ region, however, and the sensitivity maps for $[\widetilde{u}-\widetilde{d}](x \gtrsim 1)$
are qualitatively similar to those shown in Figs.~\ref{fig:quasi}(b) and (d).

In addition, one can examine the qPDF sensitivity for shallower values of $|x|\! \sim\! 0$,
which we illustrate in Fig.~\ref{fig:quasi_smallx} to further explicate the sensitivity
dependence on $x$ of the qPDF. In Fig.~\ref{fig:quasi_smallx}, we plot $[\widetilde{u}-\widetilde{d}](x=0.-05)$
(a) and $[\widetilde{u}-\widetilde{d}](x=0.05)$ (b), both for $P_z = 1.5$ GeV.
Compared with $[\widetilde{u}-\widetilde{d}](x=-0.5)$ shown in Fig.~\ref{fig:quasi}(c), we
find a significant enhancement in the sensitivity of the smaller-$x$ fixed-target and E866 ratio
data to $[\widetilde{u}-\widetilde{d}](x=-0.05)$ in Fig.~\ref{fig:quasi_smallx}(a); this shift is accompanied
by a moderate redistribution in the sensitivity of the $W$-production data about $x \sim 10^{-2}$ in $(x,\mu)$-space.
On the other hand, in moving from $x=-0.05 \to 0.05$, changes to the data pulls on
$[\widetilde{u}-\widetilde{d}](x=0.05)$ plotted in Fig.~\ref{fig:quasi_smallx}(b)
occur mostly for the $W$- and $Z$-production data, especially at larger rapidities, corresponding to
$x \gtrsim 10^{-1}$ and $x \lesssim 10^{-3}$, while the sensitivities of the data in the fixed-target
region show comparatively weaker dependence on $x$ of the qPDF in the $|x|\!\sim\!0$ region. The
qPDF sensitivities of the full CTEQ-TEA data therefore exhibit a complex dependence upon $x$ of the qPDF
that implies the importance of experimental information from diverse channels in improving phenomenological
benchmarks for future qPDF lattice calculations.
%
% - - - - - - - - - - - - - - - - - - - - - - - - - - - - - - - - - - - - - - - - - - - - - -
%

\begin{figure}[b]
\hspace*{-0.75cm}
\includegraphics[scale=0.59]{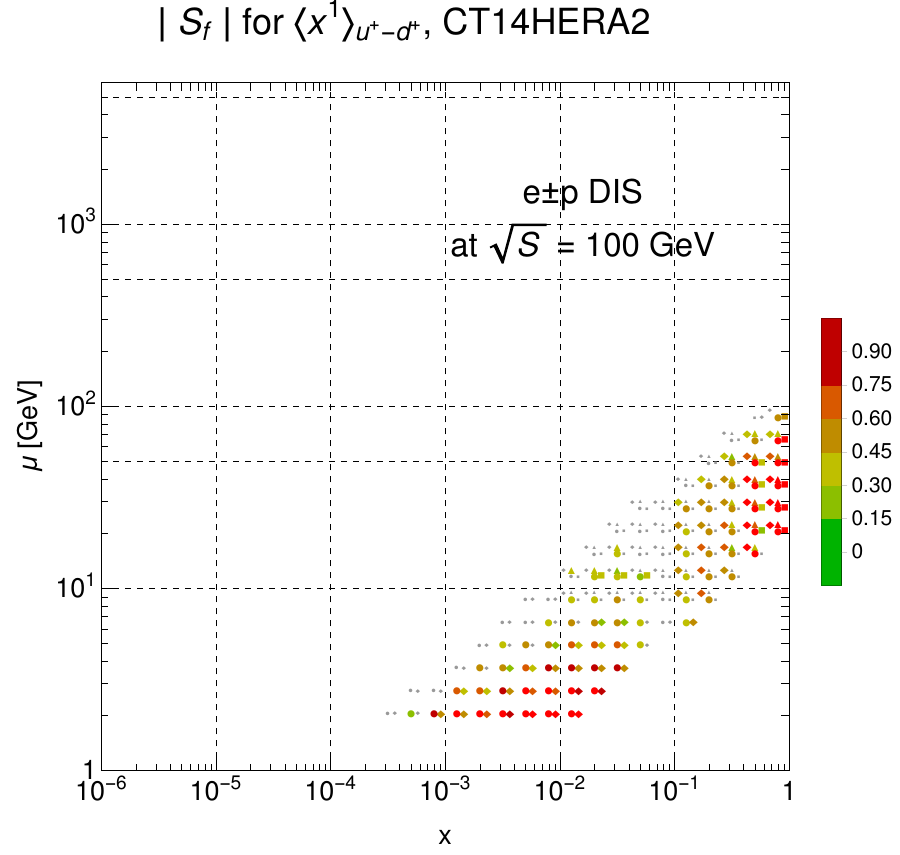} \ \
\includegraphics[scale=0.59]{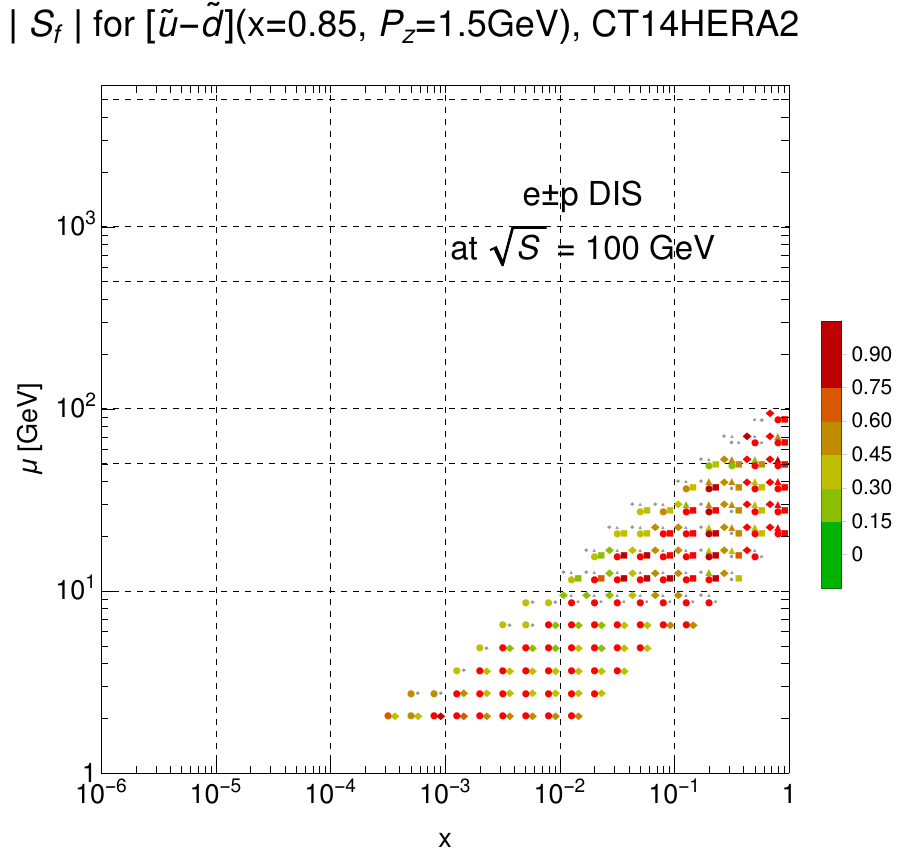}
	\caption{Sensitivity of pseudodata for the inclusive DIS of $e^\pm$ on unpolarized
	protons at $\sqrt{s} = 100$ GeV to the first Mellin moment of the isovector PDF combination $\langle x \rangle_{u^{+}-d^{+}}$ (left)
	at an $\overline{\mathrm{MS}}$ scale of $\mu = 2$ GeV. The right panel shows the sensitivity to the high-$x$ behavior of
	the quasi-distribution for the same isovector PDF, $[\widetilde{u}-\widetilde{d}](x,P_z,\widetilde{\mu})$ for $P_z=1.5$ GeV,
	$\mu = 3$ GeV, and $\widetilde{\mu}$ taken from Ref.~\cite{Liu:2018uuj} computed according to Eq.~(\ref{eq:match}).
        The plotted symbols characterize the specific channel as NC $e^-p$ (disks); NC $e^+p$ (diamonds); CC $e^-p$ (squares); CC $e^+p$ (triangles).
	}
\label{fig:EIC} 
\end{figure}

\section{Motivation for future experiments}
\label{sec:EIC}
A number of
futuristic machines have either been proposed or planned with a stated aim (among other physics motivations)
of disentangling the collinear structure of hadronic states, including various futuristic
hadron-collider experiments like the HL-LHC \cite{Apollinari:2015bam} and, {\it e.g.}, the
AFTER@CERN proposal \cite{Brodsky:2012vg}. Among these proposals,
a number of lepton-hadron colliders
have been advocated, especially a future US-based electron-ion collider
(EIC) \cite{Accardi:2012qut,Boer:2011fh,Abeyratne:2012ah,Aschenauer:2014cki} and a
lepton-nucleon/nucleus variant of the LHC, the Large Hadron-Electron Collider
(LHeC) \cite{AbelleiraFernandez:2012cc}. An EIC, in particular,
is most likely to serve the dedicated role
of a hadron tomography machine, given its high-luminosity coverage of
the crucial few-GeV quark-hadron transition  
region in the kinematical parameter space. An EIC would
enjoy unprecedented facility in
unfolding the nucleon's collinear and transverse structure at scales adjacent
to the nucleon mass, $\gtrsim\! M$. The science output of an EIC would greatly
build upon the JLab12 program \cite{Dudek:2012vr}, while the kinematic
coverage of an EIC would be particularly favorable for constraining
the quantities accessible in next-generation lattice QCD calculations. 

As a simple illustration of the potential of a future DIS program, we compute the sensitivity maps that result
from implementing a set of pseudodata into the \texttt{PDFSense} framework and examining
our impact metrics for the first moment of the isovector distribution $\langle x \rangle_{u^{+}-d^{+}}$
and the high-$x$ behavior of the $P_z = 1.5$ GeV isovector quark quasi-distribution
$[\widetilde{u}-\widetilde{d}](x,P_z,\widetilde{\mu})$. To avoid marrying our predictions
to the specifics of a particular experimental proposal, we consider an archetypal machine that measures the reduced cross section $\sigma_r (x, Q^2)$ via inclusive
$e^\pm$ scattering on an unpolarized proton target. For this example, cross section pseudodata
are produced in bins specified by a realistic Monte Carlo event generator
about the CT14HERA2 NNLO theoretical prediction with a
Gaussian smearing function of standard deviation equal to the an assumed uncorrelated
error taken from the Monte Carlo simulation. Theoretical predictions are for the reduced
cross section measured in $e^\pm\,p$ scattering at $\sqrt{s} = 100$ GeV in both neutral- and charge-current interactions.
For this illustration, statistical uncertainties are based upon assumed integrated
luminosities of $\mathcal{L} = 100\, \mathrm{fb}^{-1}$ in $e^-p$ scattering and
$\mathcal{L} = 10\, \mathrm{fb}^{-1}$ for $e^+p$ events.

Fig.~\ref{fig:EIC} estimates the potential impact such a lepton-nucleon
collider might have on the above-noted lattice computable quantities:
in the left panel, the first moment, $\langle x \rangle_{u^+-d^+}$, of the isovector
quark distribution, and, in the right panel, the large-$x$ quasi-PDF matched from the CT14HERA2 NNLO
NNLO PDF set according to Eq.~(\ref{eq:match}). In both panels, physical channels
for the inclusive DIS process are explicitly represented by unique symbols; these
are NC $e^-p$ (disks); NC $e^+p$ (diamonds); CC $e^-p$ (squares); CC $e^+p$ (triangles).

Fig.~\ref{fig:EIC} indicates that measurements at a high-luminosity lepton-nucleon collider
can considerably improve the constraints on both quantities considered. In particular, they supply
very substantial sensitivities across the range of $x$ of the
data set, with especially large predicted impacts for $x\! \gtrsim\! 0.1$ as well
as the $x\! \lesssim\! 0.01$ regions. A notable feature of this information is the separation that
emerges illustrating the crucial role of both electron and positron probes:
once separated among channels, a prominent
effect is the important role of the charged current (CC)-mediated
positron-nucleon scattering ($e^{+}p$); this impact is very pronounced at large $x\! \gtrsim\! 0.1$. At the EIC, the sensitivities at $x>0.01$ arise from both the NC and CC channels, while at $x<0.001$, the sensitivities
mainly come from NC $e^{+}p$ and $e^{-}p$ channels.

\begin{figure}[p]
\vspace{-0.5cm}
\includegraphics[scale=0.4]{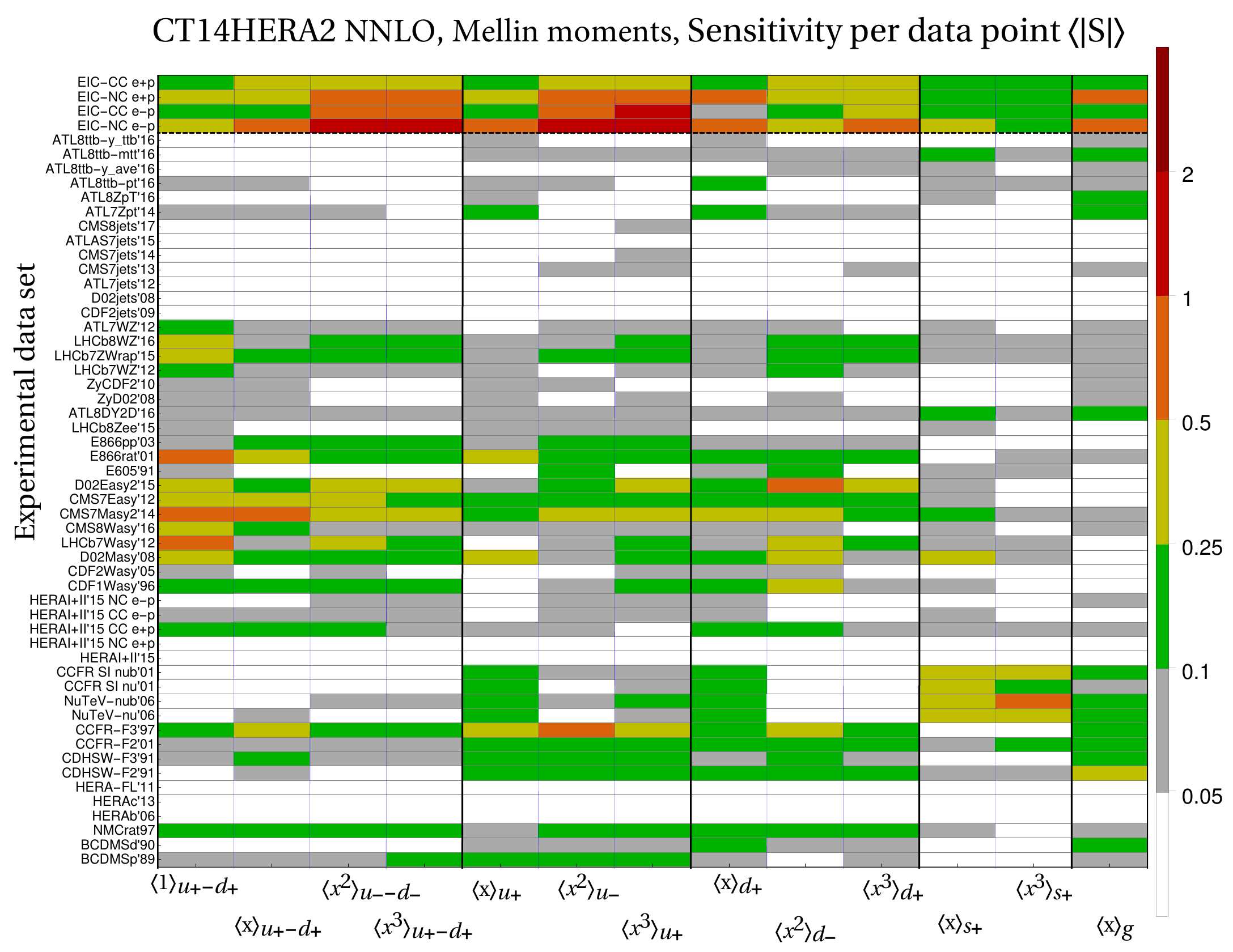} \
\includegraphics[scale=0.4]{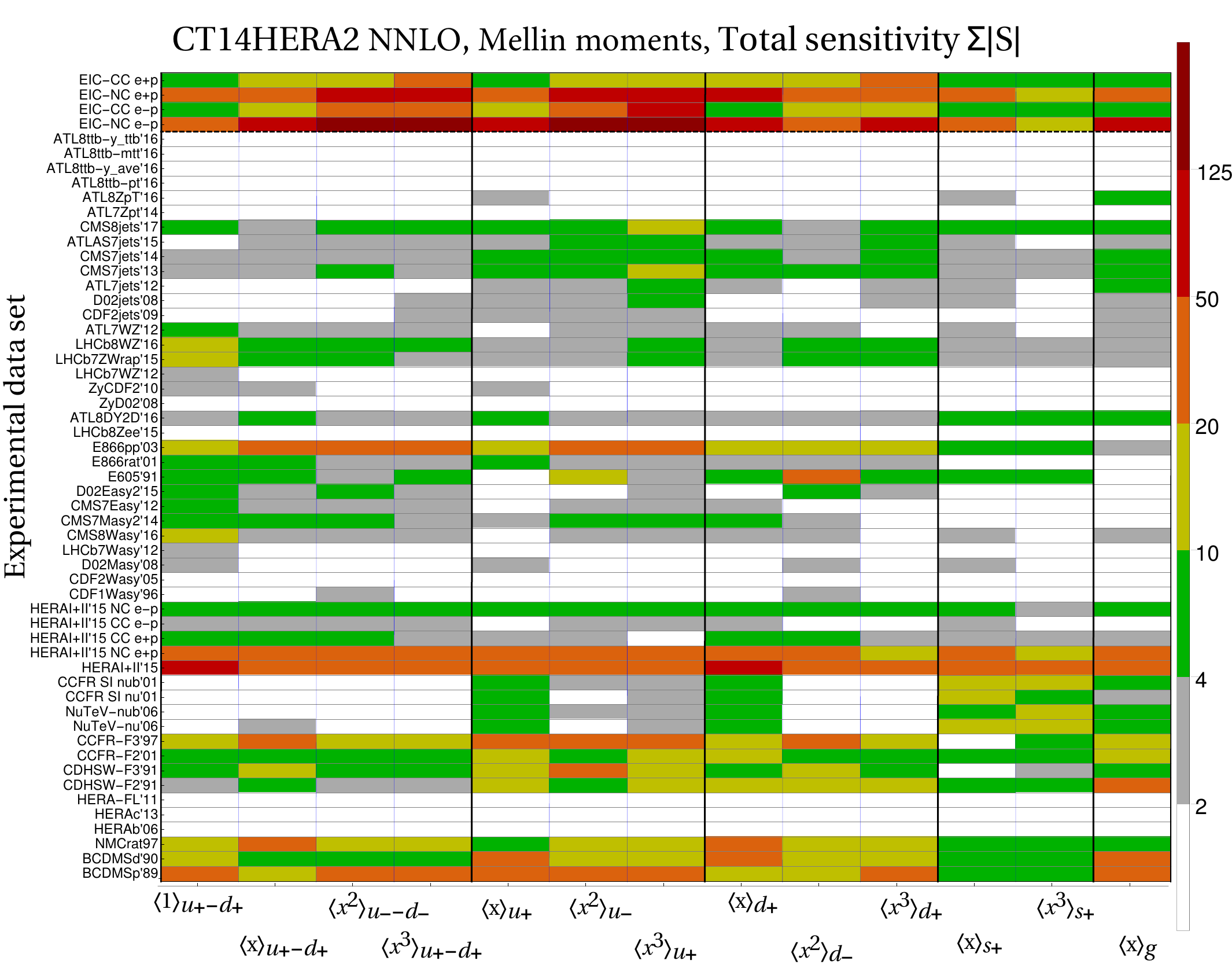}
\caption{
A graphical representation of the sensitivity of each of the constituent experiments
contributing to the CTEQ-TEA data set. The grids summarize the point-averaged (upper panel) and
summed or total (lower panel) sensitivities of each experimental data set to each moment for several flavor combinations
of strong interest; the color of the cell encodes the magnitude of the combined
sensitivity for that particular moment. In addition, we also include in the rightmost rows the sensitivities obtained for
pseudodata consistent with a future EIC-like lepton-nucleon collider experiment in the inclusive, unpolarized
sector.
	}
\label{fig:rank_plot} 
\end{figure}

\begin{figure}[p]
\vspace{-0.5cm}
\includegraphics[scale=0.5]{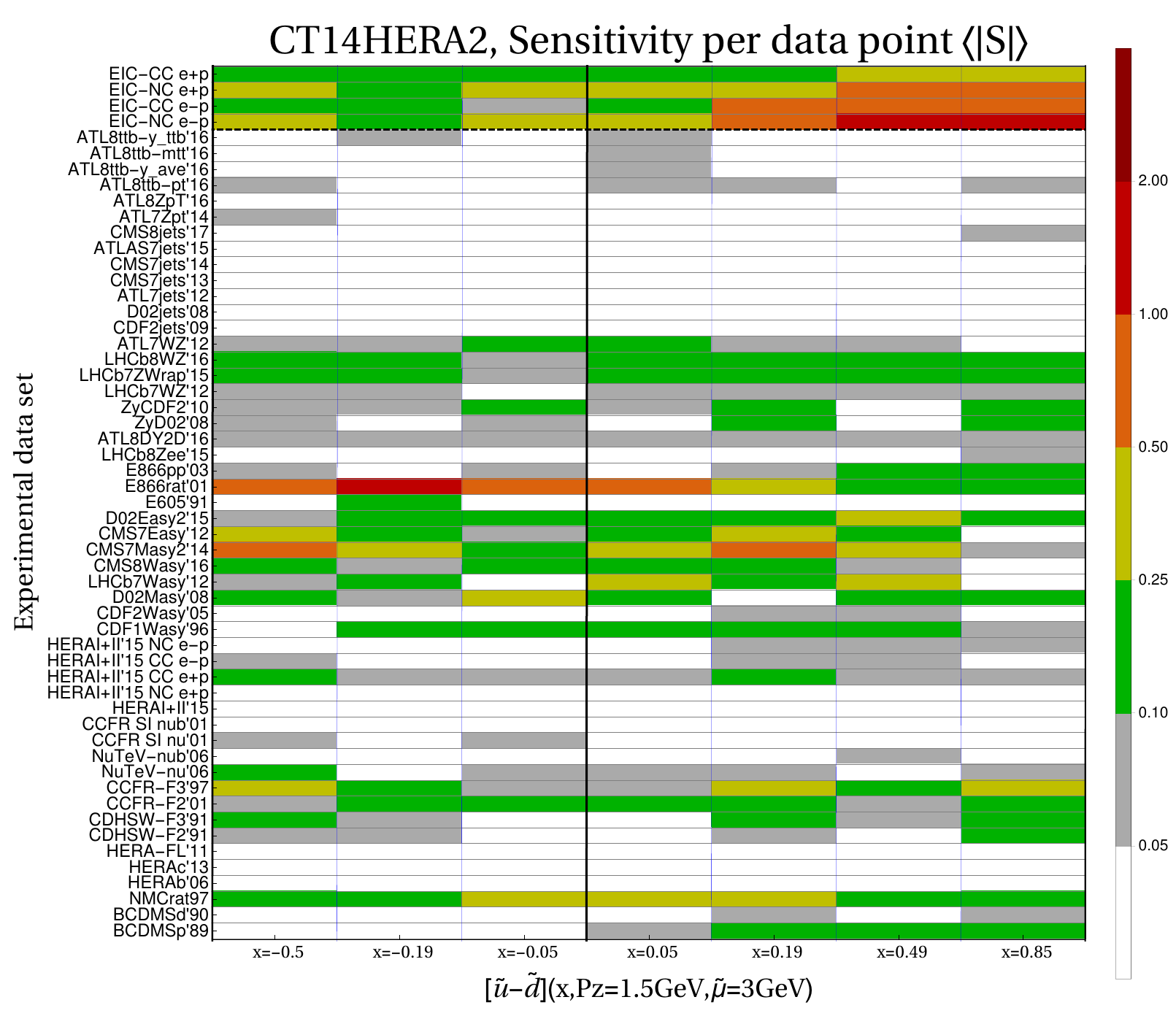} \
\includegraphics[scale=0.5]{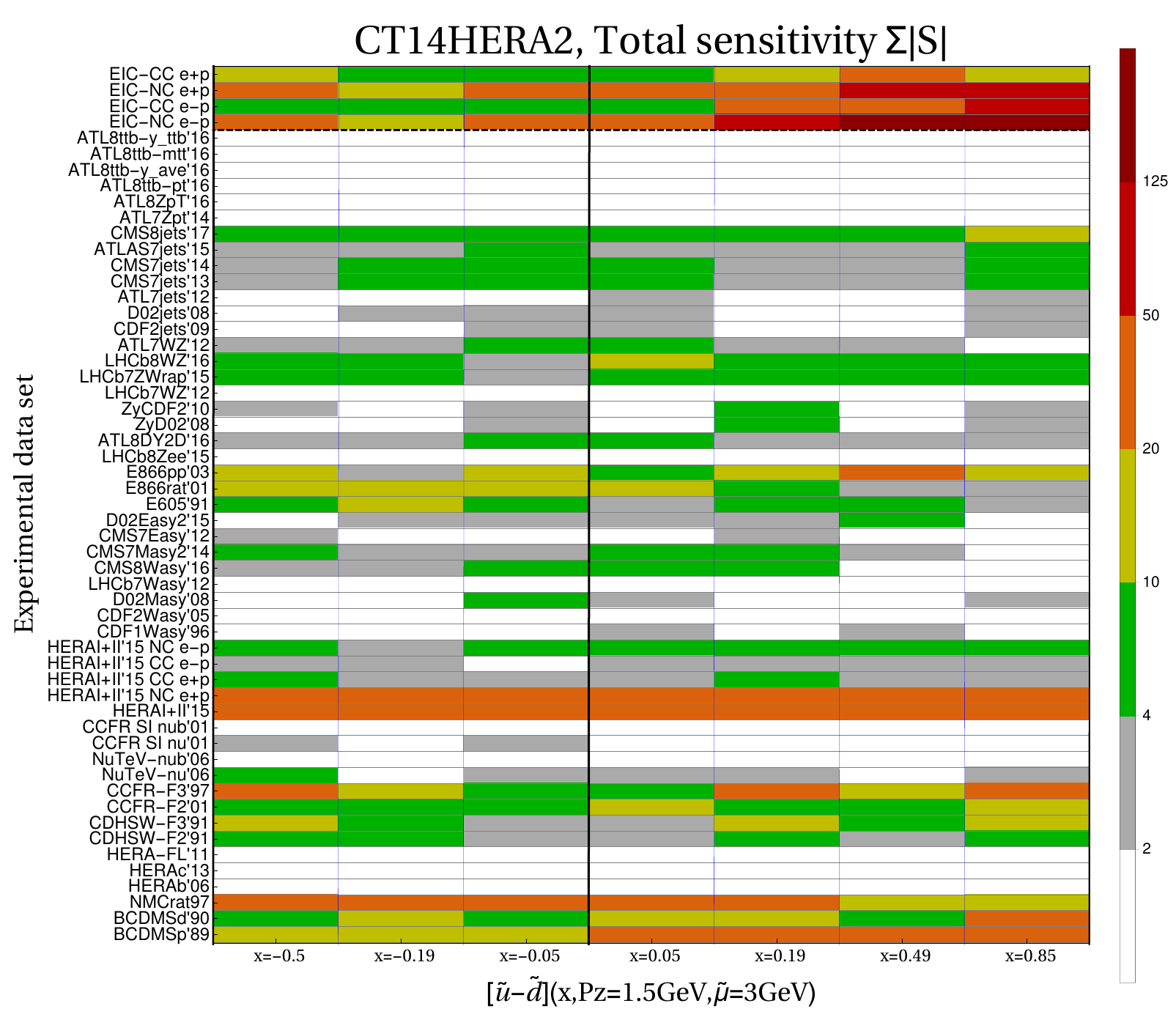} 
\caption{
	As Fig.~\ref{fig:rank_plot}, but in this case illustrating the per-datum (upper
	panel) and aggregated (lower panel) sensitivities of the experiments within the CTEQ-TEA set to specific
	$x$ regions (indicated at the bottom) of the isovector
	quasi-distribution at $P_z = 1.5$ GeV. As in Fig.~\ref{fig:rank_plot}, we again
	show the evaluations for pseudodata corresponding to a future EIC-like machine at the
	top.
	}
\label{fig:rank_plot_qPDF} 
\end{figure}

The $|S_f|(x_i,\mu_i)$ sensitivity maps appearing throughout this analysis, including
those in Fig.~\ref{fig:EIC} for the EIC pseudodata, are complemented by the companion plots
shown in Figs.~\ref{fig:rank_plot} and \ref{fig:rank_plot_qPDF}. These figures
integrate the information displayed in the sensitivity maps for the PDF moments
(Fig.~\ref{fig:rank_plot}) and qPDFs (Fig.~\ref{fig:rank_plot_qPDF})
experiment-by-experiment, thereby summarizing
the point-averaged sensitivities $\langle | S_f | \rangle$ for all of the
CTEQ-TEA experiments and EIC pseudodata considered in this analysis, as well as
the corresponding aggregated sensitivities, $\sum_{i \in N_\mathit{pt}} |S^i_f|$.
These companion plots encapsulate the related information summarized in Tables~\ref{tab:EXP_1}-\ref{tab:EXP_3}
of Appendix~\ref{sec:append}. These tables sort the CTEQ-TEA experiments
in descending order of their total sensitivity (A, B, C) and point-averaged
sensitivity (1, 2, 3). The conventions that define these A-B-C, and 1-2-3
categories are defined in the caption of Table~\ref{tab:EXP_1}. Information
on the PDF moments for additional flavor combinations, as well as the numerical
values for the total and aggregated sensitivities, can be found on the companion
webpage, Ref.~\cite{Website}.

It is worth noting that there is
often a closer correspondence between experiments highlighted in
sensitivity maps like Fig.~\ref{fig:mom_iso1} and those identified
in the grid plots for the point-averaged sensitivities $\langle | S_f | \rangle$,
{\it i.e.}, the upper panels of Figs.~\ref{fig:rank_plot} and \ref{fig:rank_plot_qPDF}.
For this reason, in discussing our numerical results as illustrated
by the following sensitivity maps, we summarize the highest impact experiments
according to complementary considerations of those data sets that enjoy sizable
per-datum sensitivities to the PDF moments, and those that may not in general
possess high-impact points taken in isolation, but are nonetheless predicted
to have a large aggregated impact --- often by merit of the large number of
experimental data points $N_\mathit{pt}$ they have.

\clearpage

Our recurring observation in this article has been that, in the
CTEQ-TEA global analysis, 
the experimental data involving nuclear targets affords  critical, and, in many cases, leading, information on essentially
all PDF moments analyzed in Sec.~\ref{sec:moments}. This is similarly true of the isovector
qPDF examined in Sec.~\ref{sec:quasi}. Details of the nuclear binding at work in the deuteron, for instance,
are relevant for a number of the CTEQ-TEA sets, including BCDMSd'90, NMCrat'97, and E866rat'01. On the other hand,
heavier nuclear systems were probed in several other fixed-target experiments, especially those involving
$\nu$DIS; these include CDHSW (both $F_2$ and $F_3$ sets, measured on Fe), the inclusive CCFR and
semi-inclusive dimuon data from NuTeV and CCFR (all also measured on Fe), and the E605 fixed-target $pA$
Drell-Yan measurements (Cu target). In multiple instances --- for example, in the impact plots for the
strangeness moments $\langle x^{1,3} \rangle_{s^+}$, the $C$-odd combinations $\langle x^2 \rangle_{u^-,d^-}$,
and even the gluon total momentum $\langle x \rangle_g$ --- these experiments represent the first, second, or
third most influential information by the aggregated or point-averaged sensitivity, or both.

Present phenomenological constraints, particularly at large $x$, are therefore strongly dependent on data for which
nuclear corrections are an important consideration. These corrections are
imperfectly known, and often dependent on model treatments or an assumption that that
nuclear correction effects are simply absorbed into extracted PDF uncertainties.
An EIC would be well-poised to address these issues by performing detailed studies
of nuclear medium effects.
%

%
% - - - - - - - - - - - - - - - - - - - - - - - - - - - - - - - - - - - - - - - - - - - - - -
%
%
%
\section{Implementation of lattice data in QCD analyses}
\label{sec:implem}
In the foregoing sections we have analyzed various empirical constraints
upon {\it individual} lattice QCD observables which are either presently accessible or expected to
be in the foreseeable future. These experimental data were taken either from the CTEQ-TEA high-energy data set
or generated as hypothetical pseudodata recorded at an EIC-like $e^\pm p$ DIS collider. The main purpose of this
exploration was to identify the experimental
processes and measurements that will impose the strongest
constraints on the lattice-calculable quantities dependent on the PDFs,
and that thus can serve as stringent phenomenological benchmarks to help
lattice calculations reach their maturity.
But to this latter point, is it possible to proceed in the direction
converse to the one taken by asking: how might the multifaceted results of lattice QCD constrain the $x$ dependence of collinear
PDFs fitted in future global analyses?
Given the complexity of the multichannel information on the Mellin
moments and qPDFs furnished by the lattice, its
inclusion in the upcoming PDF fits may produce a plethora of
nontrivial constraints on the underlying parametrizations.
Rather than attempting to disentangle the many potential effects from
the implementation of lattice data into a global fit, we will use
\texttt{PDFSense} to investigate 
some general properties, such as the typical momentum fractions
constrained by including a specific Mellin moment.

\begin{figure}[t]
\hspace*{-0.75cm}
\includegraphics[scale=0.59]{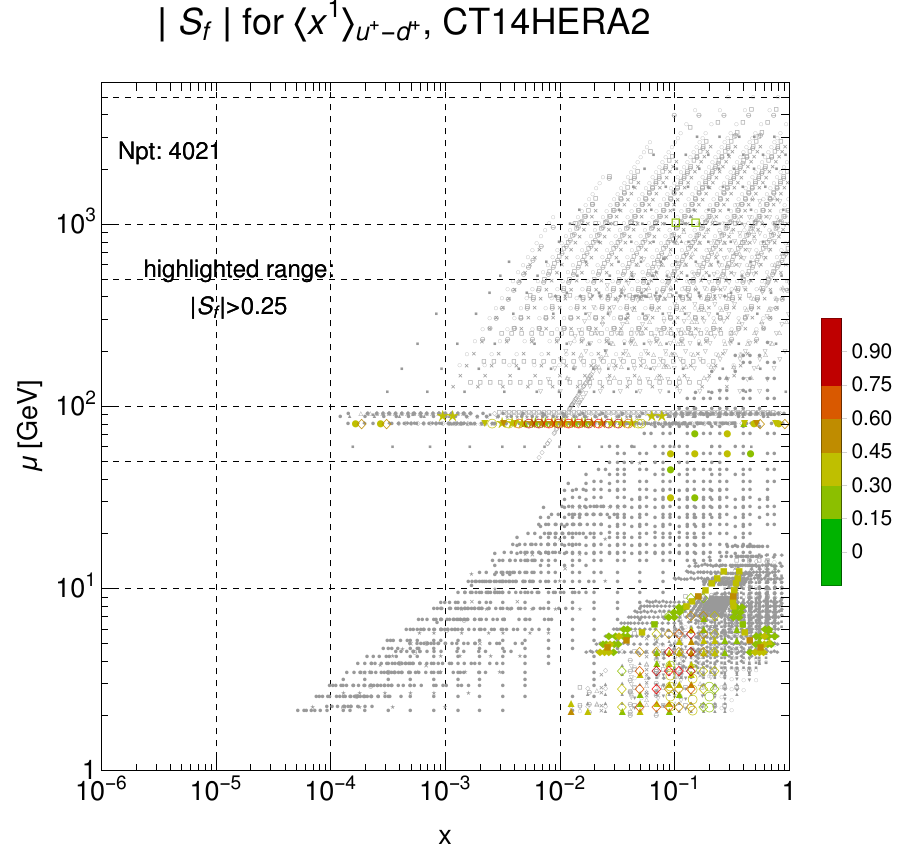}
\ \ \                        
\includegraphics[scale=0.59]{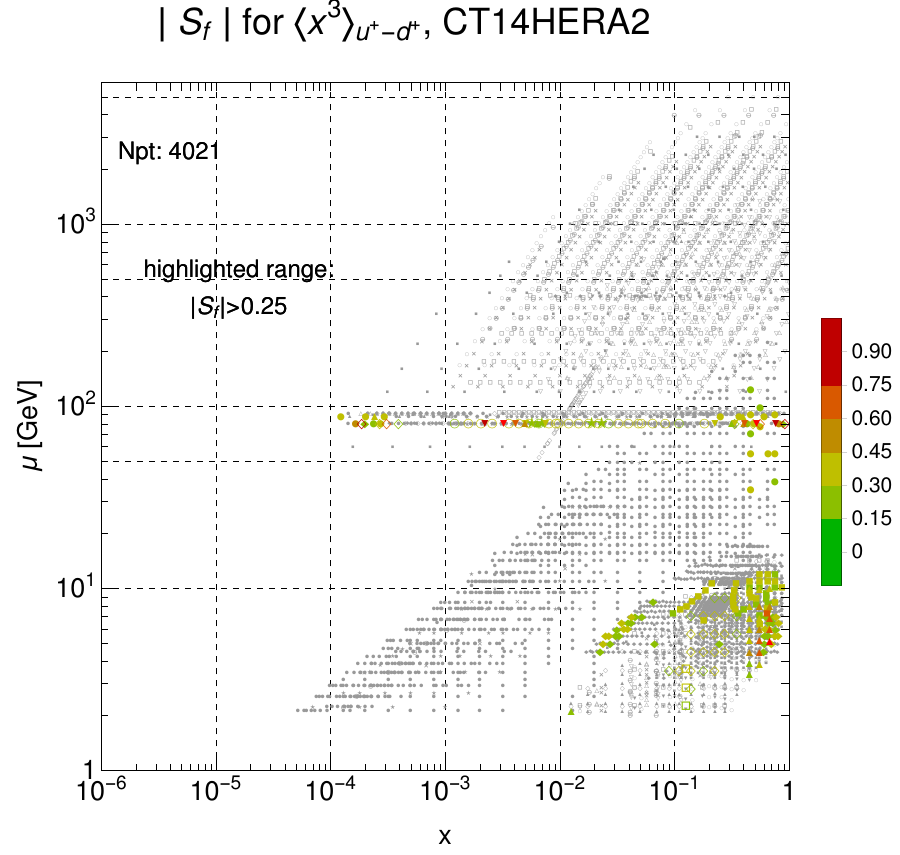}
\caption{
The two lowest lattice-accessible moments of the isovector PDF combination $u^+\!-\!d^+$.
	}
\label{fig:isovec} 
\end{figure}

\begin{figure*}
\hspace*{-0.5cm}
\includegraphics[scale=0.78]{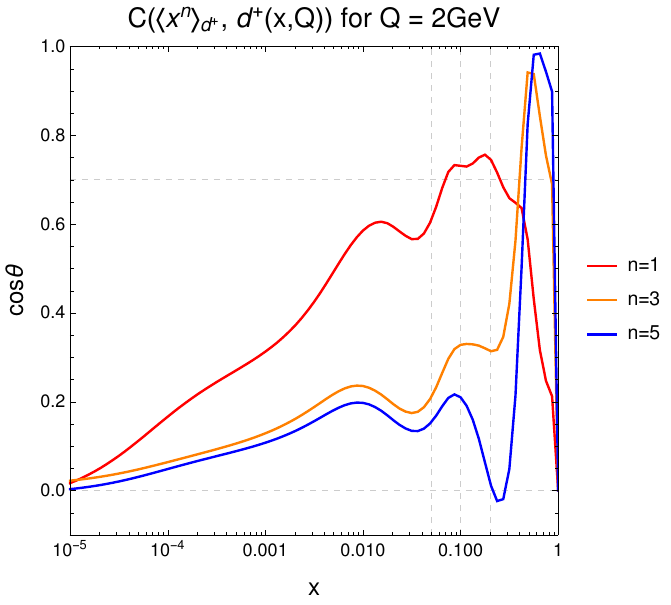} \ \ \ 
\includegraphics[scale=0.78]{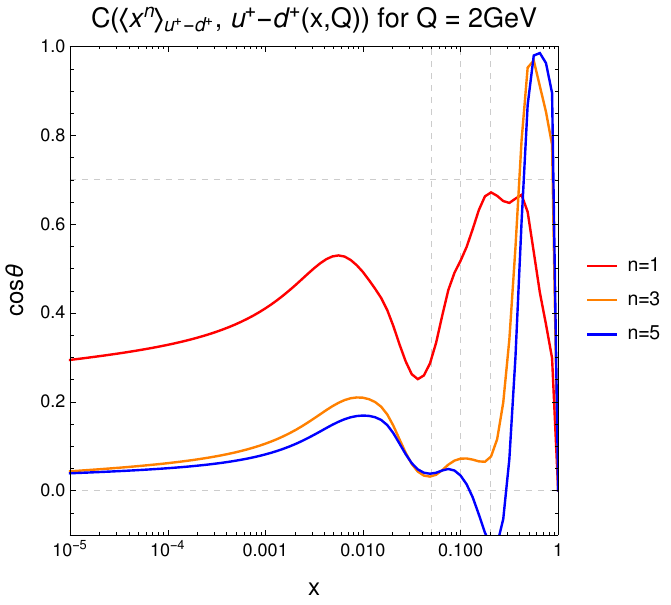}
	\caption{The correlation between the Mellin moments and their corresponding PDFs for the $d^+$ (left panel)
	as well as the isovector $u^+\! -\! d^+$ (right panel) moments plotted in Fig.~\ref{fig:isovec}. As done elsewhere
	for the Mellin moment calculations, these are shown for $\mu_F = 2$ GeV. For each flavor,
	we plot the $x$-dependent correlation between the PDF and its integrated moment $\langle x^n \rangle_{q^+}$,
	for $n = 1, 3$, and $5$.
	}
\label{fig:corr-mell} 
\end{figure*}

We again consider the moments of the $\mathrm{SU}(2)$ isovector
distribution $u-d$, in this case, contrasting the two lowest moments of the
$q^+$-type distribution,
$\langle x \rangle_{u^+-d^+}$ and $\langle x^3 \rangle_{u^+-d^+}$, for which we
plot the $|S_f|(x,\mu)$ sensitivity map in the left and right panels of
Fig.~\ref{fig:isovec}, respectively.
We clarify that in Fig.~\ref{fig:mom_iso1} of Sec.~\ref{sec:q-moms}, we examined
$\langle x \rangle_{u^+-d^+}$ and $\langle x^2 \rangle_{u^--d^-}$, but here
we directly examine the effect of incrementing the order of the Mellin
moment on a specific
flavor/charge combination for the purpose of showcasing the relationship
between the order and the associated $x$ dependence of the constraints imposed by
data.
In moving from the first moment characterized in Fig.~\ref{fig:mom_iso1} and surrounding
text to the third, we find a notable reduction in the point-averaged sensitivity, $\langle | S_f | \rangle$,
of the leading experiment, which remains CMS7Masy2'14 (0.342), but is now immediately
succeeded by D02Easy2'15 (0.307). The other leading experiments by per-datum sensitivity also
remain E866rat'01 (0.225), CMS7Easy'12 (0.203), and CCFR-F3'97 (0.187), but with significant
decreases in their values of $\langle | S_f | \rangle$ for $\langle x^3 \rangle_{u^+-d^+}$.
By inspecting the total sensitivities, an important reordering of the experimental
hierarchy becomes evident. In this case, BCDMSp'89 (39.4), with its large share of measurements
at high $x$ and $\mu\! \sim\! 10$ GeV, displaces the combined HERA data, HERAI+II'15 (34.9),
in terms of total pull. The important experiments identified by their aggregated sensitivities
to $\langle x \rangle_{u^+-d^+}$ continue to place strong constraints upon $\langle x^3 \rangle_{u^+-d^+}$,
with these being E866pp'03 (24.4), CCFR-F3'97 (16.1), and NMCrat'97 (15.5). For these data, however, there is
a salient rightward shift toward higher $x$ in the $(x,\mu)$ space displayed in Fig.~\ref{fig:isovec}.
As a straightforward metric to quantify the distribution over $x$ of the sensitivities $|S_f|(x,\mu)$ plotted
in Fig.~\ref{fig:isovec}, we may evaluate an ensemble average
\begin{equation}
\overline{x}_{|S_f|} = { \sum_i x_i |S^i_f| \over \sum_i |S^i_f| }\ ,
\label{eq:COG}
\end{equation}
where the sum $i$ runs over data points in the CTEQ-TEA set. On the basis
of this metric and the panels of Fig.~\ref{fig:isovec}, a relationship emerges
between the order $n$ of the PDF moment $\langle x^n \rangle_{q^\pm}$
and the kinematics of the most constraining data in the global analysis, with PDF moments
of higher order being constrained more strongly by data recorded at higher $x$.
For the first and third isovector moments plotted in
Fig.~\ref{fig:isovec}, we obtain a systematic increase in $\overline{x}_{|S_f|}$
as the order of the moment is enlarged, finding a shift from $\overline{x}_{|S_f|} = 0.193$
for $\langle x \rangle_{u^+-d^+}$ to $\overline{x}_{|S_f|} = 0.286$ for $\langle x^3 \rangle_{u^+-d^+}$.
Similar relationships are observed between $\overline{x}_{|S_f|}$ as given by
Eq.~(\ref{eq:COG}) and the Mellin moments of other PDF flavors and combinations.
It is possible to further unravel the observed $x$ dependence in $|S_f|(x,\mu)$ by considering the correlation defined
in Eq.~(\ref{eq:corr-def}). In Fig.~\ref{fig:corr-mell}, we plot the $x$-dependent correlation between
the PDF and its corresponding Mellin moment for two examples --- the lowest three lattice-accessible moments
of the $d^+$ distribution (left panel) and the same information for the isovector $u^+\! -\! d^+$. Across both
panels, we observe the same qualitative $x$ dependence in the correlation as the order $n$ of the
Mellin moment $\langle x^n \rangle_{q^+}$ is increased. Specifically, while the lowest $n=1$ moment is significantly
correlated with its PDF's $x$ dependence over a wide range of $x$, peaking near $x\! \sim\! 0.1$, this correlation
vanishes rapidly at highest $x$. On the other hand, the PDF correlations with higher moments are rather different, in this
case being quite modest, especially for the highest $n=5$ moment, over most of the plotted range before becoming very
large, $C_f\! \sim\! 1$, at $x\! \gtrsim\! 0.3$. In fact, this behavior was reflected in
Fig.~\ref{fig:isovec}, which demonstrated the sensitivity shift in $|S_f|(x,\mu)$ to favor many of the
large-$x$ data as the Mellin moment is increased. Taken in conjunction with the correlation results shown
in Fig.~\ref{fig:corr-mell}, we may infer that the sensitivity of high-$x$ data to higher moments
follows from an underlying reciprocal relation that connects the high-$x$ behavior of PDFs to
their higher-order Mellin moments.
The observation that moments $\langle x^n \rangle_{q^\pm}$ of successively higher order ($n \ge 1$)
are increasingly sensitive to the PDFs' large-$x$ behavior provides an impetus to seek alternative 
moment-weighting functions which may be sensitive to low $x$.
One possible choice would be successively higher moments of distributions
smeared with polynomials in the difference, $(1-x)^n$, {\it i.e.}, $\langle(1-x)^{n}\rangle_{q^\pm}$.
In principle, information on the $\langle(1-x)^{n}\rangle_{q^\pm}$ moments may 
be of use for constraining PDFs in the region of small $x$, where they
must be integrable in the limit $x\rightarrow0$ to ensure $\langle(1-x)^{n}\rangle_{q}$
are well-defined. In fact, since the polynomial expansion of $(1-x)^{n}$ is a linear combination
in $x^{n}$, {\it i.e.},
\begin{align}
\langle(1-x)^{n}\rangle_{q}=\sum_{k=0}^{n}\ &C_{k}^{n}\ (-1)^{k}\ \langle x^{k}\rangle_{q} \nonumber \\
	                                   &C_{k}^{n}\equiv\frac{n!}{k!(n-k)!}\ ,
\label{eq:bin}
\end{align}
results for select moments, $\langle x^{n} \rangle$, provided by the lattice
might not only help constrain PDF behavior at successively larger $x$, but might also provide
useful supplementary information to assist efforts in the context of QCD global analyses to
determine PDF behavior at lower $x$. We clarify this point with a specific example. As argued
in Sec.~\ref{sec:SFs}, the low-$x$ behavior of the isovector difference $u^+(x)\! -\! d^+(x)$
in the integrand of Eq.~(\ref{eq:Gott}) depends primarily upon the sea-quark flavor
asymmetry $\bar{d}-\bar{u}$. By the logic of Eq.~(\ref{eq:bin}), a useful quantity with
sensitivity to small-$x$ variations in $u^+(x)\! -\! d^+(x)$ might be the weighted-moment,
\begin{align}
\langle(1-x)^{3}\rangle_{u^+-d^+}= 
   \langle 1 \rangle_{u^+-d^+}
-3 \langle x \rangle_{u^+-d^+}^\star
+3 \langle x^{2} \rangle_{u^+-d^+}
-  \langle x^{3} \rangle_{u^+-d^+}^\star\ ,
\end{align}
in which the two ``starred'' $\langle x^{1,3} \rangle$ terms above can in principle be directly
informed by lattice calculations. While successively higher moments $\langle x^{n} \rangle$
are increasingly correlated with PDF behavior at the largest $x \to 1$, the right panel of
Fig.~\ref{fig:corr-mell} suggests that, through the underlying parametrization, the
$\langle x^{n} \rangle$ moments may nonetheless have significant correlation with the
behavior of $u^+(x)\! -\! d^+(x)$ for $x \lesssim 10^{-2}$ or smaller --- especially for
$n=1$. As a result, future PDF analyses seeking to unlock low-$x$ densities may benefit
from lattice QCD constraints to select Mellin moments, which, with sufficient precision,
might help reduce parametrization dependence and supplement fits to low-$x$ data.

We therefore stress that it will be crucial for future phenomenological analyses to leverage
a plurality of lattice results to most strongly constrain their likelihood functions or parameter spaces. The necessity of doing so
is apparent from our results on the moments of $\langle x^n \rangle_{u^+-d^+}$, in which
higher moments ($n > 1$) will be useful in coordination with the leading total momentum fractions
$\langle x \rangle_{u^+-d^+}$ to maximize constraints over wide reaches of $x$. We also emphasize
the fact that the $|S_f|(x,\mu)$ plots imply a synergistic relationship between high-energy and lattice
data, with, for example, higher-order moments being especially valuable in informing fitted PDFs' $x$
dependence near $x\! \approx\! 1$, where high-precision data can be technically difficult
to obtain. Similar logic will place a high premium on lattice results obtained using the quasi-PDF approach.
%
%
% - - - - - - - - - - - - - - - - - - - - - - - - - - - - - - - - - - - - - - - - - - - - - -
%
%
\section{Conclusion}
\label{sec:conc}
As the number of observables accessible to lattice QCD continues to grow,
the necessity for PDF phenomenologists to grapple with the resulting output will
be increasingly unavoidable.
The chief message of this article is that this represents an opportunity to form
a potentially powerfully synergy between phenomenological PDF analyses and lattice QCD.
This synergy will be grounded in the ability of PDF phenomenologists to drive improvements in lattice
calculations with benchmarks informed by high-energy data, while the lattice provides informative
constraints in kinematical regions that are otherwise challenging to constrain empirically.
Before the envisioned relationship can be fully realized, however, both communities must establish
a common basis for comparing results from lattice QCD and global fits --- a challenging undertaking
given the complex contemporary landscape of lattice calculations and global fits, which involve
a patchwork of theoretical settings, systematic assumptions, and, in the case of QCD analyses, empirical
data sets. The \texttt{PDFSense} technology deployed in this article provides a standardized framework in which
apples-to-apples evaluations of the pulls of experimental information on lattice-calculable
quantities are possible. This fact suggests one avenue for assessing the empirical origins of
phenomenological predictions of lattice data, and a path forward for improving them.

While some studies have investigated the result of selecting
assortments of lattice data for inclusion into a global analysis,
in this work we have gone the other direction and
examined the constraints data place on a quantities which have been computed
on the lattice. In the process, we have established several primary findings:

\begin{itemize}

\item
We have demonstrated the correspondence between phenomenological predictions for specific physical
measurements and the importance of experimental information for benchmarking
lattice calculations. Conversely, our impact study in the form of the sensitivity maps for the various lattice
observables illustrates those regions of parameter space where improved lattice data can be expected
to have a driving impact on PDF studies. For instance, precise lattice data on $\langle x \rangle_{d^+}$
could improve knowledge of $d^+(x)$ in the high-$x$ region, given the sensitivity map of
Fig.~\ref{fig:Mellin_ud+moments} (right panel) and the correlations plotted in Fig.~\ref{fig:corr-mell}.
Similar logic applies to the moments of other flavors.

\item
In Sec.~\ref{sec:moments}, we found that that most moments presently accessible to the lattice are
mainly constrained by a small collection of high-impact experiments --- for many of the light quark Mellin
moments, for instance, a combination of HERA and fixed-target DIS data are especially decisive, as
illustrated for the  $u^+\!-\!d^+$, $u^+,\,d^+$ and $s^+$ moments appearing in Figs.~\ref{fig:mom_iso1}, \ref{fig:Mellin_ud+moments},
and \ref{fig:Mellin_s}, respectively.  The overall sensitivities of the CTEQ-TEA experiments to the quantities considered
in this analysis are summarized in Figs.~\ref{fig:rank_plot}
and~\ref{fig:rank_plot_qPDF}, as well as Tables~\ref{tab:EXP_1}-\ref{tab:EXP_3}. The CTEQ-TEA data pulls
differ substantially for even vs.~odd Mellin moments, due to the $C$-oddness vs.~evenness
of the associated quark distributions, seen by comparing the sensitivity maps of
Fig.~\ref{fig:Mellin_ud+moments} against those of Fig.~\ref{fig:Mellin_ud-moments}.
		
\item
We have also observed systematic tendencies in the sensitivities
of high-energy data to Mellin moments, including a robust connection between the order $n$ of the Mellin
moments $\langle x^n \rangle_{q^\pm}$ of quark distributions and regions of $x$ of the PDFs shown
in Fig.~\ref{fig:isovec}. This connection
is `bidirectional' in the sense that experimental information at higher $x$ are likely to exert stronger
pulls on higher-order Mellin moments, while lattice information on the higher-order moments may potentially
constrain the high-$x$ behavior of fitted PDFs. We conclude the eventual implementation
of lattice data into QCD analyses will benefit from the inclusion of Mellin moments
of various orders and parton flavors as well as knowledge gained from quasi-PDFs (qPDFs) to constrain PDFs'
$x$ dependence as widely as possible.

\item
We have for the first time studied in Sec.~\ref{sec:quasi} the driving constraints from
high-energy data on calculations of the $P_z$-dependent qPDFs. In so doing, we have illustrated the direct
link between qPDFs as theoretical quantities and the empirical information upon which calculations of the
matched qPDFs from phenomenological distributions depend. One intriguing consequence of this is the possibility of
more thoroughly constraining the $P_z$ dependence of the phenomenologically-matched qPDFs with, {\it e.g.}, DIS measurements concentrated at
large $x$ shown in Fig.~\ref{fig:quasi}. Reductions to the PDF uncertainties of the high-$x$ qPDFs could help drive theoretical
improvements in the LaMET formalism used to extract PDFs from lattice data.

\item
We can assess the potential impact of future experiments like
the high-luminosity lepton-nucleon collider considered in Fig.~\ref{fig:EIC}, and we conclude
that data from such a machine would be extremely beneficial for advancing the phenomenology of
PDF moments and matched qPDFs.
The constraints on Mellin moments typically arise from DIS data on fixed {\it nuclear} targets, such that
nuclear corrections may be an important effect in PDF extractions involving these data. As one example, Sec.~\ref{sec:gluon}
showed that the gluon Mellin moment $\langle x \rangle_g$ receives its largest
constraints from DIS data measured on iron nuclei (CCFR, CDHSW), which are known to have a different
preference for the large-$x$ gluon than HERA DIS information on the proton. A
future lepton-nucleus collider would enlighten such dependence of $\langle x \rangle_g$
and other moments on the nuclear environment.
\end{itemize}

As gains continue to be made on the complementary fronts of lattice theory and QCD analysis in the coming
years, the analysis carried out in this work will be of value to guide phenomenologists and lattice
practitioners in fully leveraging the synergy between their fields to improve our knowledge of
hadron structure.
Finally, we note that a comprehensive set of results have been collected at the public
URL in Ref.~\cite{Website}. While this collection includes many of the calculations shown
in this manuscript, a range of other sensitivity maps and related computations omitted
here for brevity are also shown.
%
% - - - - - - - - - - - - - - - - - - - - - - - - - - - - - - - - - - - - - - - - - - - - - -
%

%
\section{Acknowledgments}
We are grateful to Yu-Sheng Liu for his assistance regarding the computation of
matched quasi-distributions from phenomenological fits. We also thank Wally Melnitchouk,
Kostas Orginos, Rik Yoshida, and our CTEQ-TEA colleagues for helpful exchanges during the
completion of this manuscript. This work was supported by the U.S.~Department of Energy
under Grant No.~DE-SC0010129. T.~J.~Hobbs acknowledges support from an EIC Center Fellowship.
%
%
% - - - - - - - - - - - - - - - - - - - - - - - - - - - - - - - - - - - - - - - - - - - - - -
%

\appendix

\section{Appendix: Tabulated sensitivities}
\label{sec:append}
Similar to Table VI of the \texttt{PDFSense} paper in Ref.~\cite{Wang:2018heo}, the Appendix gives in
Tables~\ref{tab:rank_1}--\ref{tab:rank_3} a series of ranking tables of the Mellin moments $\langle x^n \rangle$.
Additional tables for other moments and for the isovector quasi-PDF at several values
of $x$ and $P_z$ appear in the companion website given as Ref.~\cite{Website}.

\begin{table}
\begin{tabular}{|l|c|lr|c|}
\hline
\textbf{Experiment name}  &  \textbf{CT\ ID\#}  & \textbf{Dataset details}  &  & $N_{\mathit{pt}}$ \tabularnewline
\hline
 BCDMSp'89      &  101  & BCDMS $F_{2}^{p}$  & \cite{Benvenuti:1989rh}  & 337 \tabularnewline
\hline
 BCDMSd'90      &  102  & BCDMS $F_{2}^{d}$  & \cite{Benvenuti:1989fm}  & 250 \tabularnewline
\hline
 NMCrat'97     &  104  & NMC $F_{2}^{d}/F_{2}^{p}$  & \cite{Arneodo:1996qe}  & 123 \tabularnewline
\hline
 CDHSW-F2'91     &  108  & CDHSW $F_{2}^{p}$  & \cite{Berge:1989hr}  & 85 \tabularnewline
\hline
 CDHSW-F3'91     &  109  & CDHSW $F_{3}^{p}$  & \cite{Berge:1989hr}  & 96 \tabularnewline
\hline
 CCFR-F2'01      &  110  & CCFR $F_{2}^{p}$  & \cite{Yang:2000ju}  & 69 \tabularnewline
\hline
 CCFR-F3'97      &  111  & CCFR $xF_{3}^{p}$  & \cite{Seligman:1997mc}  & 86 \tabularnewline
\hline
 NuTeV-nu'06    &  124  & NuTeV $\nu\mu\mu$ SIDIS  & \cite{Mason:2006qa}  & 38 \tabularnewline
\hline
 NuTeV-nub'06    &  125  & NuTeV $\bar{\nu}\mu\mu$ SIDIS  & \cite{Mason:2006qa}  & 33 \tabularnewline
\hline
 CCFR SI nu'01   &  126  & CCFR $\nu\mu\mu$ SIDIS  & \cite{Goncharov:2001qe}  & 40 \tabularnewline
\hline
 CCFR SI nub'01  &  127  & CCFR $\bar{\nu}\mu\mu$ SIDIS  & \cite{Goncharov:2001qe}  & 38 \tabularnewline
\hline
 HERAb'06       &  145  & H1 $\sigma_{r}^{b}$ ($57.4\mbox{ pb}^{-1}$)  & \cite{Aktas:2004az}\cite{Aktas:2005iw}  & 10 \tabularnewline
\hline
 HERAc'13       &  147  & Combined HERA charm production ($1.504\mbox{ fb}^{-1}$)  & \cite{Abramowicz:1900rp}  & 47 \tabularnewline
\hline
 HERAI+II'15     &  160  & HERA1+2 Combined NC and CC DIS ($1\mbox{ fb}^{-1}$)  & \cite{Abramowicz:2015mha}  & 1120 \tabularnewline
\hline
 HERA-FL'11      &  169  & H1 $F_{L}$ ($121.6\mbox{ pb}^{-1}$)  & \cite{Collaboration:2010ry}  & 9 \tabularnewline
\hline
\end{tabular}\caption{Experimental data sets considered as part of CT14HERA2 NNLO and included
in this analysis: deep-inelastic scattering. We point out that the numbering scheme (CT ID\#)
included in this and subsequent tables follows the standard CTEQ labeling system
with, {\it e.g.}, Expt.~IDs of the form 1XX representing DIS experiments, {\it etc.}
The HERA combined data set HERAI+II'15 consists of both neutral-current (NC) and charge-current
(CC) scattering events.
\label{tab:EXP_1} }
\end{table}

\begin{table}
\begin{tabular}{|l|c|lr|c|}
\hline
\textbf{Experiment name}  &  \textbf{CT\ ID\#}  & \textbf{Dataset details}  &  & $N_{\mathit{pt}}$ \tabularnewline
\hline
\hline
 E605'91      &   201   & E605 DY  & \cite{Moreno:1990sf}  & 119 \tabularnewline
\hline
 E866rat'01 &   203   & E866 DY, $\sigma_{pd}/(2\sigma_{pp})$  & \cite{Towell:2001nh}  & 15 \tabularnewline
\hline
 E866pp'03    &   204   & E866 DY, $Q^{3}d^{2}\sigma_{pp}/(dQdx_{F})$  & \cite{Webb:2003ps}  & 184 \tabularnewline
\hline
 CDF1Wasy'96 &   225   & CDF Run-1 $A_{e}(\eta^{e})$ ($110\mbox{ pb}^{-1}$)  & \cite{Abe:1996us}  & 11 \tabularnewline
\hline
 CDF2Wasy'05 &   227   & CDF Run-2 $A_{e}(\eta^{e})$ ($170\mbox{ pb}^{-1}$)  & \cite{Acosta:2005ud}  & 11 \tabularnewline
\hline
 D02Masy'08   &   234   & D$\emptyset$~ Run-2 $A_{\mu}(\eta^{\mu})$ ($0.3\mbox{ fb}^{-1}$)  & \cite{Abazov:2007pm}  & 9 \tabularnewline
\hline
 LHCb7WZ'12   &   240   & LHCb 7 TeV $W/Z$ muon forward-$\eta$ Xsec ($35\mbox{ pb}^{-1}$)  & \cite{Aaij:2012vn}  & 14 \tabularnewline
\hline
 LHCb7Wasy'12 &   241   & LHCb 7 TeV $W$ $A_{\mu}(\eta^{\mu})$ ($35\mbox{ pb}^{-1}$)  & \cite{Aaij:2012vn}  & 5 \tabularnewline
\hline
 ZyD02'08     &   260   & D$\emptyset$~ Run-2 $Z$ $d\sigma/dy_{Z}$ ($0.4\mbox{ fb}^{-1}$)  & \cite{Abazov:2006gs}  & 28 \tabularnewline
\hline
 ZyCDF2'10    &   261   & CDF Run-2 $Z$ $d\sigma/dy_{Z}$ ($2.1\mbox{ fb}^{-1}$)  & \cite{Aaltonen:2010zza}  & 29 \tabularnewline
\hline
 CMS7Masy2'14 &   266   & CMS 7 TeV $A_{\mu}(\eta)$ ($4.7\mbox{ fb}^{-1}$)  & \cite{Chatrchyan:2013mza}  & 11 \tabularnewline
\hline
 CMS7Easy'12 &   267   & CMS 7 TeV $A_{e}(\eta)$ ($0.840\mbox{ fb}^{-1}$)  & \cite{Chatrchyan:2012xt}  & 11 \tabularnewline
\hline
 ATL7WZ'12    &   268   & ATLAS 7 TeV $W/Z$ Xsec, $A_{\mu}(\eta)$ ($35\mbox{ pb}^{-1}$)  & \cite{Aad:2011dm}  & 41 \tabularnewline
\hline
 D02Easy2'15  &   281   & D$\emptyset$~ Run-2 $A_{e}(\eta)$ ($9.7\mbox{ fb}^{-1}$)  & \cite{D0:2014kma}  & 13 \tabularnewline
\hline
 CDF2jets'09 &   504   & CDF Run-2 incl. jet ($d^2\sigma/dp_{T}^{j}dy_{j}$) ($1.13\mbox{ fb}^{-1}$)  & \cite{Aaltonen:2008eq}  & 72 \tabularnewline
\hline
 D02jets'08  &   514   & D$\emptyset$~ Run-2 incl. jet ($d^2\sigma/dp_{T}^{j}dy_{j}$) ($0.7\mbox{ fb}^{-1}$)  & \cite{Abazov:2008ae}  & 110 \tabularnewline
\hline
 ATL7jets'12 &   535   & ATLAS 7 TeV incl. jet ($d^2\sigma/dp_{T}^{j}dy_{j}$) ($35\mbox{ pb}^{-1}$)  & \cite{Aad:2011fc}  & 90 \tabularnewline
\hline
 CMS7jets'13 &   538   & CMS 7 TeV incl. jet ($d^2\sigma/dp_{T}^{j}dy_{j}$) ($5\mbox{ fb}^{-1}$)  & \cite{Chatrchyan:2012bja}  & 133 \tabularnewline
\hline
\end{tabular}\caption{Same as Table~\ref{tab:EXP_1}, showing experimental data sets for
production of vector bosons, single-inclusive jets, and $t\bar{t}$
pairs.
\label{tab:EXP_2} }
\end{table}

\begin{table}
\begin{tabular}{|l|c|lr|c|}
\hline
\textbf{Experiment name}  &  \textbf{CT\ ID\#}  & \textbf{Dataset details}  &  & $N_{\mathit{pt}}$ \tabularnewline
\hline
\hline
	\textbf{LHCb7ZWrap'15}      &  \textbf{245}   & LHCb 7 TeV Z/W muon forward-$\eta$ Xsec ($1.0\mbox{ fb}^{-1}$)  & \cite{Aaij:2015gna}  & 33 \tabularnewline
\hline
	\textbf{LHCb8Zee'15}        &  \textbf{246}   & LHCb 8 TeV Z electron forward-$\eta$ $d\sigma/dy_{Z}$ ($2.0\mbox{ fb}^{-1}$)  & \cite{Aaij:2015vua}  & 17 \tabularnewline
\hline
	\textbf{ATL7ZpT'14}         &  \textbf{247}   & ATLAS 7 TeV $d\sigma/dp_{T}^{Z}$ ($4.7\mbox{ fb}^{-1}$)  & \cite{Aad:2014xaa}  & 8 \tabularnewline
\hline
	\textbf{CMS8Wasy'16}        &  \textbf{249}   & CMS 8 TeV W muon, Xsec, $A_{\mu}(\eta^{\mu})$ ($18.8\mbox{ fb}^{-1}$)  & \cite{Khachatryan:2016pev}  & 33 \tabularnewline
\hline
	\textbf{LHCb8WZ'16}         &  \textbf{250}   & LHCb 8 TeV W/Z muon, Xsec, $A_{\mu}(\eta^{\mu})$ ($2.0\mbox{ fb}^{-1}$)  & \cite{Aaij:2015zlq}  & 42 \tabularnewline
\hline
	\textbf{ATL8DY2D'16}       &  \textbf{252}   & ATLAS 8 TeV Z ($d^{2}\sigma/d|y|_{ll}dm_{ll}$) ($20.3\mbox{ fb}^{-1}$)  & \cite{Aad:2016zzw}  & 48 \tabularnewline
\hline
	\textbf{ATL8ZpT'16}         &  \textbf{253}   & ATLAS 8 TeV ($d^{2}\sigma/dp_{T}^{Z}dm_{ll}$) ($20.3\mbox{ fb}^{-1}$)  & \cite{Aad:2015auj}  & 45 \tabularnewline
\hline
	\textbf{CMS7jets'14}       &  \textbf{542}   & CMS 7 TeV incl. jet, R=0.7, ($d^2\sigma/dp_{T}^{j}dy_{j}$) ($5\mbox{ fb}^{-1}$)  & \cite{Chatrchyan:2014gia}  & 158 \tabularnewline
\hline
	\textbf{ATLAS7jets'15}     &  \textbf{544}   & ATLAS 7 TeV incl. jet, R=0.6, ($d^2\sigma/dp_{T}^{j}dy_{j}$) ($4.5\mbox{ fb}^{-1}$)  & \cite{Aad:2014vwa}  & 140 \tabularnewline
\hline
	\textbf{CMS8jets'17}       &  \textbf{545}   & CMS 8 TeV incl. jet, R=0.7, ($d^2\sigma/dp_{T}^{j}dy_{j}$) ($19.7\mbox{ fb}^{-1}$)  & \cite{Khachatryan:2016mlc}  & 185 \tabularnewline
\hline
	\textbf{ATL8ttb-pt'16}     &  \textbf{565}   & ATLAS 8 TeV $t\overline{t}\:d\sigma/dp_{T}^{t}$ ($20.3\mbox{ fb}^{-1}$)  & \cite{Aad:2015mbv}  & 8 \tabularnewline
\hline
	\textbf{ATL8ttb-y\_ave'16} &  \textbf{566}   & ATLAS 8 TeV $t\overline{t}\:d\sigma/dy_{<t/\overline{t}>}$ ($20.3\mbox{ fb}^{-1}$)  & \cite{Aad:2015mbv}  & 5 \tabularnewline
\hline
	\textbf{ATL8ttb-mtt'16}    &  \textbf{567}   & ATLAS 8 TeV $t\overline{t}\:d\sigma/dm_{t\overline{t}}$ ($20.3\mbox{ fb}^{-1}$)  & \cite{Aad:2015mbv}  & 7 \tabularnewline
\hline
	\textbf{ATL8ttb-y\_ttb'16} &  \textbf{568}   & ATLAS 8 TeV $t\overline{t}\:d\sigma/dy_{t\overline{t}}$ ($20.3\mbox{ fb}^{-1}$)  & \cite{Aad:2015mbv}  & 5 \tabularnewline
\hline
\end{tabular}\caption{Same as Table~\ref{tab:EXP_1}, showing experimental data sets for
production of vector bosons, single-inclusive jets, and $t\bar{t}$
pairs that were not incorporated in the CT14HERA2 NNLO fit but included
in our augmented CTEQ-TEA set.
\label{tab:EXP_3} }
\end{table}

\linespread{0.8}

\begin{table}
\begin{tabular}{c|c|ccc|cc|cc|cc}
No. & \text{ID} & $N_\mathit{pt}$ & $\sum |S^E|$ & $\langle \sum |S^E|\rangle$ & $|S_{\langle x^1\rangle_{u^+}}|$ & $\langle |S_{\langle x^1\rangle_{u^+}}|\rangle $ & $|S_{\langle x^2\rangle_{u^-}}|$ & $\langle |S_{\langle x^2\rangle_{u^-}}|\rangle $ & $|S_{\langle x^3\rangle_{u^+}}|$ & $\langle |S_{\langle x^3\rangle_{u^+}}|\rangle $ \\
    &      &     &      &       &          &   &          &   &            &   \\
\noalign{\vskip -3mm}
\hline
\noalign{\vskip -1.5mm}
    &      &     &      &       &          &   &          &   &            &   \\
1   &  BCDMSp'89 & 337 & 125. & 0.123 & \text{A} & 3 & \text{A} & 3 & \text{A} & 3 \\
2   &  HERAI+II'15 & 1120. & 122. & 0.0363 & \text{A} &   & \text{A} &   & \text{A} &   \\
3   &  CCFR-F3'97 & 86 & 95.8 & 0.371 & \text{A} & 2 & \text{A} & 1 & \text{A} & 2 \\
4   &  E866pp'03 & 184 & 69.3 & 0.125 & \text{B} &   & \text{A} & 3 & \text{A} & 3 \\
5   &  BCDMSd'90 & 250 & 55.3 & 0.0737 & \text{A} &   & \text{B} &   & \text{B} &   \\
6   &  CDHSW-F3'91 & 96 & 52.7 & 0.183 & \text{B} & 3 & \text{A} & 3 & \text{B} & 3 \\
7   &  CDHSW-F2'91 & 85 & 35.7 & 0.14 & \text{B} & 3 & \text{C} & 3 & \text{B} & 3 \\
8   &  NMCrat'97 & 123 & 34.5 & 0.0934 & \text{C} &   & \text{B} & 3 & \text{B} & 3 \\
9   &  CCFR-F2'01 & 69 & 31.5 & 0.152 & \text{C} & 3 & \text{C} & 3 & \text{B} & 3 \\
10  &  CMS7jets'13 & 133 & 27.5 & 0.0689 & \text{C} &   & \text{C} &   & \text{B} &   \\
11  &  CMS8jets'17 & 185 & 25.3 & 0.0456 & \text{C} &   & \text{C} &   & \text{B} &   \\
12  &  E605'91 & 119 & 24. & 0.0671 &   &   & \text{B} & 3 &   &   \\
13  &  CMS7jets'14 & 158 & 19.5 & 0.0411 & \text{C} &   & \text{C} &   & \text{C} &   \\
14  &  ATLAS7jets'15 & 140 & 16.2 & 0.0387 &   &   & \text{C} &   & \text{C} &   \\
15  &  NuTeV-nu'06 & 38 & 12.9 & 0.113 & \text{C} & 3 &   &   &   &   \\
16  &  LHCb8WZ'16 & 42 & 12.5 & 0.0989 &   &   &   &   & \text{C} & 3 \\
17  &  CCFR SI nub'01 & 38 & 11.9 & 0.104 & \text{C} & 3 &   &   &   &   \\
18  &  NuTeV-nub'06 & 33 & 11.8 & 0.119 & \text{C} & 3 &   &   &   & 3 \\
19  &  CMS7Masy2'14 & 11 & 11.5 & 0.349 &   & 2 & \text{C} & 2 & \text{C} & 2 \\
20  &  LHCb7ZWrap'15 & 33 & 11.4 & 0.115 &   &   &   & 3 & \text{C} & 3 \\
21  &  ATL7jets'12 & 90 & 11.1 & 0.0412 &   &   &   &   & \text{C} &   \\
22  &  ATL8DY2D'16 & 48 & 10.2 & 0.0706 & \text{C} &   &   &   &   &   \\
23  &  D02jets'08 & 110 & 10.2 & 0.0308 &   &   &   &   & \text{C} &   \\
24  &  E866rat'01 & 15 & 10. & 0.223 & \text{C} & 2 &   & 3 &   & 3 \\
25  &  CCFR SI nu'01 & 40 & 9.71 & 0.0809 & \text{C} & 3 &   &   &   &   \\
26  &  CDF2jets'09 & 72 & 8.51 & 0.0394 &   &   &   &   &   &   \\
27  &  CMS8Wasy'16 & 33 & 7.18 & 0.0726 &   &   &   &   &   &   \\
28  &  ATL7WZ'12 & 41 & 6.97 & 0.0567 &   &   &   &   &   &   \\
29  &  CMS7Easy'12 & 11 & 6.81 & 0.206 &   & 3 &   & 3 &   & 3 \\
30  &  D02Easy2'15 & 13 & 6.46 & 0.166 &   &   &   & 3 &   & 2 \\
31  &  ATL8ZpT'16 & 45 & 5.21 & 0.0386 &   &   &   &   &   &   \\
32  &  ZyCDF2'10 & 29 & 4.88 & 0.0561 &   &   &   &   &   &   \\
33  &  ZyD02'08 & 28 & 4.61 & 0.0549 &   &   &   &   &   &   \\
34  &  D02Masy'08 & 9 & 4.15 & 0.154 &   & 2 &   &   &   & 3 \\
35  &  LHCb8Zee'15 & 17 & 3.32 & 0.065 &   &   &   &   &   &   \\
36  &  HERAc'13 & 47 & 3.15 & 0.0224 &   &   &   &   &   &   \\
37  &  CDF1Wasy'96 & 11 & 2.75 & 0.0835 &   &   &   &   &   & 3 \\
38  &  LHCb7WZ'12 & 14 & 2.34 & 0.0557 &   &   &   &   &   &   \\
39  &  ATL8ttb-mtt'16 & 7 & 1.2 & 0.0573 &   &   &   &   &   &   \\
40  &  ATL8ttb-pt'16 & 8 & 1.17 & 0.0489 &   &   &   &   &   &   \\
41  &  LHCb7Wasy'12 & 5 & 1.17 & 0.0781 &   &   &   &   &   & 3 \\
42  &  ATL7ZpT'14 & 8 & 1.04 & 0.0435 &   & 3 &   &   &   &   \\
43  &  CDF2Wasy'05 & 11 & 1.02 & 0.0311 &   &   &   &   &   &   \\
44  &  ATL8ttb-y\_ave'16 & 5 & 0.548 & 0.0365 &   &   &   &   &   &   \\
45  &  HERA-FL'11 & 9 & 0.508 & 0.0188 &   &   &   &   &   &   \\
46  &  ATL8ttb-y\_ttb'16 & 5 & 0.495 & 0.033 &   &   &   &   &   &   \\
47  &  HERAb'06 & 10 & 0.328 & 0.0109 &   &   &   &   &   &   \\
\hline
\end{tabular}
\caption{
The aggregated and point-averaged sensitivities to moments of the $u^\pm$ quark
distributions of the experiments in the CTEQ-TEA set, ranking according to the
conventions of Ref.~\cite{Wang:2018heo}. Here and in the subsequent tables,
we arrange the CTEQ-TEA experiments in descending order based on their summed sensitivity $\sum |S^E|$
to each of the three moments displayed in the rightmost columns. For each
moment, we award especially sensitive experiments a rank $\mathrm{A,B,C}$
or $1,2,3$ based on their total and point-averaged
sensitivities, respectively. These ranks are decided using the criteria:
$C\protect\iff|S^E_{f}|\in[4,10]$, $B\protect\iff|S^E_{f}|\in[10,20]$, $A\protect\iff|S^E_{f}|\in[20,50]$
and $A^* \protect\iff|S^E_{f}|>50$ according to the total sensitivities
for each flavor; and, analogously, $3\protect\iff\langle|S^E_{f}|\rangle\in[0.1,0.25]$,
$2\protect\iff\langle|S^E_{f}|\rangle\in[0.25,0.5]$, $1 \protect\iff\langle|S^E_{f}|\rangle\in[0.5,1]$,
and $1^* \protect\iff\langle|S^E_{f}|>1$
according to the point-averaged sensitivities. Experiments with sensitivities
falling below the lowest ranks (that is, with $|S^E_{f}|<4$ or $\langle|S^E_{f}|\rangle<0.1$)
are not awarded a rank for that moment.
	}
\label{tab:rank_1}
\end{table}

\begin{table}
\begin{tabular}{c|c|ccc|cc|cc|cc}
No. & \text{ID} & $N_\mathit{pt}$ & $\sum |S^E|$ & $\langle \sum |S^E|\rangle$ & $|S_{\langle x^1\rangle_{d^+}}|$ & $\langle |S_{\langle x^1\rangle_{d^+}}|\rangle $ & $|S_{\langle x^2\rangle_{d^-}}|$ & $\langle |S_{\langle x^2\rangle_{d^-}}|\rangle $ & $|S_{\langle x^3\rangle_{d^+}}|$ & $\langle |S_{\langle x^3\rangle_{d^+}}|\rangle $ \\
    &      &     &      &       &          &   &          &   &            &   \\
\noalign{\vskip -3mm}
\hline
\noalign{\vskip -1.5mm}
    &      &     &      &       &          &   &          &   &            &   \\
1   &   HERAI+II'15 & 1120. & 116. & 0.0346 & \text{A*} &   & \text{A} &   & \text{A} &   \\
2   &   CCFR-F3'97 & 86 & 70.3 & 0.272 & \text{B} & 3 & \text{A} & 2 & \text{B} & 3 \\
3   &   BCDMSd'90 & 250 & 58.8 & 0.0784 & \text{A} & 3 & \text{B} &   & \text{B} &   \\
4   &   NMCrat'97 & 123 & 55.6 & 0.151 & \text{A} & 3 & \text{B} & 3 & \text{B} & 3 \\
5   &   BCDMSp'89 & 337 & 53.8 & 0.0532 & \text{B} &   & \text{B} &   & \text{A} &   \\
6   &   E866pp'03 & 184 & 44.2 & 0.08 & \text{B} &   & \text{B} &   & \text{B} &   \\
7   &   CDHSW-F2'91 & 85 & 43.5 & 0.171 & \text{B} & 3 & \text{B} & 3 & \text{B} & 3 \\
8   &   E605'91 & 119 & 37.3 & 0.105 & \text{C} &   & \text{A} & 3 & \text{C} &   \\
9   &   CDHSW-F3'91 & 96 & 36.2 & 0.126 & \text{C} &   & \text{B} & 3 & \text{C} &   \\
10  &   CCFR-F2'01 & 69 & 27.4 & 0.133 & \text{B} & 3 & \text{C} & 3 & \text{C} & 3 \\
11  &   LHCb8WZ'16 & 42 & 16.4 & 0.13 &   &   & \text{C} & 3 & \text{C} & 3 \\
12  &   CMS7jets'13 & 133 & 16.2 & 0.0405 & \text{C} &   & \text{C} &   & \text{C} &   \\
13  &   CMS8jets'17 & 185 & 15.2 & 0.0273 & \text{C} &   &   &   & \text{C} &   \\
14  &   LHCb7ZWrap'15 & 33 & 14. & 0.141 &   &   & \text{C} & 3 & \text{C} & 3 \\
15  &   CMS7jets'14 & 158 & 13.4 & 0.0283 & \text{C} &   &   &   & \text{C} &   \\
16  &   D02Easy2'15 & 13 & 12. & 0.309 &   & 3 & \text{C} & 1 &   & 2 \\
17  &   NuTeV-nu'06 & 38 & 11.7 & 0.103 & \text{C} & 3 &   &   &   &   \\
18  &   CMS7Masy2'14 & 11 & 10.2 & 0.308 & \text{C} & 2 &   & 2 &   & 3 \\
19  &   ATLAS7jets'15 & 140 & 9.11 & 0.0217 &   &   &   &   & \text{C} &   \\
20  &   ATL8DY2D'16 & 48 & 8.82 & 0.0612 &   &   &   &   &   &   \\
21  &   CCFR SI nu'01 & 40 & 8.61 & 0.0718 & \text{C} & 3 &   &   &   &   \\
22  &   E866rat'01 & 15 & 8.48 & 0.188 &   & 3 &   & 3 &   & 3 \\
23  &   ATL7WZ'12 & 41 & 8.38 & 0.0681 &   &   &   &   &   &   \\
24  &   CCFR SI nub'01 & 38 & 7.62 & 0.0668 & \text{C} & 3 &   &   &   &   \\
25  &   CMS8Wasy'16 & 33 & 6.99 & 0.0706 &   &   &   &   &   &   \\
26  &   NuTeV-nub'06 & 33 & 6.68 & 0.0674 & \text{C} & 3 &   &   &   &   \\
27  &   ATL7jets'12 & 90 & 6.61 & 0.0245 &   &   &   &   &   &   \\
28  &   D02jets'08 & 110 & 6.14 & 0.0186 &   &   &   &   &   &   \\
29  &   CMS7Easy'12 & 11 & 5.72 & 0.173 &   & 3 &   & 3 &   & 3 \\
30  &   CDF1Wasy'96 & 11 & 5.12 & 0.155 &   & 3 &   & 2 &   &   \\
31  &   ATL8ZpT'16 & 45 & 4.99 & 0.037 &   &   &   &   &   &   \\
32  &   D02Masy'08 & 9 & 4.35 & 0.161 &   & 3 &   & 2 &   &   \\
33  &   ZyD02'08 & 28 & 3.8 & 0.0452 &   &   &   &   &   &   \\
34  &   LHCb7WZ'12 & 14 & 3.77 & 0.0898 &   &   &   & 3 &   &   \\
35  &   CDF2jets'09 & 72 & 3.58 & 0.0166 &   &   &   &   &   &   \\
36  &   ZyCDF2'10 & 29 & 3.06 & 0.0352 &   &   &   &   &   &   \\
37  &   LHCb7Wasy'12 & 5 & 2.78 & 0.185 &   &   &   & 2 &   & 3 \\
38  &   ATL7ZpT'14 & 8 & 2.34 & 0.0976 &   & 3 &   &   &   &   \\
39  &   HERAc'13 & 47 & 1.97 & 0.014 &   &   &   &   &   &   \\
40  &   CDF2Wasy'05 & 11 & 1.86 & 0.0565 &   &   &   &   &   &   \\
41  &   LHCb8Zee'15 & 17 & 1.63 & 0.032 &   &   &   &   &   &   \\
42  &   ATL8ttb-pt'16 & 8 & 1.47 & 0.0614 &   & 3 &   &   &   &   \\
43  &   ATL8ttb-mtt'16 & 7 & 1.37 & 0.0654 &   &   &   &   &   &   \\
44  &   ATL8ttb-y\_ave'16 & 5 & 0.776 & 0.0517 &   &   &   &   &   &   \\
45  &   ATL8ttb-y\_ttb'16 & 5 & 0.461 & 0.0307 &   &   &   &   &   &   \\
46  &   HERA-FL'11 & 9 & 0.407 & 0.0151 &   &   &   &   &   &   \\
47  &   HERAb'06 & 10 & 0.252 & 0.0084 &   &   &   &   &   &   \\
\hline
\end{tabular}
\caption{
The sensitivities to moments of the $d^\pm$ quark
distributions of the CTEQ-TEA experiments in the
CTEQ-TEA set.
	}
\label{tab:rank_2}
\end{table}

\begin{table}
\hspace*{-1.25cm} \begin{tabular}{c|c|ccc|cc|cc|cc}
No. & \text{ID} & $N_\mathit{pt}$ & $\sum |S^E|$ & $\langle \sum |S^E|\rangle$ & $|S_{\langle x^1\rangle_{u^+-d^+}}|$ & $\langle |S_{\langle x^1\rangle_{u^+-d^+}}|\rangle $ & $|S_{\langle x^2\rangle_{u^--d^-}}|$ & $\langle |S_{\langle x^2\rangle_{u^--d^-}}|\rangle $ & $|S_{\langle x^3\rangle_{u^+-d^+}}|$ & $\langle |S_{\langle x^3\rangle_{u^+-d^+}}|\rangle $ \\
    &      &     &      &       &          &   &          &   &            &   \\
\noalign{\vskip -3mm}
\hline
\noalign{\vskip -1.5mm}
    &      &     &      &       &          &   &          &   &            &   \\
1   &  HERAI+II'15 & 1120. & 109. & 0.0325 & \text{A} &   & \text{A} &   & \text{A} &   \\
2   &  BCDMSp'89 & 337 & 91.9 & 0.0909 & \text{B} &   & \text{A} &   & \text{A} & 3 \\
3   &  E866pp'03 & 184 & 67.3 & 0.122 & \text{A} & 3 & \text{A} & 3 & \text{A} & 3 \\
4   &  CCFR-F3'97 & 86 & 61.9 & 0.24 & \text{A} & 2 & \text{B} & 3 & \text{B} & 3 \\
5   &  NMCrat'97 & 123 & 60. & 0.163 & \text{A} & 3 & \text{B} & 3 & \text{B} & 3 \\
6   &  CDHSW-F3'91 & 96 & 27.4 & 0.0953 & \text{B} & 3 & \text{C} &   & \text{C} &   \\
7   &  BCDMSd'90 & 250 & 25.5 & 0.034 & \text{C} &   & \text{C} &   & \text{C} &   \\
8   &  LHCb8WZ'16 & 42 & 15.3 & 0.122 & \text{C} &   & \text{C} & 3 & \text{C} & 3 \\
9   &  CMS7Masy2'14 & 11 & 15.3 & 0.463 & \text{C} & 1 & \text{C} & 2 &   & 2 \\
10  &  CCFR-F2'01 & 69 & 15.2 & 0.0733 & \text{C} &   & \text{C} &   & \text{C} &   \\
11  &  E605'91 & 119 & 15.1 & 0.0423 & \text{C} &   &   &   & \text{C} &   \\
12  &  CMS8jets'17 & 185 & 14.1 & 0.0254 &   &   & \text{C} &   & \text{C} &   \\
13  &  LHCb7ZWrap'15 & 33 & 12.9 & 0.13 & \text{C} & 3 & \text{C} & 3 &   & 3 \\
14  &  D02Easy2'15 & 13 & 11.6 & 0.298 &   & 3 & \text{C} & 2 &   & 2 \\
15  &  CDHSW-F2'91 & 85 & 11.4 & 0.0449 & \text{C} &   &   &   &   &   \\
16  &  E866rat'01 & 15 & 11.3 & 0.251 & \text{C} & 2 &   & 3 &   & 3 \\
17  &  CMS7jets'13 & 133 & 11.2 & 0.0282 &   &   & \text{C} &   &   &   \\
18  &  ATL8DY2D'16 & 48 & 11.1 & 0.0772 & \text{C} &   &   &   &   &   \\
19  &  CMS7Easy'12 & 11 & 8.98 & 0.272 &   & 2 &   & 2 &   & 3 \\
20  &  CMS8Wasy'16 & 33 & 8.77 & 0.0886 &   & 3 &   &   &   &   \\
21  &  ATL7WZ'12 & 41 & 8.64 & 0.0703 &   &   &   &   &   &   \\
22  &  CMS7jets'14 & 158 & 8.54 & 0.018 &   &   &   &   &   &   \\
23  &  ATLAS7jets'15 & 140 & 8.07 & 0.0192 &   &   &   &   &   &   \\
24  &  CDF2jets'09 & 72 & 5.22 & 0.0241 &   &   &   &   &   &   \\
25  &  D02jets'08 & 110 & 5. & 0.0151 &   &   &   &   &   &   \\
26  &  NuTeV-nu'06 & 38 & 4.99 & 0.0437 &   &   &   &   &   &   \\
27  &  ATL7jets'12 & 90 & 4.84 & 0.0179 &   &   &   &   &   &   \\
28  &  CDF1Wasy'96 & 11 & 4.83 & 0.146 &   & 3 &   & 3 &   & 3 \\
29  &  NuTeV-nub'06 & 33 & 4.28 & 0.0432 &   &   &   &   &   &   \\
30  &  D02Masy'08 & 9 & 4.2 & 0.155 &   & 3 &   & 3 &   & 3 \\
31  &  CCFR SI nub'01 & 38 & 3.71 & 0.0325 &   &   &   &   &   &   \\
32  &  CCFR SI nu'01 & 40 & 3.61 & 0.0301 &   &   &   &   &   &   \\
33  &  ZyCDF2'10 & 29 & 3.58 & 0.0411 &   &   &   &   &   &   \\
34  &  ATL8ZpT'16 & 45 & 3.29 & 0.0243 &   &   &   &   &   &   \\
35  &  ZyD02'08 & 28 & 3.11 & 0.0371 &   &   &   &   &   &   \\
36  &  LHCb7WZ'12 & 14 & 2.88 & 0.0685 &   &   &   &   &   &   \\
37  &  LHCb7Wasy'12 & 5 & 2.47 & 0.165 &   &   &   & 2 &   & 3 \\
38  &  HERAc'13 & 47 & 1.79 & 0.0127 &   &   &   &   &   &   \\
39  &  LHCb8Zee'15 & 17 & 1.74 & 0.034 &   &   &   &   &   &   \\
40  &  CDF2Wasy'05 & 11 & 1.73 & 0.0524 &   &   &   &   &   &   \\
41  &  ATL7ZpT'14 & 8 & 1.2 & 0.05 &   &   &   &   &   &   \\
42  &  ATL8ttb-pt'16 & 8 & 0.951 & 0.0396 &   &   &   &   &   &   \\
43  &  ATL8ttb-mtt'16 & 7 & 0.517 & 0.0246 &   &   &   &   &   &   \\
44  &  ATL8ttb-y\_ave'16 & 5 & 0.236 & 0.0157 &   &   &   &   &   &   \\
45  &  ATL8ttb-y\_ttb'16 & 5 & 0.164 & 0.0109 &   &   &   &   &   &   \\
46  &  HERAb'06 & 10 & 0.162 & 0.0054 &   &   &   &   &   &   \\
47  &  HERA-FL'11 & 9 & 0.15 & 0.00554 &   &   &   &   &   &   \\
\hline
\end{tabular}
\caption{
The sensitivities to moments of the $u^\pm\! -\! d^\pm$ isovector
quark distributions of the CTEQ-TEA experiments in the
CTEQ-TEA set.
	}
\label{tab:rank_3}
\end{table}

\linespread{1.6}
%
%
% - - - - - - - - - - - - - - - - - - - - - - - - - - - - - - - - - - - - - - - - - - - - - -
%

\clearpage

\bibliography{mellin_v2}

\begin{thebibliography}{117}%
\makeatletter
\providecommand \@ifxundefined [1]{%
 \@ifx{#1\undefined}
}%
\providecommand \@ifnum [1]{%
 \ifnum #1\expandafter \@firstoftwo
 \else \expandafter \@secondoftwo
 \fi
}%
\providecommand \@ifx [1]{%
 \ifx #1\expandafter \@firstoftwo
 \else \expandafter \@secondoftwo
 \fi
}%
\providecommand \natexlab [1]{#1}%
\providecommand \enquote  [1]{``#1''}%
\providecommand \bibnamefont  [1]{#1}%
\providecommand \bibfnamefont [1]{#1}%
\providecommand \citenamefont [1]{#1}%
\providecommand \href@noop [0]{\@secondoftwo}%
\providecommand \href [0]{\begingroup \@sanitize@url \@href}%
\providecommand \@href[1]{\@@startlink{#1}\@@href}%
\providecommand \@@href[1]{\endgroup#1\@@endlink}%
\providecommand \@sanitize@url [0]{\catcode `\\12\catcode `\$12\catcode
  `\&12\catcode `\#12\catcode `\^12\catcode `\_12\catcode `\%12\relax}%
\providecommand \@@startlink[1]{}%
\providecommand \@@endlink[0]{}%
\providecommand \url  [0]{\begingroup\@sanitize@url \@url }%
\providecommand \@url [1]{\endgroup\@href {#1}{\urlprefix }}%
\providecommand \urlprefix  [0]{URL }%
\providecommand \Eprint [0]{\href }%
\providecommand \doibase [0]{http://dx.doi.org/}%
\providecommand \selectlanguage [0]{\@gobble}%
\providecommand \bibinfo  [0]{\@secondoftwo}%
\providecommand \bibfield  [0]{\@secondoftwo}%
\providecommand \translation [1]{[#1]}%
\providecommand \BibitemOpen [0]{}%
\providecommand \bibitemStop [0]{}%
\providecommand \bibitemNoStop [0]{.\EOS\space}%
\providecommand \EOS [0]{\spacefactor3000\relax}%
\providecommand \BibitemShut  [1]{\csname bibitem#1\endcsname}%
\let\auto@bib@innerbib\@empty
%</preamble>
\bibitem [{\citenamefont {Wang}\ \emph {et~al.}(2018)\citenamefont {Wang},
  \citenamefont {Hobbs}, \citenamefont {Doyle}, \citenamefont {Gao},
  \citenamefont {Hou}, \citenamefont {Nadolsky},\ and\ \citenamefont
  {Olness}}]{Wang:2018heo}%
  \BibitemOpen
  \bibfield  {author} {\bibinfo {author} {\bibfnamefont {B.-T.}\ \bibnamefont
  {Wang}}, \bibinfo {author} {\bibfnamefont {T.~J.}\ \bibnamefont {Hobbs}},
  \bibinfo {author} {\bibfnamefont {S.}~\bibnamefont {Doyle}}, \bibinfo
  {author} {\bibfnamefont {J.}~\bibnamefont {Gao}}, \bibinfo {author}
  {\bibfnamefont {T.-J.}\ \bibnamefont {Hou}}, \bibinfo {author} {\bibfnamefont
  {P.~M.}\ \bibnamefont {Nadolsky}}, \ and\ \bibinfo {author} {\bibfnamefont
  {F.~I.}\ \bibnamefont {Olness}},\ }\href {\doibase
  10.1103/PhysRevD.98.094030} {\bibfield  {journal} {\bibinfo  {journal} {Phys.
  Rev.}\ }\textbf {\bibinfo {volume} {D98}},\ \bibinfo {pages} {094030}
  (\bibinfo {year} {2018})},\ \Eprint {http://arxiv.org/abs/1803.02777}
  {arXiv:1803.02777 [hep-ph]} \BibitemShut {NoStop}%
%%CITATION = ARXIV:1803.02777;%%
\bibitem [{\citenamefont {Accardi}\ \emph
  {et~al.}(2016{\natexlab{a}})\citenamefont {Accardi}, \citenamefont {Brady},
  \citenamefont {Melnitchouk}, \citenamefont {Owens},\ and\ \citenamefont
  {Sato}}]{Accardi:2016qay}%
  \BibitemOpen
  \bibfield  {author} {\bibinfo {author} {\bibfnamefont {A.}~\bibnamefont
  {Accardi}}, \bibinfo {author} {\bibfnamefont {L.~T.}\ \bibnamefont {Brady}},
  \bibinfo {author} {\bibfnamefont {W.}~\bibnamefont {Melnitchouk}}, \bibinfo
  {author} {\bibfnamefont {J.~F.}\ \bibnamefont {Owens}}, \ and\ \bibinfo
  {author} {\bibfnamefont {N.}~\bibnamefont {Sato}},\ }\href {\doibase
  10.1103/PhysRevD.93.114017} {\bibfield  {journal} {\bibinfo  {journal} {Phys.
  Rev.}\ }\textbf {\bibinfo {volume} {D93}},\ \bibinfo {pages} {114017}
  (\bibinfo {year} {2016}{\natexlab{a}})},\ \Eprint
  {http://arxiv.org/abs/1602.03154} {arXiv:1602.03154 [hep-ph]} \BibitemShut
  {NoStop}%
%%CITATION = ARXIV:1602.03154;%%
\bibitem [{\citenamefont {Harland-Lang}\ \emph {et~al.}(2015)\citenamefont
  {Harland-Lang}, \citenamefont {Martin}, \citenamefont {Motylinski},\ and\
  \citenamefont {Thorne}}]{Harland-Lang:2014zoa}%
  \BibitemOpen
  \bibfield  {author} {\bibinfo {author} {\bibfnamefont {L.~A.}\ \bibnamefont
  {Harland-Lang}}, \bibinfo {author} {\bibfnamefont {A.~D.}\ \bibnamefont
  {Martin}}, \bibinfo {author} {\bibfnamefont {P.}~\bibnamefont {Motylinski}},
  \ and\ \bibinfo {author} {\bibfnamefont {R.~S.}\ \bibnamefont {Thorne}},\
  }\href {\doibase 10.1140/epjc/s10052-015-3397-6} {\bibfield  {journal}
  {\bibinfo  {journal} {Eur. Phys. J.}\ }\textbf {\bibinfo {volume} {C75}},\
  \bibinfo {pages} {204} (\bibinfo {year} {2015})},\ \Eprint
  {http://arxiv.org/abs/1412.3989} {arXiv:1412.3989 [hep-ph]} \BibitemShut
  {NoStop}%
%%CITATION = ARXIV:1412.3989;%%
\bibitem [{\citenamefont {Ball}\ \emph {et~al.}(2017)\citenamefont {Ball} \emph
  {et~al.}}]{Ball:2017nwa}%
  \BibitemOpen
  \bibfield  {author} {\bibinfo {author} {\bibfnamefont {R.~D.}\ \bibnamefont
  {Ball}} \emph {et~al.} (\bibinfo {collaboration} {NNPDF}),\ }\href {\doibase
  10.1140/epjc/s10052-017-5199-5} {\bibfield  {journal} {\bibinfo  {journal}
  {Eur. Phys. J.}\ }\textbf {\bibinfo {volume} {C77}},\ \bibinfo {pages} {663}
  (\bibinfo {year} {2017})},\ \Eprint {http://arxiv.org/abs/1706.00428}
  {arXiv:1706.00428 [hep-ph]} \BibitemShut {NoStop}%
%%CITATION = ARXIV:1706.00428;%%
\bibitem [{\citenamefont {Dulat}\ \emph {et~al.}(2016)\citenamefont {Dulat},
  \citenamefont {Hou}, \citenamefont {Gao}, \citenamefont {Guzzi},
  \citenamefont {Huston}, \citenamefont {Nadolsky}, \citenamefont {Pumplin},
  \citenamefont {Schmidt}, \citenamefont {Stump},\ and\ \citenamefont
  {Yuan}}]{Dulat:2015mca}%
  \BibitemOpen
  \bibfield  {author} {\bibinfo {author} {\bibfnamefont {S.}~\bibnamefont
  {Dulat}}, \bibinfo {author} {\bibfnamefont {T.-J.}\ \bibnamefont {Hou}},
  \bibinfo {author} {\bibfnamefont {J.}~\bibnamefont {Gao}}, \bibinfo {author}
  {\bibfnamefont {M.}~\bibnamefont {Guzzi}}, \bibinfo {author} {\bibfnamefont
  {J.}~\bibnamefont {Huston}}, \bibinfo {author} {\bibfnamefont
  {P.}~\bibnamefont {Nadolsky}}, \bibinfo {author} {\bibfnamefont
  {J.}~\bibnamefont {Pumplin}}, \bibinfo {author} {\bibfnamefont
  {C.}~\bibnamefont {Schmidt}}, \bibinfo {author} {\bibfnamefont
  {D.}~\bibnamefont {Stump}}, \ and\ \bibinfo {author} {\bibfnamefont {C.-P.}\
  \bibnamefont {Yuan}},\ }\href {\doibase 10.1103/PhysRevD.93.033006}
  {\bibfield  {journal} {\bibinfo  {journal} {Phys. Rev.}\ }\textbf {\bibinfo
  {volume} {D93}},\ \bibinfo {pages} {033006} (\bibinfo {year} {2016})},\
  \Eprint {http://arxiv.org/abs/1506.07443} {arXiv:1506.07443 [hep-ph]}
  \BibitemShut {NoStop}%
%%CITATION = ARXIV:1506.07443;%%
\bibitem [{\citenamefont {Hou}\ \emph {et~al.}(2017)\citenamefont {Hou},
  \citenamefont {Dulat}, \citenamefont {Gao}, \citenamefont {Guzzi},
  \citenamefont {Huston}, \citenamefont {Nadolsky}, \citenamefont {Pumplin},
  \citenamefont {Schmidt}, \citenamefont {Stump},\ and\ \citenamefont
  {Yuan}}]{Hou:2016nqm}%
  \BibitemOpen
  \bibfield  {author} {\bibinfo {author} {\bibfnamefont {T.-J.}\ \bibnamefont
  {Hou}}, \bibinfo {author} {\bibfnamefont {S.}~\bibnamefont {Dulat}}, \bibinfo
  {author} {\bibfnamefont {J.}~\bibnamefont {Gao}}, \bibinfo {author}
  {\bibfnamefont {M.}~\bibnamefont {Guzzi}}, \bibinfo {author} {\bibfnamefont
  {J.}~\bibnamefont {Huston}}, \bibinfo {author} {\bibfnamefont
  {P.}~\bibnamefont {Nadolsky}}, \bibinfo {author} {\bibfnamefont
  {J.}~\bibnamefont {Pumplin}}, \bibinfo {author} {\bibfnamefont
  {C.}~\bibnamefont {Schmidt}}, \bibinfo {author} {\bibfnamefont
  {D.}~\bibnamefont {Stump}}, \ and\ \bibinfo {author} {\bibfnamefont {C.-P.}\
  \bibnamefont {Yuan}},\ }\href {\doibase 10.1103/PhysRevD.95.034003}
  {\bibfield  {journal} {\bibinfo  {journal} {Phys. Rev.}\ }\textbf {\bibinfo
  {volume} {D95}},\ \bibinfo {pages} {034003} (\bibinfo {year} {2017})},\
  \Eprint {http://arxiv.org/abs/1609.07968} {arXiv:1609.07968 [hep-ph]}
  \BibitemShut {NoStop}%
%%CITATION = ARXIV:1609.07968;%%
\bibitem [{\citenamefont {Speth}\ and\ \citenamefont
  {Thomas}(1997)}]{Speth:1996pz}%
  \BibitemOpen
  \bibfield  {author} {\bibinfo {author} {\bibfnamefont {J.}~\bibnamefont
  {Speth}}\ and\ \bibinfo {author} {\bibfnamefont {A.~W.}\ \bibnamefont
  {Thomas}},\ }\href {\doibase 10.1007/0-306-47073-X_2} {\bibfield  {journal}
  {\bibinfo  {journal} {Adv. Nucl. Phys.}\ }\textbf {\bibinfo {volume} {24}},\
  \bibinfo {pages} {83} (\bibinfo {year} {1997})},\ \bibinfo {note}
  {[,83(1996)]}\BibitemShut {NoStop}%
%%CITATION = ANUPB,24,83;%%
\bibitem [{\citenamefont {Kumano}(1998)}]{Kumano:1997cy}%
  \BibitemOpen
  \bibfield  {author} {\bibinfo {author} {\bibfnamefont {S.}~\bibnamefont
  {Kumano}},\ }\href {\doibase 10.1016/S0370-1573(98)00016-7} {\bibfield
  {journal} {\bibinfo  {journal} {Phys. Rept.}\ }\textbf {\bibinfo {volume}
  {303}},\ \bibinfo {pages} {183} (\bibinfo {year} {1998})},\ \Eprint
  {http://arxiv.org/abs/hep-ph/9702367} {arXiv:hep-ph/9702367 [hep-ph]}
  \BibitemShut {NoStop}%
%%CITATION = HEP-PH/9702367;%%
\bibitem [{\citenamefont {Bourrely}\ \emph {et~al.}(2002)\citenamefont
  {Bourrely}, \citenamefont {Soffer},\ and\ \citenamefont
  {Buccella}}]{Bourrely:2001du}%
  \BibitemOpen
  \bibfield  {author} {\bibinfo {author} {\bibfnamefont {C.}~\bibnamefont
  {Bourrely}}, \bibinfo {author} {\bibfnamefont {J.}~\bibnamefont {Soffer}}, \
  and\ \bibinfo {author} {\bibfnamefont {F.}~\bibnamefont {Buccella}},\ }\href
  {\doibase 10.1007/s100520100855} {\bibfield  {journal} {\bibinfo  {journal}
  {Eur. Phys. J.}\ }\textbf {\bibinfo {volume} {C23}},\ \bibinfo {pages} {487}
  (\bibinfo {year} {2002})},\ \Eprint {http://arxiv.org/abs/hep-ph/0109160}
  {arXiv:hep-ph/0109160 [hep-ph]} \BibitemShut {NoStop}%
%%CITATION = HEP-PH/0109160;%%
\bibitem [{\citenamefont {Holt}\ and\ \citenamefont
  {Roberts}(2010)}]{Holt:2010vj}%
  \BibitemOpen
  \bibfield  {author} {\bibinfo {author} {\bibfnamefont {R.~J.}\ \bibnamefont
  {Holt}}\ and\ \bibinfo {author} {\bibfnamefont {C.~D.}\ \bibnamefont
  {Roberts}},\ }\href {\doibase 10.1103/RevModPhys.82.2991} {\bibfield
  {journal} {\bibinfo  {journal} {Rev. Mod. Phys.}\ }\textbf {\bibinfo {volume}
  {82}},\ \bibinfo {pages} {2991} (\bibinfo {year} {2010})},\ \Eprint
  {http://arxiv.org/abs/1002.4666} {arXiv:1002.4666 [nucl-th]} \BibitemShut
  {NoStop}%
%%CITATION = ARXIV:1002.4666;%%
\bibitem [{\citenamefont {Avakian}\ \emph {et~al.}(2010)\citenamefont
  {Avakian}, \citenamefont {Efremov}, \citenamefont {Schweitzer},\ and\
  \citenamefont {Yuan}}]{Avakian:2010br}%
  \BibitemOpen
  \bibfield  {author} {\bibinfo {author} {\bibfnamefont {H.}~\bibnamefont
  {Avakian}}, \bibinfo {author} {\bibfnamefont {A.~V.}\ \bibnamefont
  {Efremov}}, \bibinfo {author} {\bibfnamefont {P.}~\bibnamefont {Schweitzer}},
  \ and\ \bibinfo {author} {\bibfnamefont {F.}~\bibnamefont {Yuan}},\ }\href
  {\doibase 10.1103/PhysRevD.81.074035} {\bibfield  {journal} {\bibinfo
  {journal} {Phys. Rev.}\ }\textbf {\bibinfo {volume} {D81}},\ \bibinfo {pages}
  {074035} (\bibinfo {year} {2010})},\ \Eprint {http://arxiv.org/abs/1001.5467}
  {arXiv:1001.5467 [hep-ph]} \BibitemShut {NoStop}%
%%CITATION = ARXIV:1001.5467;%%
\bibitem [{\citenamefont {Bednar}\ \emph {et~al.}(2018)\citenamefont {Bednar},
  \citenamefont {Cloët},\ and\ \citenamefont {Tandy}}]{Bednar:2018htv}%
  \BibitemOpen
  \bibfield  {author} {\bibinfo {author} {\bibfnamefont {K.~D.}\ \bibnamefont
  {Bednar}}, \bibinfo {author} {\bibfnamefont {I.~C.}\ \bibnamefont {Cloët}},
  \ and\ \bibinfo {author} {\bibfnamefont {P.~C.}\ \bibnamefont {Tandy}},\
  }\href {\doibase 10.1016/j.physletb.2018.06.020} {\bibfield  {journal}
  {\bibinfo  {journal} {Phys. Lett.}\ }\textbf {\bibinfo {volume} {B782}},\
  \bibinfo {pages} {675} (\bibinfo {year} {2018})},\ \Eprint
  {http://arxiv.org/abs/1803.03656} {arXiv:1803.03656 [nucl-th]} \BibitemShut
  {NoStop}%
%%CITATION = ARXIV:1803.03656;%%
\bibitem [{\citenamefont {Hobbs}\ \emph {et~al.}(2015)\citenamefont {Hobbs},
  \citenamefont {Alberg},\ and\ \citenamefont {Miller}}]{Hobbs:2014lea}%
  \BibitemOpen
  \bibfield  {author} {\bibinfo {author} {\bibfnamefont {T.~J.}\ \bibnamefont
  {Hobbs}}, \bibinfo {author} {\bibfnamefont {M.}~\bibnamefont {Alberg}}, \
  and\ \bibinfo {author} {\bibfnamefont {G.~A.}\ \bibnamefont {Miller}},\
  }\href {\doibase 10.1103/PhysRevC.91.035205} {\bibfield  {journal} {\bibinfo
  {journal} {Phys. Rev.}\ }\textbf {\bibinfo {volume} {C91}},\ \bibinfo {pages}
  {035205} (\bibinfo {year} {2015})},\ \Eprint {http://arxiv.org/abs/1412.4871}
  {arXiv:1412.4871 [nucl-th]} \BibitemShut {NoStop}%
%%CITATION = ARXIV:1412.4871;%%
\bibitem [{\citenamefont {Melnitchouk}\ and\ \citenamefont
  {Malheiro}(1999)}]{Melnitchouk:1999mv}%
  \BibitemOpen
  \bibfield  {author} {\bibinfo {author} {\bibfnamefont {W.}~\bibnamefont
  {Melnitchouk}}\ and\ \bibinfo {author} {\bibfnamefont {M.}~\bibnamefont
  {Malheiro}},\ }\href {\doibase 10.1016/S0370-2693(99)00182-3} {\bibfield
  {journal} {\bibinfo  {journal} {Phys. Lett.}\ }\textbf {\bibinfo {volume}
  {B451}},\ \bibinfo {pages} {224} (\bibinfo {year} {1999})},\ \Eprint
  {http://arxiv.org/abs/hep-ph/9901321} {arXiv:hep-ph/9901321 [hep-ph]}
  \BibitemShut {NoStop}%
%%CITATION = HEP-PH/9901321;%%
\bibitem [{\citenamefont {Hobbs}\ \emph {et~al.}(2014)\citenamefont {Hobbs},
  \citenamefont {Londergan},\ and\ \citenamefont
  {Melnitchouk}}]{Hobbs:2013bia}%
  \BibitemOpen
  \bibfield  {author} {\bibinfo {author} {\bibfnamefont {T.~J.}\ \bibnamefont
  {Hobbs}}, \bibinfo {author} {\bibfnamefont {J.~T.}\ \bibnamefont
  {Londergan}}, \ and\ \bibinfo {author} {\bibfnamefont {W.}~\bibnamefont
  {Melnitchouk}},\ }\href {\doibase 10.1103/PhysRevD.89.074008} {\bibfield
  {journal} {\bibinfo  {journal} {Phys. Rev.}\ }\textbf {\bibinfo {volume}
  {D89}},\ \bibinfo {pages} {074008} (\bibinfo {year} {2014})},\ \Eprint
  {http://arxiv.org/abs/1311.1578} {arXiv:1311.1578 [hep-ph]} \BibitemShut
  {NoStop}%
%%CITATION = ARXIV:1311.1578;%%
\bibitem [{\citenamefont {Hobbs}\ \emph {et~al.}(2017)\citenamefont {Hobbs},
  \citenamefont {Alberg},\ and\ \citenamefont {Miller}}]{Hobbs:2017fom}%
  \BibitemOpen
  \bibfield  {author} {\bibinfo {author} {\bibfnamefont {T.~J.}\ \bibnamefont
  {Hobbs}}, \bibinfo {author} {\bibfnamefont {M.}~\bibnamefont {Alberg}}, \
  and\ \bibinfo {author} {\bibfnamefont {G.~A.}\ \bibnamefont {Miller}},\
  }\href {\doibase 10.1103/PhysRevD.96.074023} {\bibfield  {journal} {\bibinfo
  {journal} {Phys. Rev.}\ }\textbf {\bibinfo {volume} {D96}},\ \bibinfo {pages}
  {074023} (\bibinfo {year} {2017})},\ \Eprint
  {http://arxiv.org/abs/1707.06711} {arXiv:1707.06711 [hep-ph]} \BibitemShut
  {NoStop}%
%%CITATION = ARXIV:1707.06711;%%
\bibitem [{\citenamefont {Gao}\ \emph {et~al.}(2018)\citenamefont {Gao},
  \citenamefont {Harland-Lang},\ and\ \citenamefont {Rojo}}]{Gao:2017yyd}%
  \BibitemOpen
  \bibfield  {author} {\bibinfo {author} {\bibfnamefont {J.}~\bibnamefont
  {Gao}}, \bibinfo {author} {\bibfnamefont {L.}~\bibnamefont {Harland-Lang}}, \
  and\ \bibinfo {author} {\bibfnamefont {J.}~\bibnamefont {Rojo}},\ }\href
  {\doibase 10.1016/j.physrep.2018.03.002} {\bibfield  {journal} {\bibinfo
  {journal} {Phys. Rept.}\ }\textbf {\bibinfo {volume} {742}},\ \bibinfo
  {pages} {1} (\bibinfo {year} {2018})},\ \Eprint
  {http://arxiv.org/abs/1709.04922} {arXiv:1709.04922 [hep-ph]} \BibitemShut
  {NoStop}%
%%CITATION = ARXIV:1709.04922;%%
\bibitem [{\citenamefont {Kovařík}\ \emph {et~al.}(2019)\citenamefont
  {Kovařík}, \citenamefont {Nadolsky},\ and\ \citenamefont
  {Soper}}]{Kovarik:2019xvh}%
  \BibitemOpen
  \bibfield  {author} {\bibinfo {author} {\bibfnamefont {K.}~\bibnamefont
  {Kovařík}}, \bibinfo {author} {\bibfnamefont {P.~M.}\ \bibnamefont
  {Nadolsky}}, \ and\ \bibinfo {author} {\bibfnamefont {D.~E.}\ \bibnamefont
  {Soper}},\ }\href@noop {} {\  (\bibinfo {year} {2019})},\ \Eprint
  {http://arxiv.org/abs/1905.06957} {arXiv:1905.06957 [hep-ph]} \BibitemShut
  {NoStop}%
%%CITATION = ARXIV:1905.06957;%%
\bibitem [{\citenamefont {Lin}\ \emph {et~al.}(2018)\citenamefont {Lin} \emph
  {et~al.}}]{Lin:2017snn}%
  \BibitemOpen
  \bibfield  {author} {\bibinfo {author} {\bibfnamefont {H.-W.}\ \bibnamefont
  {Lin}} \emph {et~al.},\ }\href {\doibase 10.1016/j.ppnp.2018.01.007}
  {\bibfield  {journal} {\bibinfo  {journal} {Prog. Part. Nucl. Phys.}\
  }\textbf {\bibinfo {volume} {100}},\ \bibinfo {pages} {107} (\bibinfo {year}
  {2018})},\ \Eprint {http://arxiv.org/abs/1711.07916} {arXiv:1711.07916
  [hep-ph]} \BibitemShut {NoStop}%
%%CITATION = ARXIV:1711.07916;%%
\bibitem [{\citenamefont {Detmold}\ \emph
  {et~al.}(2003{\natexlab{a}})\citenamefont {Detmold}, \citenamefont
  {Melnitchouk},\ and\ \citenamefont {Thomas}}]{Detmold:2003rq}%
  \BibitemOpen
  \bibfield  {author} {\bibinfo {author} {\bibfnamefont {W.}~\bibnamefont
  {Detmold}}, \bibinfo {author} {\bibfnamefont {W.}~\bibnamefont
  {Melnitchouk}}, \ and\ \bibinfo {author} {\bibfnamefont {A.~W.}\ \bibnamefont
  {Thomas}},\ }\href {\doibase 10.1142/S0217732303012209} {\bibfield  {journal}
  {\bibinfo  {journal} {Mod. Phys. Lett.}\ }\textbf {\bibinfo {volume} {A18}},\
  \bibinfo {pages} {2681} (\bibinfo {year} {2003}{\natexlab{a}})},\ \Eprint
  {http://arxiv.org/abs/hep-lat/0310003} {arXiv:hep-lat/0310003 [hep-lat]}
  \BibitemShut {NoStop}%
%%CITATION = HEP-LAT/0310003;%%
\bibitem [{\citenamefont {Ji}(2013)}]{Ji:2013dva}%
  \BibitemOpen
  \bibfield  {author} {\bibinfo {author} {\bibfnamefont {X.}~\bibnamefont
  {Ji}},\ }\href {\doibase 10.1103/PhysRevLett.110.262002} {\bibfield
  {journal} {\bibinfo  {journal} {Phys. Rev. Lett.}\ }\textbf {\bibinfo
  {volume} {110}},\ \bibinfo {pages} {262002} (\bibinfo {year} {2013})},\
  \Eprint {http://arxiv.org/abs/1305.1539} {arXiv:1305.1539 [hep-ph]}
  \BibitemShut {NoStop}%
%%CITATION = ARXIV:1305.1539;%%
\bibitem [{\citenamefont {Radyushkin}(2017)}]{Radyushkin:2017cyf}%
  \BibitemOpen
  \bibfield  {author} {\bibinfo {author} {\bibfnamefont {A.~V.}\ \bibnamefont
  {Radyushkin}},\ }\href {\doibase 10.1103/PhysRevD.96.034025} {\bibfield
  {journal} {\bibinfo  {journal} {Phys. Rev.}\ }\textbf {\bibinfo {volume}
  {D96}},\ \bibinfo {pages} {034025} (\bibinfo {year} {2017})},\ \Eprint
  {http://arxiv.org/abs/1705.01488} {arXiv:1705.01488 [hep-ph]} \BibitemShut
  {NoStop}%
%%CITATION = ARXIV:1705.01488;%%
\bibitem [{\citenamefont {Zimmermann}(1973)}]{Zimmermann:1972tv}%
  \BibitemOpen
  \bibfield  {author} {\bibinfo {author} {\bibfnamefont {W.}~\bibnamefont
  {Zimmermann}},\ }\bibfield  {booktitle} {\emph {\bibinfo {booktitle} {{In
  *Tegernsee 1998, Quantum field theory* 278-309}}},\ }\href {\doibase
  10.1016/0003-4916(73)90430-2} {\bibfield  {journal} {\bibinfo  {journal}
  {Annals Phys.}\ }\textbf {\bibinfo {volume} {77}},\ \bibinfo {pages} {570}
  (\bibinfo {year} {1973})},\ \bibinfo {note} {[Lect. Notes
  Phys.558,278(2000)]}\BibitemShut {NoStop}%
%%CITATION = APNYA,77,570;%%
\bibitem [{\citenamefont {Wilson}(1969)}]{Wilson:1969zs}%
  \BibitemOpen
  \bibfield  {author} {\bibinfo {author} {\bibfnamefont {K.~G.}\ \bibnamefont
  {Wilson}},\ }\href {\doibase 10.1103/PhysRev.179.1499} {\bibfield  {journal}
  {\bibinfo  {journal} {Phys. Rev.}\ }\textbf {\bibinfo {volume} {179}},\
  \bibinfo {pages} {1499} (\bibinfo {year} {1969})}\BibitemShut {NoStop}%
%%CITATION = PHRVA,179,1499;%%
\bibitem [{\citenamefont {Christ}\ \emph {et~al.}(1972)\citenamefont {Christ},
  \citenamefont {Hasslacher},\ and\ \citenamefont {Mueller}}]{Christ:1972ms}%
  \BibitemOpen
  \bibfield  {author} {\bibinfo {author} {\bibfnamefont {N.~H.}\ \bibnamefont
  {Christ}}, \bibinfo {author} {\bibfnamefont {B.}~\bibnamefont {Hasslacher}},
  \ and\ \bibinfo {author} {\bibfnamefont {A.~H.}\ \bibnamefont {Mueller}},\
  }\href {\doibase 10.1103/PhysRevD.6.3543} {\bibfield  {journal} {\bibinfo
  {journal} {Phys. Rev.}\ }\textbf {\bibinfo {volume} {D6}},\ \bibinfo {pages}
  {3543} (\bibinfo {year} {1972})}\BibitemShut {NoStop}%
%%CITATION = PHRVA,D6,3543;%%
\bibitem [{\citenamefont {Gross}\ and\ \citenamefont
  {Wilczek}(1973)}]{Gross:1973ju}%
  \BibitemOpen
  \bibfield  {author} {\bibinfo {author} {\bibfnamefont {D.~J.}\ \bibnamefont
  {Gross}}\ and\ \bibinfo {author} {\bibfnamefont {F.}~\bibnamefont
  {Wilczek}},\ }\href {\doibase 10.1103/PhysRevD.8.3633} {\bibfield  {journal}
  {\bibinfo  {journal} {Phys. Rev.}\ }\textbf {\bibinfo {volume} {D8}},\
  \bibinfo {pages} {3633} (\bibinfo {year} {1973})}\BibitemShut {NoStop}%
%%CITATION = PHRVA,D8,3633;%%
\bibitem [{\citenamefont {Gross}\ and\ \citenamefont
  {Wilczek}(1974)}]{Gross:1974cs}%
  \BibitemOpen
  \bibfield  {author} {\bibinfo {author} {\bibfnamefont {D.~J.}\ \bibnamefont
  {Gross}}\ and\ \bibinfo {author} {\bibfnamefont {F.}~\bibnamefont
  {Wilczek}},\ }\href {\doibase 10.1103/PhysRevD.9.980} {\bibfield  {journal}
  {\bibinfo  {journal} {Phys. Rev.}\ }\textbf {\bibinfo {volume} {D9}},\
  \bibinfo {pages} {980} (\bibinfo {year} {1974})}\BibitemShut {NoStop}%
%%CITATION = PHRVA,D9,980;%%
\bibitem [{\citenamefont {Georgi}\ and\ \citenamefont
  {Politzer}(1974)}]{Georgi:1951sr}%
  \BibitemOpen
  \bibfield  {author} {\bibinfo {author} {\bibfnamefont {H.}~\bibnamefont
  {Georgi}}\ and\ \bibinfo {author} {\bibfnamefont {H.~D.}\ \bibnamefont
  {Politzer}},\ }\href {\doibase 10.1103/PhysRevD.9.416} {\bibfield  {journal}
  {\bibinfo  {journal} {Phys. Rev.}\ }\textbf {\bibinfo {volume} {D9}},\
  \bibinfo {pages} {416} (\bibinfo {year} {1974})}\BibitemShut {NoStop}%
%%CITATION = PHRVA,D9,416;%%
\bibitem [{\citenamefont {Georgi}\ and\ \citenamefont
  {Politzer}(1976)}]{Georgi:1976ve}%
  \BibitemOpen
  \bibfield  {author} {\bibinfo {author} {\bibfnamefont {H.}~\bibnamefont
  {Georgi}}\ and\ \bibinfo {author} {\bibfnamefont {H.~D.}\ \bibnamefont
  {Politzer}},\ }\href {\doibase 10.1103/PhysRevD.14.1829} {\bibfield
  {journal} {\bibinfo  {journal} {Phys. Rev.}\ }\textbf {\bibinfo {volume}
  {D14}},\ \bibinfo {pages} {1829} (\bibinfo {year} {1976})}\BibitemShut
  {NoStop}%
%%CITATION = PHRVA,D14,1829;%%
\bibitem [{\citenamefont {\protect{Bl\"umlein, Johannes and Bottcher,
  Helmut}}(2008)}]{Blumlein:2008kz}%
  \BibitemOpen
  \bibfield  {author} {\bibinfo {author} {\bibnamefont {\protect{Bl\"umlein,
  Johannes and Bottcher, Helmut}}},\ }\href {\doibase
  10.1016/j.physletb.2008.03.026} {\bibfield  {journal} {\bibinfo  {journal}
  {Phys. Lett.}\ }\textbf {\bibinfo {volume} {B662}},\ \bibinfo {pages} {336}
  (\bibinfo {year} {2008})},\ \Eprint {http://arxiv.org/abs/0802.0408}
  {arXiv:0802.0408 [hep-ph]} \BibitemShut {NoStop}%
%%CITATION = ARXIV:0802.0408;%%
\bibitem [{\citenamefont {Gockeler}\ \emph
  {et~al.}(1996{\natexlab{a}})\citenamefont {Gockeler}, \citenamefont
  {Horsley}, \citenamefont {Ilgenfritz}, \citenamefont {Perlt}, \citenamefont
  {Rakow}, \citenamefont {Schierholz},\ and\ \citenamefont
  {Schiller}}]{Gockeler:1995wg}%
  \BibitemOpen
  \bibfield  {author} {\bibinfo {author} {\bibfnamefont {M.}~\bibnamefont
  {Gockeler}}, \bibinfo {author} {\bibfnamefont {R.}~\bibnamefont {Horsley}},
  \bibinfo {author} {\bibfnamefont {E.-M.}\ \bibnamefont {Ilgenfritz}},
  \bibinfo {author} {\bibfnamefont {H.}~\bibnamefont {Perlt}}, \bibinfo
  {author} {\bibfnamefont {P.~E.~L.}\ \bibnamefont {Rakow}}, \bibinfo {author}
  {\bibfnamefont {G.}~\bibnamefont {Schierholz}}, \ and\ \bibinfo {author}
  {\bibfnamefont {A.}~\bibnamefont {Schiller}},\ }\href {\doibase
  10.1103/PhysRevD.53.2317} {\bibfield  {journal} {\bibinfo  {journal} {Phys.
  Rev.}\ }\textbf {\bibinfo {volume} {D53}},\ \bibinfo {pages} {2317} (\bibinfo
  {year} {1996}{\natexlab{a}})},\ \Eprint
  {http://arxiv.org/abs/hep-lat/9508004} {arXiv:hep-lat/9508004 [hep-lat]}
  \BibitemShut {NoStop}%
%%CITATION = HEP-LAT/9508004;%%
\bibitem [{\citenamefont {Detmold}\ \emph {et~al.}(2001)\citenamefont
  {Detmold}, \citenamefont {Melnitchouk},\ and\ \citenamefont
  {Thomas}}]{Detmold:2001dv}%
  \BibitemOpen
  \bibfield  {author} {\bibinfo {author} {\bibfnamefont {W.}~\bibnamefont
  {Detmold}}, \bibinfo {author} {\bibfnamefont {W.}~\bibnamefont
  {Melnitchouk}}, \ and\ \bibinfo {author} {\bibfnamefont {A.~W.}\ \bibnamefont
  {Thomas}},\ }\href {\doibase 10.1007/s1010501c0013} {\bibfield  {journal}
  {\bibinfo  {journal} {Eur. Phys. J.direct}\ }\textbf {\bibinfo {volume}
  {3}},\ \bibinfo {pages} {13} (\bibinfo {year} {2001})},\ \Eprint
  {http://arxiv.org/abs/hep-lat/0108002} {arXiv:hep-lat/0108002 [hep-lat]}
  \BibitemShut {NoStop}%
%%CITATION = HEP-LAT/0108002;%%
\bibitem [{\citenamefont {Detmold}\ \emph
  {et~al.}(2003{\natexlab{b}})\citenamefont {Detmold}, \citenamefont
  {Melnitchouk},\ and\ \citenamefont {Thomas}}]{Detmold:2003tm}%
  \BibitemOpen
  \bibfield  {author} {\bibinfo {author} {\bibfnamefont {W.}~\bibnamefont
  {Detmold}}, \bibinfo {author} {\bibfnamefont {W.}~\bibnamefont
  {Melnitchouk}}, \ and\ \bibinfo {author} {\bibfnamefont {A.~W.}\ \bibnamefont
  {Thomas}},\ }\href {\doibase 10.1103/PhysRevD.68.034025} {\bibfield
  {journal} {\bibinfo  {journal} {Phys. Rev.}\ }\textbf {\bibinfo {volume}
  {D68}},\ \bibinfo {pages} {034025} (\bibinfo {year} {2003}{\natexlab{b}})},\
  \Eprint {http://arxiv.org/abs/hep-lat/0303015} {arXiv:hep-lat/0303015
  [hep-lat]} \BibitemShut {NoStop}%
%%CITATION = HEP-LAT/0303015;%%
\bibitem [{\citenamefont {Aoki}\ \emph {et~al.}(2017)\citenamefont {Aoki} \emph
  {et~al.}}]{1607.00299}%
  \BibitemOpen
  \bibfield  {author} {\bibinfo {author} {\bibfnamefont {S.}~\bibnamefont
  {Aoki}} \emph {et~al.},\ }\href {\doibase 10.1140/epjc/s10052-016-4509-7}
  {\bibfield  {journal} {\bibinfo  {journal} {Eur. Phys. J.}\ }\textbf
  {\bibinfo {volume} {C77}},\ \bibinfo {pages} {112} (\bibinfo {year}
  {2017})},\ \Eprint {http://arxiv.org/abs/1607.00299} {arXiv:1607.00299
  [hep-lat]} \BibitemShut {NoStop}%
%%CITATION = ARXIV:1607.00299;%%
\bibitem [{\citenamefont {Gockeler}\ \emph
  {et~al.}(1996{\natexlab{b}})\citenamefont {Gockeler}, \citenamefont
  {Horsley}, \citenamefont {Ilgenfritz}, \citenamefont {Perlt}, \citenamefont
  {Rakow}, \citenamefont {Schierholz},\ and\ \citenamefont
  {Schiller}}]{Gockeler:1996mu}%
  \BibitemOpen
  \bibfield  {author} {\bibinfo {author} {\bibfnamefont {M.}~\bibnamefont
  {Gockeler}}, \bibinfo {author} {\bibfnamefont {R.}~\bibnamefont {Horsley}},
  \bibinfo {author} {\bibfnamefont {E.-M.}\ \bibnamefont {Ilgenfritz}},
  \bibinfo {author} {\bibfnamefont {H.}~\bibnamefont {Perlt}}, \bibinfo
  {author} {\bibfnamefont {P.~E.~L.}\ \bibnamefont {Rakow}}, \bibinfo {author}
  {\bibfnamefont {G.}~\bibnamefont {Schierholz}}, \ and\ \bibinfo {author}
  {\bibfnamefont {A.}~\bibnamefont {Schiller}},\ }\href {\doibase
  10.1103/PhysRevD.54.5705} {\bibfield  {journal} {\bibinfo  {journal} {Phys.
  Rev.}\ }\textbf {\bibinfo {volume} {D54}},\ \bibinfo {pages} {5705} (\bibinfo
  {year} {1996}{\natexlab{b}})},\ \Eprint
  {http://arxiv.org/abs/hep-lat/9602029} {arXiv:hep-lat/9602029 [hep-lat]}
  \BibitemShut {NoStop}%
%%CITATION = HEP-LAT/9602029;%%
\bibitem [{\citenamefont {Alexandrou}\ \emph {et~al.}(2017)\citenamefont
  {Alexandrou}, \citenamefont {Constantinou}, \citenamefont {Hadjiyiannakou},
  \citenamefont {Jansen}, \citenamefont {Kallidonis}, \citenamefont {Koutsou},
  \citenamefont {Vaquero Avilés-Casco},\ and\ \citenamefont
  {Wiese}}]{Alexandrou:2017oeh}%
  \BibitemOpen
  \bibfield  {author} {\bibinfo {author} {\bibfnamefont {C.}~\bibnamefont
  {Alexandrou}}, \bibinfo {author} {\bibfnamefont {M.}~\bibnamefont
  {Constantinou}}, \bibinfo {author} {\bibfnamefont {K.}~\bibnamefont
  {Hadjiyiannakou}}, \bibinfo {author} {\bibfnamefont {K.}~\bibnamefont
  {Jansen}}, \bibinfo {author} {\bibfnamefont {C.}~\bibnamefont {Kallidonis}},
  \bibinfo {author} {\bibfnamefont {G.}~\bibnamefont {Koutsou}}, \bibinfo
  {author} {\bibfnamefont {A.}~\bibnamefont {Vaquero Avilés-Casco}}, \ and\
  \bibinfo {author} {\bibfnamefont {C.}~\bibnamefont {Wiese}},\ }\href
  {\doibase 10.1103/PhysRevLett.119.142002} {\bibfield  {journal} {\bibinfo
  {journal} {Phys. Rev. Lett.}\ }\textbf {\bibinfo {volume} {119}},\ \bibinfo
  {pages} {142002} (\bibinfo {year} {2017})},\ \Eprint
  {http://arxiv.org/abs/1706.02973} {arXiv:1706.02973 [hep-lat]} \BibitemShut
  {NoStop}%
%%CITATION = ARXIV:1706.02973;%%
\bibitem [{\citenamefont {Deka}\ \emph {et~al.}(2009)\citenamefont {Deka},
  \citenamefont {Streuer}, \citenamefont {Doi}, \citenamefont {Dong},
  \citenamefont {Draper}, \citenamefont {Liu}, \citenamefont {Mathur},\ and\
  \citenamefont {Thomas}}]{Deka:2008xr}%
  \BibitemOpen
  \bibfield  {author} {\bibinfo {author} {\bibfnamefont {M.}~\bibnamefont
  {Deka}}, \bibinfo {author} {\bibfnamefont {T.}~\bibnamefont {Streuer}},
  \bibinfo {author} {\bibfnamefont {T.}~\bibnamefont {Doi}}, \bibinfo {author}
  {\bibfnamefont {S.~J.}\ \bibnamefont {Dong}}, \bibinfo {author}
  {\bibfnamefont {T.}~\bibnamefont {Draper}}, \bibinfo {author} {\bibfnamefont
  {K.~F.}\ \bibnamefont {Liu}}, \bibinfo {author} {\bibfnamefont
  {N.}~\bibnamefont {Mathur}}, \ and\ \bibinfo {author} {\bibfnamefont {A.~W.}\
  \bibnamefont {Thomas}},\ }\href {\doibase 10.1103/PhysRevD.79.094502}
  {\bibfield  {journal} {\bibinfo  {journal} {Phys. Rev.}\ }\textbf {\bibinfo
  {volume} {D79}},\ \bibinfo {pages} {094502} (\bibinfo {year} {2009})},\
  \Eprint {http://arxiv.org/abs/0811.1779} {arXiv:0811.1779 [hep-ph]}
  \BibitemShut {NoStop}%
%%CITATION = ARXIV:0811.1779;%%
\bibitem [{\citenamefont {Liang}\ \emph {et~al.}(2019)\citenamefont {Liang},
  \citenamefont {Sun}, \citenamefont {Yang}, \citenamefont {Draper},\ and\
  \citenamefont {Liu}}]{Liang:2019xdx}%
  \BibitemOpen
  \bibfield  {author} {\bibinfo {author} {\bibfnamefont {J.}~\bibnamefont
  {Liang}}, \bibinfo {author} {\bibfnamefont {M.}~\bibnamefont {Sun}}, \bibinfo
  {author} {\bibfnamefont {Y.-B.}\ \bibnamefont {Yang}}, \bibinfo {author}
  {\bibfnamefont {T.}~\bibnamefont {Draper}}, \ and\ \bibinfo {author}
  {\bibfnamefont {K.-F.}\ \bibnamefont {Liu}},\ }\href@noop {} {\  (\bibinfo
  {year} {2019})},\ \Eprint {http://arxiv.org/abs/1901.07526} {arXiv:1901.07526
  [hep-ph]} \BibitemShut {NoStop}%
%%CITATION = ARXIV:1901.07526;%%
\bibitem [{\citenamefont {Bali}\ \emph {et~al.}(2014)\citenamefont {Bali},
  \citenamefont {Collins}, \citenamefont {Gläßle}, \citenamefont {Göckeler},
  \citenamefont {Najjar}, \citenamefont {Rödl}, \citenamefont {Schäfer},
  \citenamefont {Schiel}, \citenamefont {Sternbeck},\ and\ \citenamefont
  {Söldner}}]{Bali:2014gha}%
  \BibitemOpen
  \bibfield  {author} {\bibinfo {author} {\bibfnamefont {G.~S.}\ \bibnamefont
  {Bali}}, \bibinfo {author} {\bibfnamefont {S.}~\bibnamefont {Collins}},
  \bibinfo {author} {\bibfnamefont {B.}~\bibnamefont {Gläßle}}, \bibinfo
  {author} {\bibfnamefont {M.}~\bibnamefont {Göckeler}}, \bibinfo {author}
  {\bibfnamefont {J.}~\bibnamefont {Najjar}}, \bibinfo {author} {\bibfnamefont
  {R.~H.}\ \bibnamefont {Rödl}}, \bibinfo {author} {\bibfnamefont
  {A.}~\bibnamefont {Schäfer}}, \bibinfo {author} {\bibfnamefont {R.~W.}\
  \bibnamefont {Schiel}}, \bibinfo {author} {\bibfnamefont {A.}~\bibnamefont
  {Sternbeck}}, \ and\ \bibinfo {author} {\bibfnamefont {W.}~\bibnamefont
  {Söldner}},\ }\href {\doibase 10.1103/PhysRevD.90.074510} {\bibfield
  {journal} {\bibinfo  {journal} {Phys. Rev.}\ }\textbf {\bibinfo {volume}
  {D90}},\ \bibinfo {pages} {074510} (\bibinfo {year} {2014})},\ \Eprint
  {http://arxiv.org/abs/1408.6850} {arXiv:1408.6850 [hep-lat]} \BibitemShut
  {NoStop}%
%%CITATION = ARXIV:1408.6850;%%
\bibitem [{\citenamefont {Green}\ \emph {et~al.}(2014)\citenamefont {Green},
  \citenamefont {Engelhardt}, \citenamefont {Krieg}, \citenamefont {Negele},
  \citenamefont {Pochinsky},\ and\ \citenamefont {Syritsyn}}]{Green:2012ud}%
  \BibitemOpen
  \bibfield  {author} {\bibinfo {author} {\bibfnamefont {J.~R.}\ \bibnamefont
  {Green}}, \bibinfo {author} {\bibfnamefont {M.}~\bibnamefont {Engelhardt}},
  \bibinfo {author} {\bibfnamefont {S.}~\bibnamefont {Krieg}}, \bibinfo
  {author} {\bibfnamefont {J.~W.}\ \bibnamefont {Negele}}, \bibinfo {author}
  {\bibfnamefont {A.~V.}\ \bibnamefont {Pochinsky}}, \ and\ \bibinfo {author}
  {\bibfnamefont {S.~N.}\ \bibnamefont {Syritsyn}},\ }\href {\doibase
  10.1016/j.physletb.2014.05.075} {\bibfield  {journal} {\bibinfo  {journal}
  {Phys. Lett.}\ }\textbf {\bibinfo {volume} {B734}},\ \bibinfo {pages} {290}
  (\bibinfo {year} {2014})},\ \Eprint {http://arxiv.org/abs/1209.1687}
  {arXiv:1209.1687 [hep-lat]} \BibitemShut {NoStop}%
%%CITATION = ARXIV:1209.1687;%%
\bibitem [{\citenamefont {Gockeler}\ \emph {et~al.}(2005)\citenamefont
  {Gockeler}, \citenamefont {Horsley}, \citenamefont {Pleiter}, \citenamefont
  {Rakow},\ and\ \citenamefont {Schierholz}}]{Gockeler:2004wp}%
  \BibitemOpen
  \bibfield  {author} {\bibinfo {author} {\bibfnamefont {M.}~\bibnamefont
  {Gockeler}}, \bibinfo {author} {\bibfnamefont {R.}~\bibnamefont {Horsley}},
  \bibinfo {author} {\bibfnamefont {D.}~\bibnamefont {Pleiter}}, \bibinfo
  {author} {\bibfnamefont {P.~E.~L.}\ \bibnamefont {Rakow}}, \ and\ \bibinfo
  {author} {\bibfnamefont {G.}~\bibnamefont {Schierholz}} (\bibinfo
  {collaboration} {QCDSF}),\ }\href {\doibase 10.1103/PhysRevD.71.114511}
  {\bibfield  {journal} {\bibinfo  {journal} {Phys. Rev.}\ }\textbf {\bibinfo
  {volume} {D71}},\ \bibinfo {pages} {114511} (\bibinfo {year} {2005})},\
  \Eprint {http://arxiv.org/abs/hep-ph/0410187} {arXiv:hep-ph/0410187 [hep-ph]}
  \BibitemShut {NoStop}%
%%CITATION = HEP-PH/0410187;%%
\bibitem [{\citenamefont {Dolgov}\ \emph {et~al.}(2002)\citenamefont {Dolgov}
  \emph {et~al.}}]{Dolgov:2002zm}%
  \BibitemOpen
  \bibfield  {author} {\bibinfo {author} {\bibfnamefont {D.}~\bibnamefont
  {Dolgov}} \emph {et~al.} (\bibinfo {collaboration} {LHPC, TXL}),\ }\href
  {\doibase 10.1103/PhysRevD.66.034506} {\bibfield  {journal} {\bibinfo
  {journal} {Phys. Rev.}\ }\textbf {\bibinfo {volume} {D66}},\ \bibinfo {pages}
  {034506} (\bibinfo {year} {2002})},\ \Eprint
  {http://arxiv.org/abs/hep-lat/0201021} {arXiv:hep-lat/0201021 [hep-lat]}
  \BibitemShut {NoStop}%
%%CITATION = HEP-LAT/0201021;%%
\bibitem [{\citenamefont {Hou}\ \emph {et~al.}(2021)\citenamefont {Hou} \emph
  {et~al.}}]{Hou:2019efy}%
  \BibitemOpen
  \bibfield  {author} {\bibinfo {author} {\bibfnamefont {T.-J.}\ \bibnamefont
  {Hou}} \emph {et~al.},\ }\href {\doibase 10.1103/PhysRevD.103.014013}
  {\bibfield  {journal} {\bibinfo  {journal} {Phys. Rev. D}\ }\textbf {\bibinfo
  {volume} {103}},\ \bibinfo {pages} {014013} (\bibinfo {year} {2021})},\
  \Eprint {http://arxiv.org/abs/1912.10053} {arXiv:1912.10053 [hep-ph]}
  \BibitemShut {NoStop}%
\bibitem [{\citenamefont {Pumplin}\ \emph {et~al.}(2002)\citenamefont
  {Pumplin}, \citenamefont {Stump}, \citenamefont {Huston}, \citenamefont
  {Lai}, \citenamefont {Nadolsky},\ and\ \citenamefont
  {Tung}}]{Pumplin:2002vw}%
  \BibitemOpen
  \bibfield  {author} {\bibinfo {author} {\bibfnamefont {J.}~\bibnamefont
  {Pumplin}}, \bibinfo {author} {\bibfnamefont {D.~R.}\ \bibnamefont {Stump}},
  \bibinfo {author} {\bibfnamefont {J.}~\bibnamefont {Huston}}, \bibinfo
  {author} {\bibfnamefont {H.~L.}\ \bibnamefont {Lai}}, \bibinfo {author}
  {\bibfnamefont {P.~M.}\ \bibnamefont {Nadolsky}}, \ and\ \bibinfo {author}
  {\bibfnamefont {W.~K.}\ \bibnamefont {Tung}},\ }\href {\doibase
  10.1088/1126-6708/2002/07/012} {\bibfield  {journal} {\bibinfo  {journal}
  {JHEP}\ }\textbf {\bibinfo {volume} {07}},\ \bibinfo {pages} {012} (\bibinfo
  {year} {2002})},\ \Eprint {http://arxiv.org/abs/hep-ph/0201195}
  {arXiv:hep-ph/0201195 [hep-ph]} \BibitemShut {NoStop}%
%%CITATION = HEP-PH/0201195;%%
\bibitem [{\citenamefont {Pumplin}\ \emph {et~al.}(2001)\citenamefont
  {Pumplin}, \citenamefont {Stump}, \citenamefont {Brock}, \citenamefont
  {Casey}, \citenamefont {Huston}, \citenamefont {Kalk}, \citenamefont {Lai},\
  and\ \citenamefont {Tung}}]{Pumplin:2001ct}%
  \BibitemOpen
  \bibfield  {author} {\bibinfo {author} {\bibfnamefont {J.}~\bibnamefont
  {Pumplin}}, \bibinfo {author} {\bibfnamefont {D.}~\bibnamefont {Stump}},
  \bibinfo {author} {\bibfnamefont {R.}~\bibnamefont {Brock}}, \bibinfo
  {author} {\bibfnamefont {D.}~\bibnamefont {Casey}}, \bibinfo {author}
  {\bibfnamefont {J.}~\bibnamefont {Huston}}, \bibinfo {author} {\bibfnamefont
  {J.}~\bibnamefont {Kalk}}, \bibinfo {author} {\bibfnamefont {H.~L.}\
  \bibnamefont {Lai}}, \ and\ \bibinfo {author} {\bibfnamefont {W.~K.}\
  \bibnamefont {Tung}},\ }\href {\doibase 10.1103/PhysRevD.65.014013}
  {\bibfield  {journal} {\bibinfo  {journal} {Phys. Rev.}\ }\textbf {\bibinfo
  {volume} {D65}},\ \bibinfo {pages} {014013} (\bibinfo {year} {2001})},\
  \Eprint {http://arxiv.org/abs/hep-ph/0101032} {arXiv:hep-ph/0101032 [hep-ph]}
  \BibitemShut {NoStop}%
%%CITATION = HEP-PH/0101032;%%
\bibitem [{\citenamefont {Stump}\ \emph {et~al.}(2001)\citenamefont {Stump},
  \citenamefont {Pumplin}, \citenamefont {Brock}, \citenamefont {Casey},
  \citenamefont {Huston}, \citenamefont {Kalk}, \citenamefont {Lai},\ and\
  \citenamefont {Tung}}]{Stump:2001gu}%
  \BibitemOpen
  \bibfield  {author} {\bibinfo {author} {\bibfnamefont {D.}~\bibnamefont
  {Stump}}, \bibinfo {author} {\bibfnamefont {J.}~\bibnamefont {Pumplin}},
  \bibinfo {author} {\bibfnamefont {R.}~\bibnamefont {Brock}}, \bibinfo
  {author} {\bibfnamefont {D.}~\bibnamefont {Casey}}, \bibinfo {author}
  {\bibfnamefont {J.}~\bibnamefont {Huston}}, \bibinfo {author} {\bibfnamefont
  {J.}~\bibnamefont {Kalk}}, \bibinfo {author} {\bibfnamefont {H.~L.}\
  \bibnamefont {Lai}}, \ and\ \bibinfo {author} {\bibfnamefont {W.~K.}\
  \bibnamefont {Tung}},\ }\href {\doibase 10.1103/PhysRevD.65.014012}
  {\bibfield  {journal} {\bibinfo  {journal} {Phys. Rev.}\ }\textbf {\bibinfo
  {volume} {D65}},\ \bibinfo {pages} {014012} (\bibinfo {year} {2001})},\
  \Eprint {http://arxiv.org/abs/hep-ph/0101051} {arXiv:hep-ph/0101051 [hep-ph]}
  \BibitemShut {NoStop}%
%%CITATION = HEP-PH/0101051;%%
\bibitem [{\citenamefont {Pumplin}(2010)}]{Pumplin:2009sc}%
  \BibitemOpen
  \bibfield  {author} {\bibinfo {author} {\bibfnamefont {J.}~\bibnamefont
  {Pumplin}},\ }\href {\doibase 10.1103/PhysRevD.81.074010} {\bibfield
  {journal} {\bibinfo  {journal} {Phys. Rev.}\ }\textbf {\bibinfo {volume}
  {D81}},\ \bibinfo {pages} {074010} (\bibinfo {year} {2010})},\ \Eprint
  {http://arxiv.org/abs/0909.0268} {arXiv:0909.0268 [hep-ph]} \BibitemShut
  {NoStop}%
%%CITATION = ARXIV:0909.0268;%%
\bibitem [{\citenamefont {Aoki}\ \emph {et~al.}(2010)\citenamefont {Aoki},
  \citenamefont {Blum}, \citenamefont {Lin}, \citenamefont {Ohta},
  \citenamefont {Sasaki}, \citenamefont {Tweedie}, \citenamefont {Zanotti},\
  and\ \citenamefont {Yamazaki}}]{Aoki:2010xg}%
  \BibitemOpen
  \bibfield  {author} {\bibinfo {author} {\bibfnamefont {Y.}~\bibnamefont
  {Aoki}}, \bibinfo {author} {\bibfnamefont {T.}~\bibnamefont {Blum}}, \bibinfo
  {author} {\bibfnamefont {H.-W.}\ \bibnamefont {Lin}}, \bibinfo {author}
  {\bibfnamefont {S.}~\bibnamefont {Ohta}}, \bibinfo {author} {\bibfnamefont
  {S.}~\bibnamefont {Sasaki}}, \bibinfo {author} {\bibfnamefont
  {R.}~\bibnamefont {Tweedie}}, \bibinfo {author} {\bibfnamefont
  {J.}~\bibnamefont {Zanotti}}, \ and\ \bibinfo {author} {\bibfnamefont
  {T.}~\bibnamefont {Yamazaki}},\ }\href {\doibase 10.1103/PhysRevD.82.014501}
  {\bibfield  {journal} {\bibinfo  {journal} {Phys. Rev.}\ }\textbf {\bibinfo
  {volume} {D82}},\ \bibinfo {pages} {014501} (\bibinfo {year} {2010})},\
  \Eprint {http://arxiv.org/abs/1003.3387} {arXiv:1003.3387 [hep-lat]}
  \BibitemShut {NoStop}%
%%CITATION = ARXIV:1003.3387;%%
\bibitem [{\citenamefont {Lin}\ \emph {et~al.}(2015)\citenamefont {Lin},
  \citenamefont {Chen}, \citenamefont {Cohen},\ and\ \citenamefont
  {Ji}}]{Lin:2014zya}%
  \BibitemOpen
  \bibfield  {author} {\bibinfo {author} {\bibfnamefont {H.-W.}\ \bibnamefont
  {Lin}}, \bibinfo {author} {\bibfnamefont {J.-W.}\ \bibnamefont {Chen}},
  \bibinfo {author} {\bibfnamefont {S.~D.}\ \bibnamefont {Cohen}}, \ and\
  \bibinfo {author} {\bibfnamefont {X.}~\bibnamefont {Ji}},\ }\href {\doibase
  10.1103/PhysRevD.91.054510} {\bibfield  {journal} {\bibinfo  {journal} {Phys.
  Rev.}\ }\textbf {\bibinfo {volume} {D91}},\ \bibinfo {pages} {054510}
  (\bibinfo {year} {2015})},\ \Eprint {http://arxiv.org/abs/1402.1462}
  {arXiv:1402.1462 [hep-ph]} \BibitemShut {NoStop}%
%%CITATION = ARXIV:1402.1462;%%
\bibitem [{\citenamefont {Chen}\ \emph {et~al.}(2018)\citenamefont {Chen},
  \citenamefont {Ishikawa}, \citenamefont {Jin}, \citenamefont {Lin},
  \citenamefont {Yang}, \citenamefont {Zhang},\ and\ \citenamefont
  {Zhao}}]{Chen:2017mzz}%
  \BibitemOpen
  \bibfield  {author} {\bibinfo {author} {\bibfnamefont {J.-W.}\ \bibnamefont
  {Chen}}, \bibinfo {author} {\bibfnamefont {T.}~\bibnamefont {Ishikawa}},
  \bibinfo {author} {\bibfnamefont {L.}~\bibnamefont {Jin}}, \bibinfo {author}
  {\bibfnamefont {H.-W.}\ \bibnamefont {Lin}}, \bibinfo {author} {\bibfnamefont
  {Y.-B.}\ \bibnamefont {Yang}}, \bibinfo {author} {\bibfnamefont {J.-H.}\
  \bibnamefont {Zhang}}, \ and\ \bibinfo {author} {\bibfnamefont
  {Y.}~\bibnamefont {Zhao}},\ }\href {\doibase 10.1103/PhysRevD.97.014505}
  {\bibfield  {journal} {\bibinfo  {journal} {Phys. Rev.}\ }\textbf {\bibinfo
  {volume} {D97}},\ \bibinfo {pages} {014505} (\bibinfo {year} {2018})},\
  \Eprint {http://arxiv.org/abs/1706.01295} {arXiv:1706.01295 [hep-lat]}
  \BibitemShut {NoStop}%
%%CITATION = ARXIV:1706.01295;%%
\bibitem [{\citenamefont {Gupta}\ \emph {et~al.}(2018)\citenamefont {Gupta},
  \citenamefont {Jang}, \citenamefont {Yoon}, \citenamefont {Lin},
  \citenamefont {Cirigliano},\ and\ \citenamefont
  {Bhattacharya}}]{Gupta:2018qil}%
  \BibitemOpen
  \bibfield  {author} {\bibinfo {author} {\bibfnamefont {R.}~\bibnamefont
  {Gupta}}, \bibinfo {author} {\bibfnamefont {Y.-C.}\ \bibnamefont {Jang}},
  \bibinfo {author} {\bibfnamefont {B.}~\bibnamefont {Yoon}}, \bibinfo {author}
  {\bibfnamefont {H.-W.}\ \bibnamefont {Lin}}, \bibinfo {author} {\bibfnamefont
  {V.}~\bibnamefont {Cirigliano}}, \ and\ \bibinfo {author} {\bibfnamefont
  {T.}~\bibnamefont {Bhattacharya}},\ }\href {\doibase
  10.1103/PhysRevD.98.034503} {\bibfield  {journal} {\bibinfo  {journal} {Phys.
  Rev.}\ }\textbf {\bibinfo {volume} {D98}},\ \bibinfo {pages} {034503}
  (\bibinfo {year} {2018})},\ \Eprint {http://arxiv.org/abs/1806.09006}
  {arXiv:1806.09006 [hep-lat]} \BibitemShut {NoStop}%
%%CITATION = ARXIV:1806.09006;%%
\bibitem [{\citenamefont {Liu}\ \emph {et~al.}(2018)\citenamefont {Liu},
  \citenamefont {Chen}, \citenamefont {Jin}, \citenamefont {Lin}, \citenamefont
  {Yang}, \citenamefont {Zhang},\ and\ \citenamefont {Zhao}}]{Liu:2018uuj}%
  \BibitemOpen
  \bibfield  {author} {\bibinfo {author} {\bibfnamefont {Y.-S.}\ \bibnamefont
  {Liu}}, \bibinfo {author} {\bibfnamefont {J.-W.}\ \bibnamefont {Chen}},
  \bibinfo {author} {\bibfnamefont {L.}~\bibnamefont {Jin}}, \bibinfo {author}
  {\bibfnamefont {H.-W.}\ \bibnamefont {Lin}}, \bibinfo {author} {\bibfnamefont
  {Y.-B.}\ \bibnamefont {Yang}}, \bibinfo {author} {\bibfnamefont {J.-H.}\
  \bibnamefont {Zhang}}, \ and\ \bibinfo {author} {\bibfnamefont
  {Y.}~\bibnamefont {Zhao}},\ }\href@noop {} {\  (\bibinfo {year} {2018})},\
  \Eprint {http://arxiv.org/abs/1807.06566} {arXiv:1807.06566 [hep-lat]}
  \BibitemShut {NoStop}%
%%CITATION = ARXIV:1807.06566;%%
\bibitem [{\citenamefont {Alexandrou}\ \emph {et~al.}(2018)\citenamefont
  {Alexandrou}, \citenamefont {Cichy}, \citenamefont {Constantinou},
  \citenamefont {Jansen}, \citenamefont {Scapellato},\ and\ \citenamefont
  {Steffens}}]{Alexandrou:2018pbm}%
  \BibitemOpen
  \bibfield  {author} {\bibinfo {author} {\bibfnamefont {C.}~\bibnamefont
  {Alexandrou}}, \bibinfo {author} {\bibfnamefont {K.}~\bibnamefont {Cichy}},
  \bibinfo {author} {\bibfnamefont {M.}~\bibnamefont {Constantinou}}, \bibinfo
  {author} {\bibfnamefont {K.}~\bibnamefont {Jansen}}, \bibinfo {author}
  {\bibfnamefont {A.}~\bibnamefont {Scapellato}}, \ and\ \bibinfo {author}
  {\bibfnamefont {F.}~\bibnamefont {Steffens}},\ }\href {\doibase
  10.1103/PhysRevLett.121.112001} {\bibfield  {journal} {\bibinfo  {journal}
  {Phys. Rev. Lett.}\ }\textbf {\bibinfo {volume} {121}},\ \bibinfo {pages}
  {112001} (\bibinfo {year} {2018})},\ \Eprint
  {http://arxiv.org/abs/1803.02685} {arXiv:1803.02685 [hep-lat]} \BibitemShut
  {NoStop}%
%%CITATION = ARXIV:1803.02685;%%
\bibitem [{\citenamefont {Zhang}\ \emph {et~al.}(2019)\citenamefont {Zhang},
  \citenamefont {Chen}, \citenamefont {Jin}, \citenamefont {Lin}, \citenamefont
  {Schäfer},\ and\ \citenamefont {Zhao}}]{Chen:2018fwa}%
  \BibitemOpen
  \bibfield  {author} {\bibinfo {author} {\bibfnamefont {J.-H.}\ \bibnamefont
  {Zhang}}, \bibinfo {author} {\bibfnamefont {J.-W.}\ \bibnamefont {Chen}},
  \bibinfo {author} {\bibfnamefont {L.}~\bibnamefont {Jin}}, \bibinfo {author}
  {\bibfnamefont {H.-W.}\ \bibnamefont {Lin}}, \bibinfo {author} {\bibfnamefont
  {A.}~\bibnamefont {Schäfer}}, \ and\ \bibinfo {author} {\bibfnamefont
  {Y.}~\bibnamefont {Zhao}},\ }\href {\doibase 10.1103/PhysRevD.100.034505}
  {\bibfield  {journal} {\bibinfo  {journal} {Phys. Rev.}\ }\textbf {\bibinfo
  {volume} {D100}},\ \bibinfo {pages} {034505} (\bibinfo {year} {2019})},\
  \Eprint {http://arxiv.org/abs/1804.01483} {arXiv:1804.01483 [hep-lat]}
  \BibitemShut {NoStop}%
%%CITATION = ARXIV:1804.01483;%%
\bibitem [{\citenamefont {Sufian}\ \emph {et~al.}(2019)\citenamefont {Sufian},
  \citenamefont {Karpie}, \citenamefont {Egerer}, \citenamefont {Orginos},
  \citenamefont {Qiu},\ and\ \citenamefont {Richards}}]{Sufian:2019bol}%
  \BibitemOpen
  \bibfield  {author} {\bibinfo {author} {\bibfnamefont {R.~S.}\ \bibnamefont
  {Sufian}}, \bibinfo {author} {\bibfnamefont {J.}~\bibnamefont {Karpie}},
  \bibinfo {author} {\bibfnamefont {C.}~\bibnamefont {Egerer}}, \bibinfo
  {author} {\bibfnamefont {K.}~\bibnamefont {Orginos}}, \bibinfo {author}
  {\bibfnamefont {J.-W.}\ \bibnamefont {Qiu}}, \ and\ \bibinfo {author}
  {\bibfnamefont {D.~G.}\ \bibnamefont {Richards}},\ }\href@noop {} {\
  (\bibinfo {year} {2019})},\ \Eprint {http://arxiv.org/abs/1901.03921}
  {arXiv:1901.03921 [hep-lat]} \BibitemShut {NoStop}%
%%CITATION = ARXIV:1901.03921;%%
\bibitem [{\citenamefont {Oehm}\ \emph {et~al.}(2019)\citenamefont {Oehm},
  \citenamefont {Alexandrou}, \citenamefont {Constantinou}, \citenamefont
  {Jansen}, \citenamefont {Koutsou}, \citenamefont {Kostrzewa}, \citenamefont
  {Steffens}, \citenamefont {Urbach},\ and\ \citenamefont
  {Zafeiropoulos}}]{Oehm:2018jvm}%
  \BibitemOpen
  \bibfield  {author} {\bibinfo {author} {\bibfnamefont {M.}~\bibnamefont
  {Oehm}}, \bibinfo {author} {\bibfnamefont {C.}~\bibnamefont {Alexandrou}},
  \bibinfo {author} {\bibfnamefont {M.}~\bibnamefont {Constantinou}}, \bibinfo
  {author} {\bibfnamefont {K.}~\bibnamefont {Jansen}}, \bibinfo {author}
  {\bibfnamefont {G.}~\bibnamefont {Koutsou}}, \bibinfo {author} {\bibfnamefont
  {B.}~\bibnamefont {Kostrzewa}}, \bibinfo {author} {\bibfnamefont
  {F.}~\bibnamefont {Steffens}}, \bibinfo {author} {\bibfnamefont
  {C.}~\bibnamefont {Urbach}}, \ and\ \bibinfo {author} {\bibfnamefont
  {S.}~\bibnamefont {Zafeiropoulos}},\ }\href {\doibase
  10.1103/PhysRevD.99.014508} {\bibfield  {journal} {\bibinfo  {journal} {Phys.
  Rev.}\ }\textbf {\bibinfo {volume} {D99}},\ \bibinfo {pages} {014508}
  (\bibinfo {year} {2019})},\ \Eprint {http://arxiv.org/abs/1810.09743}
  {arXiv:1810.09743 [hep-lat]} \BibitemShut {NoStop}%
%%CITATION = ARXIV:1810.09743;%%
\bibitem [{\citenamefont {Hobbs}(2018)}]{Hobbs:2017xtq}%
  \BibitemOpen
  \bibfield  {author} {\bibinfo {author} {\bibfnamefont {T.~J.}\ \bibnamefont
  {Hobbs}},\ }\href {\doibase 10.1103/PhysRevD.97.054028} {\bibfield  {journal}
  {\bibinfo  {journal} {Phys. Rev.}\ }\textbf {\bibinfo {volume} {D97}},\
  \bibinfo {pages} {054028} (\bibinfo {year} {2018})},\ \Eprint
  {http://arxiv.org/abs/1708.05463} {arXiv:1708.05463 [hep-ph]} \BibitemShut
  {NoStop}%
%%CITATION = ARXIV:1708.05463;%%
\bibitem [{Web()}]{Website}%
  \BibitemOpen
  \href@noop {} {}\bibinfo {howpublished}
  {\url{http://metapdf.hepforge.org/PDFSense/Lattice/}}\BibitemShut {NoStop}%
\bibitem [{\citenamefont {Thomas}(1983)}]{Thomas:1983fh}%
  \BibitemOpen
  \bibfield  {author} {\bibinfo {author} {\bibfnamefont {A.~W.}\ \bibnamefont
  {Thomas}},\ }\href {\doibase 10.1016/0370-2693(83)90026-6} {\bibfield
  {journal} {\bibinfo  {journal} {Phys. Lett.}\ }\textbf {\bibinfo {volume}
  {126B}},\ \bibinfo {pages} {97} (\bibinfo {year} {1983})}\BibitemShut
  {NoStop}%
%%CITATION = PHLTA,126B,97;%%
\bibitem [{\citenamefont {Signal}\ \emph {et~al.}(1991)\citenamefont {Signal},
  \citenamefont {Schreiber},\ and\ \citenamefont {Thomas}}]{Signal:1991ug}%
  \BibitemOpen
  \bibfield  {author} {\bibinfo {author} {\bibfnamefont {A.~I.}\ \bibnamefont
  {Signal}}, \bibinfo {author} {\bibfnamefont {A.~W.}\ \bibnamefont
  {Schreiber}}, \ and\ \bibinfo {author} {\bibfnamefont {A.~W.}\ \bibnamefont
  {Thomas}},\ }\href {\doibase 10.1142/S0217732391000233} {\bibfield  {journal}
  {\bibinfo  {journal} {Mod. Phys. Lett.}\ }\textbf {\bibinfo {volume} {A6}},\
  \bibinfo {pages} {271} (\bibinfo {year} {1991})}\BibitemShut {NoStop}%
%%CITATION = MPLAE,A6,271;%%
\bibitem [{\citenamefont {Schreiber}\ \emph {et~al.}(1992)\citenamefont
  {Schreiber}, \citenamefont {Mulders}, \citenamefont {Signal},\ and\
  \citenamefont {Thomas}}]{Schreiber:1991qx}%
  \BibitemOpen
  \bibfield  {author} {\bibinfo {author} {\bibfnamefont {A.~W.}\ \bibnamefont
  {Schreiber}}, \bibinfo {author} {\bibfnamefont {P.~J.}\ \bibnamefont
  {Mulders}}, \bibinfo {author} {\bibfnamefont {A.~I.}\ \bibnamefont {Signal}},
  \ and\ \bibinfo {author} {\bibfnamefont {A.~W.}\ \bibnamefont {Thomas}},\
  }\href {\doibase 10.1103/PhysRevD.45.3069} {\bibfield  {journal} {\bibinfo
  {journal} {Phys. Rev.}\ }\textbf {\bibinfo {volume} {D45}},\ \bibinfo {pages}
  {3069} (\bibinfo {year} {1992})}\BibitemShut {NoStop}%
%%CITATION = PHRVA,D45,3069;%%
\bibitem [{\citenamefont {Alberg}\ and\ \citenamefont
  {Miller}(2012)}]{Alberg:2012wr}%
  \BibitemOpen
  \bibfield  {author} {\bibinfo {author} {\bibfnamefont {M.}~\bibnamefont
  {Alberg}}\ and\ \bibinfo {author} {\bibfnamefont {G.~A.}\ \bibnamefont
  {Miller}},\ }\href {\doibase 10.1103/PhysRevLett.108.172001} {\bibfield
  {journal} {\bibinfo  {journal} {Phys. Rev. Lett.}\ }\textbf {\bibinfo
  {volume} {108}},\ \bibinfo {pages} {172001} (\bibinfo {year} {2012})},\
  \Eprint {http://arxiv.org/abs/1201.4184} {arXiv:1201.4184 [nucl-th]}
  \BibitemShut {NoStop}%
%%CITATION = ARXIV:1201.4184;%%
\bibitem [{\citenamefont {Salamu}\ \emph {et~al.}(2015)\citenamefont {Salamu},
  \citenamefont {Ji}, \citenamefont {Melnitchouk},\ and\ \citenamefont
  {Wang}}]{Salamu:2014pka}%
  \BibitemOpen
  \bibfield  {author} {\bibinfo {author} {\bibfnamefont {Y.}~\bibnamefont
  {Salamu}}, \bibinfo {author} {\bibfnamefont {C.-R.}\ \bibnamefont {Ji}},
  \bibinfo {author} {\bibfnamefont {W.}~\bibnamefont {Melnitchouk}}, \ and\
  \bibinfo {author} {\bibfnamefont {P.}~\bibnamefont {Wang}},\ }\href {\doibase
  10.1103/PhysRevLett.114.122001} {\bibfield  {journal} {\bibinfo  {journal}
  {Phys. Rev. Lett.}\ }\textbf {\bibinfo {volume} {114}},\ \bibinfo {pages}
  {122001} (\bibinfo {year} {2015})},\ \Eprint {http://arxiv.org/abs/1409.5885}
  {arXiv:1409.5885 [hep-ph]} \BibitemShut {NoStop}%
%%CITATION = ARXIV:1409.5885;%%
\bibitem [{\citenamefont {Amaudruz}\ \emph {et~al.}(1991)\citenamefont
  {Amaudruz} \emph {et~al.}}]{Amaudruz:1991at}%
  \BibitemOpen
  \bibfield  {author} {\bibinfo {author} {\bibfnamefont {P.}~\bibnamefont
  {Amaudruz}} \emph {et~al.} (\bibinfo {collaboration} {New Muon}),\ }\href
  {\doibase 10.1103/PhysRevLett.66.2712} {\bibfield  {journal} {\bibinfo
  {journal} {Phys. Rev. Lett.}\ }\textbf {\bibinfo {volume} {66}},\ \bibinfo
  {pages} {2712} (\bibinfo {year} {1991})}\BibitemShut {NoStop}%
%%CITATION = PRLTA,66,2712;%%
\bibitem [{\citenamefont {Arneodo}\ \emph {et~al.}(1994)\citenamefont {Arneodo}
  \emph {et~al.}}]{Arneodo:1994sh}%
  \BibitemOpen
  \bibfield  {author} {\bibinfo {author} {\bibfnamefont {M.}~\bibnamefont
  {Arneodo}} \emph {et~al.} (\bibinfo {collaboration} {New Muon}),\ }\href
  {\doibase 10.1103/PhysRevD.50.R1} {\bibfield  {journal} {\bibinfo  {journal}
  {Phys. Rev.}\ }\textbf {\bibinfo {volume} {D50}},\ \bibinfo {pages} {R1}
  (\bibinfo {year} {1994})}\BibitemShut {NoStop}%
%%CITATION = PHRVA,D50,R1;%%
\bibitem [{\citenamefont {Gottfried}(1967)}]{Gottfried:1967kk}%
  \BibitemOpen
  \bibfield  {author} {\bibinfo {author} {\bibfnamefont {K.}~\bibnamefont
  {Gottfried}},\ }\href {\doibase 10.1103/PhysRevLett.18.1174} {\bibfield
  {journal} {\bibinfo  {journal} {Phys. Rev. Lett.}\ }\textbf {\bibinfo
  {volume} {18}},\ \bibinfo {pages} {1174} (\bibinfo {year}
  {1967})}\BibitemShut {NoStop}%
%%CITATION = PRLTA,18,1174;%%
\bibitem [{\citenamefont {Martinelli}\ \emph {et~al.}(1995)\citenamefont
  {Martinelli}, \citenamefont {Pittori}, \citenamefont {Sachrajda},
  \citenamefont {Testa},\ and\ \citenamefont {Vladikas}}]{Martinelli:1994ty}%
  \BibitemOpen
  \bibfield  {author} {\bibinfo {author} {\bibfnamefont {G.}~\bibnamefont
  {Martinelli}}, \bibinfo {author} {\bibfnamefont {C.}~\bibnamefont {Pittori}},
  \bibinfo {author} {\bibfnamefont {C.~T.}\ \bibnamefont {Sachrajda}}, \bibinfo
  {author} {\bibfnamefont {M.}~\bibnamefont {Testa}}, \ and\ \bibinfo {author}
  {\bibfnamefont {A.}~\bibnamefont {Vladikas}},\ }\href {\doibase
  10.1016/0550-3213(95)00126-D} {\bibfield  {journal} {\bibinfo  {journal}
  {Nucl. Phys.}\ }\textbf {\bibinfo {volume} {B445}},\ \bibinfo {pages} {81}
  (\bibinfo {year} {1995})},\ \Eprint {http://arxiv.org/abs/hep-lat/9411010}
  {arXiv:hep-lat/9411010 [hep-lat]} \BibitemShut {NoStop}%
%%CITATION = HEP-LAT/9411010;%%
\bibitem [{\citenamefont {Constantinou}\ and\ \citenamefont
  {Panagopoulos}(2017)}]{Constantinou:2017sej}%
  \BibitemOpen
  \bibfield  {author} {\bibinfo {author} {\bibfnamefont {M.}~\bibnamefont
  {Constantinou}}\ and\ \bibinfo {author} {\bibfnamefont {H.}~\bibnamefont
  {Panagopoulos}},\ }\href {\doibase 10.1103/PhysRevD.96.054506} {\bibfield
  {journal} {\bibinfo  {journal} {Phys. Rev.}\ }\textbf {\bibinfo {volume}
  {D96}},\ \bibinfo {pages} {054506} (\bibinfo {year} {2017})},\ \Eprint
  {http://arxiv.org/abs/1705.11193} {arXiv:1705.11193 [hep-lat]} \BibitemShut
  {NoStop}%
%%CITATION = ARXIV:1705.11193;%%
\bibitem [{\citenamefont {Apollinari}\ \emph {et~al.}(2015)\citenamefont
  {Apollinari}, \citenamefont {Béjar~Alonso}, \citenamefont {Brüning},
  \citenamefont {Lamont},\ and\ \citenamefont {Rossi}}]{Apollinari:2015bam}%
  \BibitemOpen
  \bibfield  {author} {\bibinfo {author} {\bibfnamefont {G.}~\bibnamefont
  {Apollinari}}, \bibinfo {author} {\bibfnamefont {I.}~\bibnamefont
  {Béjar~Alonso}}, \bibinfo {author} {\bibfnamefont {O.}~\bibnamefont
  {Brüning}}, \bibinfo {author} {\bibfnamefont {M.}~\bibnamefont {Lamont}}, \
  and\ \bibinfo {author} {\bibfnamefont {L.}~\bibnamefont {Rossi}},\ }\href
  {\doibase 10.5170/CERN-2015-005} {\  (\bibinfo {year} {2015}),\
  10.5170/CERN-2015-005}\BibitemShut {NoStop}%
%%CITATION = CERN-2015-005;%%
\bibitem [{\citenamefont {Brodsky}\ \emph {et~al.}(2013)\citenamefont
  {Brodsky}, \citenamefont {Fleuret}, \citenamefont {Hadjidakis},\ and\
  \citenamefont {Lansberg}}]{Brodsky:2012vg}%
  \BibitemOpen
  \bibfield  {author} {\bibinfo {author} {\bibfnamefont {S.~J.}\ \bibnamefont
  {Brodsky}}, \bibinfo {author} {\bibfnamefont {F.}~\bibnamefont {Fleuret}},
  \bibinfo {author} {\bibfnamefont {C.}~\bibnamefont {Hadjidakis}}, \ and\
  \bibinfo {author} {\bibfnamefont {J.~P.}\ \bibnamefont {Lansberg}},\ }\href
  {\doibase 10.1016/j.physrep.2012.10.001} {\bibfield  {journal} {\bibinfo
  {journal} {Phys. Rept.}\ }\textbf {\bibinfo {volume} {522}},\ \bibinfo
  {pages} {239} (\bibinfo {year} {2013})},\ \Eprint
  {http://arxiv.org/abs/1202.6585} {arXiv:1202.6585 [hep-ph]} \BibitemShut
  {NoStop}%
%%CITATION = ARXIV:1202.6585;%%
\bibitem [{\citenamefont {Accardi}\ \emph
  {et~al.}(2016{\natexlab{b}})\citenamefont {Accardi} \emph
  {et~al.}}]{Accardi:2012qut}%
  \BibitemOpen
  \bibfield  {author} {\bibinfo {author} {\bibfnamefont {A.}~\bibnamefont
  {Accardi}} \emph {et~al.},\ }\href {\doibase 10.1140/epja/i2016-16268-9}
  {\bibfield  {journal} {\bibinfo  {journal} {Eur. Phys. J.}\ }\textbf
  {\bibinfo {volume} {A52}},\ \bibinfo {pages} {268} (\bibinfo {year}
  {2016}{\natexlab{b}})},\ \Eprint {http://arxiv.org/abs/1212.1701}
  {arXiv:1212.1701 [nucl-ex]} \BibitemShut {NoStop}%
%%CITATION = ARXIV:1212.1701;%%
\bibitem [{\citenamefont {Boer}\ \emph {et~al.}(2011)\citenamefont {Boer} \emph
  {et~al.}}]{Boer:2011fh}%
  \BibitemOpen
  \bibfield  {author} {\bibinfo {author} {\bibfnamefont {D.}~\bibnamefont
  {Boer}} \emph {et~al.},\ }\href@noop {} {\  (\bibinfo {year} {2011})},\
  \Eprint {http://arxiv.org/abs/1108.1713} {arXiv:1108.1713 [nucl-th]}
  \BibitemShut {NoStop}%
%%CITATION = ARXIV:1108.1713;%%
\bibitem [{\citenamefont {Abeyratne}\ \emph {et~al.}(2012)\citenamefont
  {Abeyratne} \emph {et~al.}}]{Abeyratne:2012ah}%
  \BibitemOpen
  \bibfield  {author} {\bibinfo {author} {\bibfnamefont {S.}~\bibnamefont
  {Abeyratne}} \emph {et~al.},\ }\href@noop {} {\  (\bibinfo {year} {2012})},\
  \Eprint {http://arxiv.org/abs/1209.0757} {arXiv:1209.0757 [physics.acc-ph]}
  \BibitemShut {NoStop}%
%%CITATION = ARXIV:1209.0757;%%
\bibitem [{\citenamefont {Aschenauer}\ \emph {et~al.}(2014)\citenamefont
  {Aschenauer} \emph {et~al.}}]{Aschenauer:2014cki}%
  \BibitemOpen
  \bibfield  {author} {\bibinfo {author} {\bibfnamefont {E.~C.}\ \bibnamefont
  {Aschenauer}} \emph {et~al.},\ }\href@noop {} {\  (\bibinfo {year} {2014})},\
  \Eprint {http://arxiv.org/abs/1409.1633} {arXiv:1409.1633 [physics.acc-ph]}
  \BibitemShut {NoStop}%
%%CITATION = ARXIV:1409.1633;%%
\bibitem [{\citenamefont {Abelleira~Fernandez}\ \emph
  {et~al.}(2012)\citenamefont {Abelleira~Fernandez} \emph
  {et~al.}}]{AbelleiraFernandez:2012cc}%
  \BibitemOpen
  \bibfield  {author} {\bibinfo {author} {\bibfnamefont {J.~L.}\ \bibnamefont
  {Abelleira~Fernandez}} \emph {et~al.} (\bibinfo {collaboration} {LHeC Study
  Group}),\ }\href {\doibase 10.1088/0954-3899/39/7/075001} {\bibfield
  {journal} {\bibinfo  {journal} {J. Phys.}\ }\textbf {\bibinfo {volume}
  {G39}},\ \bibinfo {pages} {075001} (\bibinfo {year} {2012})},\ \Eprint
  {http://arxiv.org/abs/1206.2913} {arXiv:1206.2913 [physics.acc-ph]}
  \BibitemShut {NoStop}%
%%CITATION = ARXIV:1206.2913;%%
\bibitem [{\citenamefont {Dudek}\ \emph {et~al.}(2012)\citenamefont {Dudek}
  \emph {et~al.}}]{Dudek:2012vr}%
  \BibitemOpen
  \bibfield  {author} {\bibinfo {author} {\bibfnamefont {J.}~\bibnamefont
  {Dudek}} \emph {et~al.},\ }\href {\doibase 10.1140/epja/i2012-12187-1}
  {\bibfield  {journal} {\bibinfo  {journal} {Eur. Phys. J.}\ }\textbf
  {\bibinfo {volume} {A48}},\ \bibinfo {pages} {187} (\bibinfo {year}
  {2012})},\ \Eprint {http://arxiv.org/abs/1208.1244} {arXiv:1208.1244
  [hep-ex]} \BibitemShut {NoStop}%
%%CITATION = ARXIV:1208.1244;%%
\bibitem [{\citenamefont {Benvenuti}\ \emph {et~al.}(1989)\citenamefont
  {Benvenuti} \emph {et~al.}}]{Benvenuti:1989rh}%
  \BibitemOpen
  \bibfield  {author} {\bibinfo {author} {\bibfnamefont {A.~C.}\ \bibnamefont
  {Benvenuti}} \emph {et~al.} (\bibinfo {collaboration} {BCDMS}),\ }\href
  {\doibase 10.1016/0370-2693(89)91637-7} {\bibfield  {journal} {\bibinfo
  {journal} {Phys. Lett.}\ }\textbf {\bibinfo {volume} {B223}},\ \bibinfo
  {pages} {485} (\bibinfo {year} {1989})}\BibitemShut {NoStop}%
%%CITATION = PHLTA,B223,485;%%
\bibitem [{\citenamefont {Benvenuti}\ \emph {et~al.}(1990)\citenamefont
  {Benvenuti} \emph {et~al.}}]{Benvenuti:1989fm}%
  \BibitemOpen
  \bibfield  {author} {\bibinfo {author} {\bibfnamefont {A.~C.}\ \bibnamefont
  {Benvenuti}} \emph {et~al.} (\bibinfo {collaboration} {BCDMS}),\ }\href
  {\doibase 10.1016/0370-2693(90)91231-Y} {\bibfield  {journal} {\bibinfo
  {journal} {Phys. Lett.}\ }\textbf {\bibinfo {volume} {B237}},\ \bibinfo
  {pages} {592} (\bibinfo {year} {1990})}\BibitemShut {NoStop}%
%%CITATION = PHLTA,B237,592;%%
\bibitem [{\citenamefont {Arneodo}\ \emph {et~al.}(1997)\citenamefont {Arneodo}
  \emph {et~al.}}]{Arneodo:1996qe}%
  \BibitemOpen
  \bibfield  {author} {\bibinfo {author} {\bibfnamefont {M.}~\bibnamefont
  {Arneodo}} \emph {et~al.} (\bibinfo {collaboration} {New Muon}),\ }\href
  {\doibase 10.1016/S0550-3213(96)00538-X} {\bibfield  {journal} {\bibinfo
  {journal} {Nucl. Phys.}\ }\textbf {\bibinfo {volume} {B483}},\ \bibinfo
  {pages} {3} (\bibinfo {year} {1997})},\ \Eprint
  {http://arxiv.org/abs/hep-ph/9610231} {arXiv:hep-ph/9610231 [hep-ph]}
  \BibitemShut {NoStop}%
%%CITATION = HEP-PH/9610231;%%
\bibitem [{\citenamefont {Berge}\ \emph {et~al.}(1991)\citenamefont {Berge}
  \emph {et~al.}}]{Berge:1989hr}%
  \BibitemOpen
  \bibfield  {author} {\bibinfo {author} {\bibfnamefont {J.~P.}\ \bibnamefont
  {Berge}} \emph {et~al.},\ }\href {\doibase 10.1007/BF01555493} {\bibfield
  {journal} {\bibinfo  {journal} {Z. Phys.}\ }\textbf {\bibinfo {volume}
  {C49}},\ \bibinfo {pages} {187} (\bibinfo {year} {1991})}\BibitemShut
  {NoStop}%
%%CITATION = ZEPYA,C49,187;%%
\bibitem [{\citenamefont {Yang}\ \emph {et~al.}(2001)\citenamefont {Yang} \emph
  {et~al.}}]{Yang:2000ju}%
  \BibitemOpen
  \bibfield  {author} {\bibinfo {author} {\bibfnamefont {U.-K.}\ \bibnamefont
  {Yang}} \emph {et~al.} (\bibinfo {collaboration} {CCFR/NuTeV}),\ }\href
  {\doibase 10.1103/PhysRevLett.86.2742} {\bibfield  {journal} {\bibinfo
  {journal} {Phys. Rev. Lett.}\ }\textbf {\bibinfo {volume} {86}},\ \bibinfo
  {pages} {2742} (\bibinfo {year} {2001})},\ \Eprint
  {http://arxiv.org/abs/hep-ex/0009041} {arXiv:hep-ex/0009041 [hep-ex]}
  \BibitemShut {NoStop}%
%%CITATION = HEP-EX/0009041;%%
\bibitem [{\citenamefont {Seligman}\ \emph {et~al.}(1997)\citenamefont
  {Seligman} \emph {et~al.}}]{Seligman:1997mc}%
  \BibitemOpen
  \bibfield  {author} {\bibinfo {author} {\bibfnamefont {W.~G.}\ \bibnamefont
  {Seligman}} \emph {et~al.},\ }\href {\doibase 10.1103/PhysRevLett.79.1213}
  {\bibfield  {journal} {\bibinfo  {journal} {Phys. Rev. Lett.}\ }\textbf
  {\bibinfo {volume} {79}},\ \bibinfo {pages} {1213} (\bibinfo {year}
  {1997})},\ \Eprint {http://arxiv.org/abs/hep-ex/9701017}
  {arXiv:hep-ex/9701017 [hep-ex]} \BibitemShut {NoStop}%
%%CITATION = HEP-EX/9701017;%%
\bibitem [{\citenamefont {Mason}(2006)}]{Mason:2006qa}%
  \BibitemOpen
  \bibfield  {author} {\bibinfo {author} {\bibfnamefont {D.~A.}\ \bibnamefont
  {Mason}},\ }\emph {\bibinfo {title} {{Measurement of the strange -
  antistrange asymmetry at NLO in QCD from NuTeV dimuon data}}},\ \href
  {\doibase 10.2172/879078} {Ph.D. thesis},\ \bibinfo  {school} {Oregon U.}
  (\bibinfo {year} {2006})\BibitemShut {NoStop}%
%%CITATION = FERMILAB-THESIS-2006-01;%%
\bibitem [{\citenamefont {Goncharov}\ \emph {et~al.}(2001)\citenamefont
  {Goncharov} \emph {et~al.}}]{Goncharov:2001qe}%
  \BibitemOpen
  \bibfield  {author} {\bibinfo {author} {\bibfnamefont {M.}~\bibnamefont
  {Goncharov}} \emph {et~al.} (\bibinfo {collaboration} {NuTeV}),\ }\href
  {\doibase 10.1103/PhysRevD.64.112006} {\bibfield  {journal} {\bibinfo
  {journal} {Phys. Rev.}\ }\textbf {\bibinfo {volume} {D64}},\ \bibinfo {pages}
  {112006} (\bibinfo {year} {2001})},\ \Eprint
  {http://arxiv.org/abs/hep-ex/0102049} {arXiv:hep-ex/0102049 [hep-ex]}
  \BibitemShut {NoStop}%
%%CITATION = HEP-EX/0102049;%%
\bibitem [{\citenamefont {Aktas}\ \emph {et~al.}(2005)\citenamefont {Aktas}
  \emph {et~al.}}]{Aktas:2004az}%
  \BibitemOpen
  \bibfield  {author} {\bibinfo {author} {\bibfnamefont {A.}~\bibnamefont
  {Aktas}} \emph {et~al.} (\bibinfo {collaboration} {H1}),\ }\href {\doibase
  10.1140/epjc/s2005-02154-8} {\bibfield  {journal} {\bibinfo  {journal} {Eur.
  Phys. J.}\ }\textbf {\bibinfo {volume} {C40}},\ \bibinfo {pages} {349}
  (\bibinfo {year} {2005})},\ \Eprint {http://arxiv.org/abs/hep-ex/0411046}
  {arXiv:hep-ex/0411046 [hep-ex]} \BibitemShut {NoStop}%
%%CITATION = HEP-EX/0411046;%%
\bibitem [{\citenamefont {Aktas}\ \emph {et~al.}(2006)\citenamefont {Aktas}
  \emph {et~al.}}]{Aktas:2005iw}%
  \BibitemOpen
  \bibfield  {author} {\bibinfo {author} {\bibfnamefont {A.}~\bibnamefont
  {Aktas}} \emph {et~al.} (\bibinfo {collaboration} {H1}),\ }\href {\doibase
  10.1140/epjc/s2005-02415-6} {\bibfield  {journal} {\bibinfo  {journal} {Eur.
  Phys. J.}\ }\textbf {\bibinfo {volume} {C45}},\ \bibinfo {pages} {23}
  (\bibinfo {year} {2006})},\ \Eprint {http://arxiv.org/abs/hep-ex/0507081}
  {arXiv:hep-ex/0507081 [hep-ex]} \BibitemShut {NoStop}%
%%CITATION = HEP-EX/0507081;%%
\bibitem [{\citenamefont {Abramowicz}\ \emph {et~al.}(2013)\citenamefont
  {Abramowicz} \emph {et~al.}}]{Abramowicz:1900rp}%
  \BibitemOpen
  \bibfield  {author} {\bibinfo {author} {\bibfnamefont {H.}~\bibnamefont
  {Abramowicz}} \emph {et~al.} (\bibinfo {collaboration} {ZEUS, H1}),\ }\href
  {\doibase 10.1140/epjc/s10052-013-2311-3} {\bibfield  {journal} {\bibinfo
  {journal} {Eur. Phys. J.}\ }\textbf {\bibinfo {volume} {C73}},\ \bibinfo
  {pages} {2311} (\bibinfo {year} {2013})},\ \Eprint
  {http://arxiv.org/abs/1211.1182} {arXiv:1211.1182 [hep-ex]} \BibitemShut
  {NoStop}%
%%CITATION = ARXIV:1211.1182;%%
\bibitem [{\citenamefont {Abramowicz}\ \emph {et~al.}(2015)\citenamefont
  {Abramowicz} \emph {et~al.}}]{Abramowicz:2015mha}%
  \BibitemOpen
  \bibfield  {author} {\bibinfo {author} {\bibfnamefont {H.}~\bibnamefont
  {Abramowicz}} \emph {et~al.} (\bibinfo {collaboration} {ZEUS, H1}),\ }\href
  {\doibase 10.1140/epjc/s10052-015-3710-4} {\bibfield  {journal} {\bibinfo
  {journal} {Eur. Phys. J.}\ }\textbf {\bibinfo {volume} {C75}},\ \bibinfo
  {pages} {580} (\bibinfo {year} {2015})},\ \Eprint
  {http://arxiv.org/abs/1506.06042} {arXiv:1506.06042 [hep-ex]} \BibitemShut
  {NoStop}%
%%CITATION = ARXIV:1506.06042;%%
\bibitem [{\citenamefont {Aaron}\ \emph {et~al.}(2011)\citenamefont {Aaron}
  \emph {et~al.}}]{Collaboration:2010ry}%
  \BibitemOpen
  \bibfield  {author} {\bibinfo {author} {\bibfnamefont {F.~D.}\ \bibnamefont
  {Aaron}} \emph {et~al.} (\bibinfo {collaboration} {H1}),\ }\href {\doibase
  10.1140/epjc/s10052-011-1579-4} {\bibfield  {journal} {\bibinfo  {journal}
  {Eur. Phys. J.}\ }\textbf {\bibinfo {volume} {C71}},\ \bibinfo {pages} {1579}
  (\bibinfo {year} {2011})},\ \Eprint {http://arxiv.org/abs/1012.4355}
  {arXiv:1012.4355 [hep-ex]} \BibitemShut {NoStop}%
%%CITATION = ARXIV:1012.4355;%%
\bibitem [{\citenamefont {Moreno}\ \emph {et~al.}(1991)\citenamefont {Moreno}
  \emph {et~al.}}]{Moreno:1990sf}%
  \BibitemOpen
  \bibfield  {author} {\bibinfo {author} {\bibfnamefont {G.}~\bibnamefont
  {Moreno}} \emph {et~al.},\ }\href {\doibase 10.1103/PhysRevD.43.2815}
  {\bibfield  {journal} {\bibinfo  {journal} {Phys. Rev.}\ }\textbf {\bibinfo
  {volume} {D43}},\ \bibinfo {pages} {2815} (\bibinfo {year}
  {1991})}\BibitemShut {NoStop}%
%%CITATION = PHRVA,D43,2815;%%
\bibitem [{\citenamefont {Towell}\ \emph {et~al.}(2001)\citenamefont {Towell}
  \emph {et~al.}}]{Towell:2001nh}%
  \BibitemOpen
  \bibfield  {author} {\bibinfo {author} {\bibfnamefont {R.~S.}\ \bibnamefont
  {Towell}} \emph {et~al.} (\bibinfo {collaboration} {NuSea}),\ }\href
  {\doibase 10.1103/PhysRevD.64.052002} {\bibfield  {journal} {\bibinfo
  {journal} {Phys. Rev.}\ }\textbf {\bibinfo {volume} {D64}},\ \bibinfo {pages}
  {052002} (\bibinfo {year} {2001})},\ \Eprint
  {http://arxiv.org/abs/hep-ex/0103030} {arXiv:hep-ex/0103030 [hep-ex]}
  \BibitemShut {NoStop}%
%%CITATION = HEP-EX/0103030;%%
\bibitem [{\citenamefont {Webb}\ \emph {et~al.}(2003)\citenamefont {Webb} \emph
  {et~al.}}]{Webb:2003ps}%
  \BibitemOpen
  \bibfield  {author} {\bibinfo {author} {\bibfnamefont {J.~C.}\ \bibnamefont
  {Webb}} \emph {et~al.} (\bibinfo {collaboration} {NuSea}),\ }\href@noop {} {\
   (\bibinfo {year} {2003})},\ \Eprint {http://arxiv.org/abs/hep-ex/0302019}
  {arXiv:hep-ex/0302019 [hep-ex]} \BibitemShut {NoStop}%
%%CITATION = HEP-EX/0302019;%%
\bibitem [{\citenamefont {Abe}\ \emph {et~al.}(1996)\citenamefont {Abe} \emph
  {et~al.}}]{Abe:1996us}%
  \BibitemOpen
  \bibfield  {author} {\bibinfo {author} {\bibfnamefont {F.}~\bibnamefont
  {Abe}} \emph {et~al.} (\bibinfo {collaboration} {CDF}),\ }\href {\doibase
  10.1103/PhysRevLett.77.2616} {\bibfield  {journal} {\bibinfo  {journal}
  {Phys. Rev. Lett.}\ }\textbf {\bibinfo {volume} {77}},\ \bibinfo {pages}
  {2616} (\bibinfo {year} {1996})}\BibitemShut {NoStop}%
%%CITATION = PRLTA,77,2616;%%
\bibitem [{\citenamefont {Acosta}\ \emph {et~al.}(2005)\citenamefont {Acosta}
  \emph {et~al.}}]{Acosta:2005ud}%
  \BibitemOpen
  \bibfield  {author} {\bibinfo {author} {\bibfnamefont {D.}~\bibnamefont
  {Acosta}} \emph {et~al.} (\bibinfo {collaboration} {CDF}),\ }\href {\doibase
  10.1103/PhysRevD.71.051104} {\bibfield  {journal} {\bibinfo  {journal} {Phys.
  Rev.}\ }\textbf {\bibinfo {volume} {D71}},\ \bibinfo {pages} {051104}
  (\bibinfo {year} {2005})},\ \Eprint {http://arxiv.org/abs/hep-ex/0501023}
  {arXiv:hep-ex/0501023 [hep-ex]} \BibitemShut {NoStop}%
%%CITATION = HEP-EX/0501023;%%
\bibitem [{\citenamefont {Abazov}\ \emph
  {et~al.}(2008{\natexlab{a}})\citenamefont {Abazov} \emph
  {et~al.}}]{Abazov:2007pm}%
  \BibitemOpen
  \bibfield  {author} {\bibinfo {author} {\bibfnamefont {V.~M.}\ \bibnamefont
  {Abazov}} \emph {et~al.} (\bibinfo {collaboration} {D0}),\ }\href {\doibase
  10.1103/PhysRevD.77.011106} {\bibfield  {journal} {\bibinfo  {journal} {Phys.
  Rev.}\ }\textbf {\bibinfo {volume} {D77}},\ \bibinfo {pages} {011106}
  (\bibinfo {year} {2008}{\natexlab{a}})},\ \Eprint
  {http://arxiv.org/abs/0709.4254} {arXiv:0709.4254 [hep-ex]} \BibitemShut
  {NoStop}%
%%CITATION = ARXIV:0709.4254;%%
\bibitem [{\citenamefont {Aaij}\ \emph {et~al.}(2012)\citenamefont {Aaij} \emph
  {et~al.}}]{Aaij:2012vn}%
  \BibitemOpen
  \bibfield  {author} {\bibinfo {author} {\bibfnamefont {R.}~\bibnamefont
  {Aaij}} \emph {et~al.} (\bibinfo {collaboration} {LHCb}),\ }\href {\doibase
  10.1007/JHEP06(2012)058} {\bibfield  {journal} {\bibinfo  {journal} {JHEP}\
  }\textbf {\bibinfo {volume} {06}},\ \bibinfo {pages} {058} (\bibinfo {year}
  {2012})},\ \Eprint {http://arxiv.org/abs/1204.1620} {arXiv:1204.1620
  [hep-ex]} \BibitemShut {NoStop}%
%%CITATION = ARXIV:1204.1620;%%
\bibitem [{\citenamefont {Abazov}\ \emph
  {et~al.}(2008{\natexlab{b}})\citenamefont {Abazov} \emph
  {et~al.}}]{Abazov:2006gs}%
  \BibitemOpen
  \bibfield  {author} {\bibinfo {author} {\bibfnamefont {V.~M.}\ \bibnamefont
  {Abazov}} \emph {et~al.} (\bibinfo {collaboration} {D0}),\ }\href {\doibase
  10.1016/j.physletb.2007.10.046} {\bibfield  {journal} {\bibinfo  {journal}
  {Phys. Lett.}\ }\textbf {\bibinfo {volume} {B658}},\ \bibinfo {pages} {112}
  (\bibinfo {year} {2008}{\natexlab{b}})},\ \Eprint
  {http://arxiv.org/abs/hep-ex/0608052} {arXiv:hep-ex/0608052 [hep-ex]}
  \BibitemShut {NoStop}%
%%CITATION = HEP-EX/0608052;%%
\bibitem [{\citenamefont {Aaltonen}\ \emph {et~al.}(2010)\citenamefont
  {Aaltonen} \emph {et~al.}}]{Aaltonen:2010zza}%
  \BibitemOpen
  \bibfield  {author} {\bibinfo {author} {\bibfnamefont {T.~A.}\ \bibnamefont
  {Aaltonen}} \emph {et~al.} (\bibinfo {collaboration} {CDF}),\ }\href
  {\doibase 10.1016/j.physletb.2010.06.043} {\bibfield  {journal} {\bibinfo
  {journal} {Phys. Lett.}\ }\textbf {\bibinfo {volume} {B692}},\ \bibinfo
  {pages} {232} (\bibinfo {year} {2010})},\ \Eprint
  {http://arxiv.org/abs/0908.3914} {arXiv:0908.3914 [hep-ex]} \BibitemShut
  {NoStop}%
%%CITATION = ARXIV:0908.3914;%%
\bibitem [{\citenamefont {Chatrchyan}\ \emph
  {et~al.}(2014{\natexlab{a}})\citenamefont {Chatrchyan} \emph
  {et~al.}}]{Chatrchyan:2013mza}%
  \BibitemOpen
  \bibfield  {author} {\bibinfo {author} {\bibfnamefont {S.}~\bibnamefont
  {Chatrchyan}} \emph {et~al.} (\bibinfo {collaboration} {CMS}),\ }\href
  {\doibase 10.1103/PhysRevD.90.032004} {\bibfield  {journal} {\bibinfo
  {journal} {Phys. Rev.}\ }\textbf {\bibinfo {volume} {D90}},\ \bibinfo {pages}
  {032004} (\bibinfo {year} {2014}{\natexlab{a}})},\ \Eprint
  {http://arxiv.org/abs/1312.6283} {arXiv:1312.6283 [hep-ex]} \BibitemShut
  {NoStop}%
%%CITATION = ARXIV:1312.6283;%%
\bibitem [{\citenamefont {Chatrchyan}\ \emph {et~al.}(2012)\citenamefont
  {Chatrchyan} \emph {et~al.}}]{Chatrchyan:2012xt}%
  \BibitemOpen
  \bibfield  {author} {\bibinfo {author} {\bibfnamefont {S.}~\bibnamefont
  {Chatrchyan}} \emph {et~al.} (\bibinfo {collaboration} {CMS}),\ }\href
  {\doibase 10.1103/PhysRevLett.109.111806} {\bibfield  {journal} {\bibinfo
  {journal} {Phys. Rev. Lett.}\ }\textbf {\bibinfo {volume} {109}},\ \bibinfo
  {pages} {111806} (\bibinfo {year} {2012})},\ \Eprint
  {http://arxiv.org/abs/1206.2598} {arXiv:1206.2598 [hep-ex]} \BibitemShut
  {NoStop}%
%%CITATION = ARXIV:1206.2598;%%
\bibitem [{\citenamefont {Aad}\ \emph {et~al.}(2012{\natexlab{a}})\citenamefont
  {Aad} \emph {et~al.}}]{Aad:2011dm}%
  \BibitemOpen
  \bibfield  {author} {\bibinfo {author} {\bibfnamefont {G.}~\bibnamefont
  {Aad}} \emph {et~al.} (\bibinfo {collaboration} {ATLAS}),\ }\href {\doibase
  10.1103/PhysRevD.85.072004} {\bibfield  {journal} {\bibinfo  {journal} {Phys.
  Rev.}\ }\textbf {\bibinfo {volume} {D85}},\ \bibinfo {pages} {072004}
  (\bibinfo {year} {2012}{\natexlab{a}})},\ \Eprint
  {http://arxiv.org/abs/1109.5141} {arXiv:1109.5141 [hep-ex]} \BibitemShut
  {NoStop}%
%%CITATION = ARXIV:1109.5141;%%
\bibitem [{\citenamefont {Abazov}\ \emph {et~al.}(2015)\citenamefont {Abazov}
  \emph {et~al.}}]{D0:2014kma}%
  \BibitemOpen
  \bibfield  {author} {\bibinfo {author} {\bibfnamefont {V.~M.}\ \bibnamefont
  {Abazov}} \emph {et~al.} (\bibinfo {collaboration} {D0}),\ }\href {\doibase
  10.1103/PhysRevD.91.032007, 10.1103/PhysRevD.91.079901} {\bibfield  {journal}
  {\bibinfo  {journal} {Phys. Rev.}\ }\textbf {\bibinfo {volume} {D91}},\
  \bibinfo {pages} {032007} (\bibinfo {year} {2015})},\ \bibinfo {note}
  {[Erratum: Phys. Rev.D91,no.7,079901(2015)]},\ \Eprint
  {http://arxiv.org/abs/1412.2862} {arXiv:1412.2862 [hep-ex]} \BibitemShut
  {NoStop}%
%%CITATION = ARXIV:1412.2862;%%
\bibitem [{\citenamefont {Aaltonen}\ \emph {et~al.}(2008)\citenamefont
  {Aaltonen} \emph {et~al.}}]{Aaltonen:2008eq}%
  \BibitemOpen
  \bibfield  {author} {\bibinfo {author} {\bibfnamefont {T.}~\bibnamefont
  {Aaltonen}} \emph {et~al.} (\bibinfo {collaboration} {CDF}),\ }\href
  {\doibase 10.1103/PhysRevD.79.119902, 10.1103/PhysRevD.78.052006} {\bibfield
  {journal} {\bibinfo  {journal} {Phys. Rev.}\ }\textbf {\bibinfo {volume}
  {D78}},\ \bibinfo {pages} {052006} (\bibinfo {year} {2008})},\ \bibinfo
  {note} {[Erratum: Phys. Rev.D79,119902(2009)]},\ \Eprint
  {http://arxiv.org/abs/0807.2204} {arXiv:0807.2204 [hep-ex]} \BibitemShut
  {NoStop}%
%%CITATION = ARXIV:0807.2204;%%
\bibitem [{\citenamefont {Abazov}\ \emph
  {et~al.}(2008{\natexlab{c}})\citenamefont {Abazov} \emph
  {et~al.}}]{Abazov:2008ae}%
  \BibitemOpen
  \bibfield  {author} {\bibinfo {author} {\bibfnamefont {V.~M.}\ \bibnamefont
  {Abazov}} \emph {et~al.} (\bibinfo {collaboration} {D0}),\ }\href {\doibase
  10.1103/PhysRevLett.101.062001} {\bibfield  {journal} {\bibinfo  {journal}
  {Phys. Rev. Lett.}\ }\textbf {\bibinfo {volume} {101}},\ \bibinfo {pages}
  {062001} (\bibinfo {year} {2008}{\natexlab{c}})},\ \Eprint
  {http://arxiv.org/abs/0802.2400} {arXiv:0802.2400 [hep-ex]} \BibitemShut
  {NoStop}%
%%CITATION = ARXIV:0802.2400;%%
\bibitem [{\citenamefont {Aad}\ \emph {et~al.}(2012{\natexlab{b}})\citenamefont
  {Aad} \emph {et~al.}}]{Aad:2011fc}%
  \BibitemOpen
  \bibfield  {author} {\bibinfo {author} {\bibfnamefont {G.}~\bibnamefont
  {Aad}} \emph {et~al.} (\bibinfo {collaboration} {ATLAS}),\ }\href {\doibase
  10.1103/PhysRevD.86.014022} {\bibfield  {journal} {\bibinfo  {journal} {Phys.
  Rev.}\ }\textbf {\bibinfo {volume} {D86}},\ \bibinfo {pages} {014022}
  (\bibinfo {year} {2012}{\natexlab{b}})},\ \Eprint
  {http://arxiv.org/abs/1112.6297} {arXiv:1112.6297 [hep-ex]} \BibitemShut
  {NoStop}%
%%CITATION = ARXIV:1112.6297;%%
\bibitem [{\citenamefont {Chatrchyan}\ \emph {et~al.}(2013)\citenamefont
  {Chatrchyan} \emph {et~al.}}]{Chatrchyan:2012bja}%
  \BibitemOpen
  \bibfield  {author} {\bibinfo {author} {\bibfnamefont {S.}~\bibnamefont
  {Chatrchyan}} \emph {et~al.} (\bibinfo {collaboration} {CMS}),\ }\href
  {\doibase 10.1103/PhysRevD.87.112002, 10.1103/PhysRevD.87.119902} {\bibfield
  {journal} {\bibinfo  {journal} {Phys. Rev.}\ }\textbf {\bibinfo {volume}
  {D87}},\ \bibinfo {pages} {112002} (\bibinfo {year} {2013})},\ \bibinfo
  {note} {[Erratum: Phys. Rev.D87,no.11,119902(2013)]},\ \Eprint
  {http://arxiv.org/abs/1212.6660} {arXiv:1212.6660 [hep-ex]} \BibitemShut
  {NoStop}%
%%CITATION = ARXIV:1212.6660;%%
\bibitem [{\citenamefont {Aaij}\ \emph
  {et~al.}(2015{\natexlab{a}})\citenamefont {Aaij} \emph
  {et~al.}}]{Aaij:2015gna}%
  \BibitemOpen
  \bibfield  {author} {\bibinfo {author} {\bibfnamefont {R.}~\bibnamefont
  {Aaij}} \emph {et~al.} (\bibinfo {collaboration} {LHCb}),\ }\href {\doibase
  10.1007/JHEP08(2015)039} {\bibfield  {journal} {\bibinfo  {journal} {JHEP}\
  }\textbf {\bibinfo {volume} {08}},\ \bibinfo {pages} {039} (\bibinfo {year}
  {2015}{\natexlab{a}})},\ \Eprint {http://arxiv.org/abs/1505.07024}
  {arXiv:1505.07024 [hep-ex]} \BibitemShut {NoStop}%
%%CITATION = ARXIV:1505.07024;%%
\bibitem [{\citenamefont {Aaij}\ \emph
  {et~al.}(2015{\natexlab{b}})\citenamefont {Aaij} \emph
  {et~al.}}]{Aaij:2015vua}%
  \BibitemOpen
  \bibfield  {author} {\bibinfo {author} {\bibfnamefont {R.}~\bibnamefont
  {Aaij}} \emph {et~al.} (\bibinfo {collaboration} {LHCb}),\ }\href {\doibase
  10.1007/JHEP05(2015)109} {\bibfield  {journal} {\bibinfo  {journal} {JHEP}\
  }\textbf {\bibinfo {volume} {05}},\ \bibinfo {pages} {109} (\bibinfo {year}
  {2015}{\natexlab{b}})},\ \Eprint {http://arxiv.org/abs/1503.00963}
  {arXiv:1503.00963 [hep-ex]} \BibitemShut {NoStop}%
%%CITATION = ARXIV:1503.00963;%%
\bibitem [{\citenamefont {Aad}\ \emph {et~al.}(2014)\citenamefont {Aad} \emph
  {et~al.}}]{Aad:2014xaa}%
  \BibitemOpen
  \bibfield  {author} {\bibinfo {author} {\bibfnamefont {G.}~\bibnamefont
  {Aad}} \emph {et~al.} (\bibinfo {collaboration} {ATLAS}),\ }\href {\doibase
  10.1007/JHEP09(2014)145} {\bibfield  {journal} {\bibinfo  {journal} {JHEP}\
  }\textbf {\bibinfo {volume} {09}},\ \bibinfo {pages} {145} (\bibinfo {year}
  {2014})},\ \Eprint {http://arxiv.org/abs/1406.3660} {arXiv:1406.3660
  [hep-ex]} \BibitemShut {NoStop}%
%%CITATION = ARXIV:1406.3660;%%
\bibitem [{\citenamefont {Khachatryan}\ \emph {et~al.}(2016)\citenamefont
  {Khachatryan} \emph {et~al.}}]{Khachatryan:2016pev}%
  \BibitemOpen
  \bibfield  {author} {\bibinfo {author} {\bibfnamefont {V.}~\bibnamefont
  {Khachatryan}} \emph {et~al.} (\bibinfo {collaboration} {CMS}),\ }\href
  {\doibase 10.1140/epjc/s10052-016-4293-4} {\bibfield  {journal} {\bibinfo
  {journal} {Eur. Phys. J.}\ }\textbf {\bibinfo {volume} {C76}},\ \bibinfo
  {pages} {469} (\bibinfo {year} {2016})},\ \Eprint
  {http://arxiv.org/abs/1603.01803} {arXiv:1603.01803 [hep-ex]} \BibitemShut
  {NoStop}%
%%CITATION = ARXIV:1603.01803;%%
\bibitem [{\citenamefont {Aaij}\ \emph {et~al.}(2016)\citenamefont {Aaij} \emph
  {et~al.}}]{Aaij:2015zlq}%
  \BibitemOpen
  \bibfield  {author} {\bibinfo {author} {\bibfnamefont {R.}~\bibnamefont
  {Aaij}} \emph {et~al.} (\bibinfo {collaboration} {LHCb}),\ }\href {\doibase
  10.1007/JHEP01(2016)155} {\bibfield  {journal} {\bibinfo  {journal} {JHEP}\
  }\textbf {\bibinfo {volume} {01}},\ \bibinfo {pages} {155} (\bibinfo {year}
  {2016})},\ \Eprint {http://arxiv.org/abs/1511.08039} {arXiv:1511.08039
  [hep-ex]} \BibitemShut {NoStop}%
%%CITATION = ARXIV:1511.08039;%%
\bibitem [{\citenamefont {Aad}\ \emph {et~al.}(2016{\natexlab{a}})\citenamefont
  {Aad} \emph {et~al.}}]{Aad:2016zzw}%
  \BibitemOpen
  \bibfield  {author} {\bibinfo {author} {\bibfnamefont {G.}~\bibnamefont
  {Aad}} \emph {et~al.} (\bibinfo {collaboration} {ATLAS}),\ }\href {\doibase
  10.1007/JHEP08(2016)009} {\bibfield  {journal} {\bibinfo  {journal} {JHEP}\
  }\textbf {\bibinfo {volume} {08}},\ \bibinfo {pages} {009} (\bibinfo {year}
  {2016}{\natexlab{a}})},\ \Eprint {http://arxiv.org/abs/1606.01736}
  {arXiv:1606.01736 [hep-ex]} \BibitemShut {NoStop}%
%%CITATION = ARXIV:1606.01736;%%
\bibitem [{\citenamefont {Aad}\ \emph {et~al.}(2016{\natexlab{b}})\citenamefont
  {Aad} \emph {et~al.}}]{Aad:2015auj}%
  \BibitemOpen
  \bibfield  {author} {\bibinfo {author} {\bibfnamefont {G.}~\bibnamefont
  {Aad}} \emph {et~al.} (\bibinfo {collaboration} {ATLAS}),\ }\href {\doibase
  10.1140/epjc/s10052-016-4070-4} {\bibfield  {journal} {\bibinfo  {journal}
  {Eur. Phys. J.}\ }\textbf {\bibinfo {volume} {C76}},\ \bibinfo {pages} {291}
  (\bibinfo {year} {2016}{\natexlab{b}})},\ \Eprint
  {http://arxiv.org/abs/1512.02192} {arXiv:1512.02192 [hep-ex]} \BibitemShut
  {NoStop}%
%%CITATION = ARXIV:1512.02192;%%
\bibitem [{\citenamefont {Chatrchyan}\ \emph
  {et~al.}(2014{\natexlab{b}})\citenamefont {Chatrchyan} \emph
  {et~al.}}]{Chatrchyan:2014gia}%
  \BibitemOpen
  \bibfield  {author} {\bibinfo {author} {\bibfnamefont {S.}~\bibnamefont
  {Chatrchyan}} \emph {et~al.} (\bibinfo {collaboration} {CMS}),\ }\href
  {\doibase 10.1103/PhysRevD.90.072006} {\bibfield  {journal} {\bibinfo
  {journal} {Phys. Rev.}\ }\textbf {\bibinfo {volume} {D90}},\ \bibinfo {pages}
  {072006} (\bibinfo {year} {2014}{\natexlab{b}})},\ \Eprint
  {http://arxiv.org/abs/1406.0324} {arXiv:1406.0324 [hep-ex]} \BibitemShut
  {NoStop}%
%%CITATION = ARXIV:1406.0324;%%
\bibitem [{\citenamefont {Aad}\ \emph {et~al.}(2015)\citenamefont {Aad} \emph
  {et~al.}}]{Aad:2014vwa}%
  \BibitemOpen
  \bibfield  {author} {\bibinfo {author} {\bibfnamefont {G.}~\bibnamefont
  {Aad}} \emph {et~al.} (\bibinfo {collaboration} {ATLAS}),\ }\href {\doibase
  10.1007/JHEP02(2015)153, 10.1007/JHEP09(2015)141} {\bibfield  {journal}
  {\bibinfo  {journal} {JHEP}\ }\textbf {\bibinfo {volume} {02}},\ \bibinfo
  {pages} {153} (\bibinfo {year} {2015})},\ \bibinfo {note} {[Erratum:
  JHEP09,141(2015)]},\ \Eprint {http://arxiv.org/abs/1410.8857}
  {arXiv:1410.8857 [hep-ex]} \BibitemShut {NoStop}%
%%CITATION = ARXIV:1410.8857;%%
\bibitem [{\citenamefont {Khachatryan}\ \emph {et~al.}(2017)\citenamefont
  {Khachatryan} \emph {et~al.}}]{Khachatryan:2016mlc}%
  \BibitemOpen
  \bibfield  {author} {\bibinfo {author} {\bibfnamefont {V.}~\bibnamefont
  {Khachatryan}} \emph {et~al.} (\bibinfo {collaboration} {CMS}),\ }\href
  {\doibase 10.1007/JHEP03(2017)156} {\bibfield  {journal} {\bibinfo  {journal}
  {JHEP}\ }\textbf {\bibinfo {volume} {03}},\ \bibinfo {pages} {156} (\bibinfo
  {year} {2017})},\ \Eprint {http://arxiv.org/abs/1609.05331} {arXiv:1609.05331
  [hep-ex]} \BibitemShut {NoStop}%
%%CITATION = ARXIV:1609.05331;%%
\bibitem [{\citenamefont {Aad}\ \emph {et~al.}(2016{\natexlab{c}})\citenamefont
  {Aad} \emph {et~al.}}]{Aad:2015mbv}%
  \BibitemOpen
  \bibfield  {author} {\bibinfo {author} {\bibfnamefont {G.}~\bibnamefont
  {Aad}} \emph {et~al.} (\bibinfo {collaboration} {ATLAS}),\ }\href {\doibase
  10.1140/epjc/s10052-016-4366-4} {\bibfield  {journal} {\bibinfo  {journal}
  {Eur. Phys. J.}\ }\textbf {\bibinfo {volume} {C76}},\ \bibinfo {pages} {538}
  (\bibinfo {year} {2016}{\natexlab{c}})},\ \Eprint
  {http://arxiv.org/abs/1511.04716} {arXiv:1511.04716 [hep-ex]} \BibitemShut
  {NoStop}%
%%CITATION = ARXIV:1511.04716;%%
\end{thebibliography}%

%
% - - - - - - - - - - - - - - - - - - - - - - - - - - - - - - - - - - - - - - - - - - - - - -
%

\clearpage{}
\end{document}